\documentclass[twocolumn]{autart}    % Enable this line and disable the 
\usepackage{amsmath,amsfonts}
\usepackage{algorithmic}
\usepackage{algorithm}
\usepackage{array}
\usepackage[caption=false,font=normalsize,labelfont=sf,textfont=sf]{subfig}
\usepackage{textcomp}
\usepackage{stfloats}
\usepackage{url}
\usepackage{booktabs}
\usepackage{tabularx}
\usepackage{verbatim}
\usepackage{graphicx}
\usepackage{graphics}
\usepackage{cite}
\usepackage{nomencl}
\makenomenclature
\usepackage{enumitem}
\usepackage{comment}
\usepackage{xcolor}

\usepackage{mathtools}

\usepackage{tabstackengine}
\usepackage{scalerel}
\usepackage{stackengine}
\usepackage{accents}
\usepackage[T1]{fontenc}
\usepackage[utf8]{inputenc}
\usepackage[spanish,english]{babel}
\usepackage{hyperref}
\hypersetup{
    colorlinks,
    citecolor=blue,
    linkcolor=red,    
    urlcolor=red,
}
\newcommand{\ut}[1]{\underaccent{\tilde}{#1}}
\renewcommand{\vec}[1]{\ut{#1}}

\newcommand{\oprocendsymbol}{\hbox{$\bullet$}} % \diamond means proof not included
\newcommand{\oprocend}{\relax\ifmmode\else\unskip\hfill\fi\oprocendsymbol}

\newcommand{\new}[1]{{\color{black}#1\color{black}}} %Modification marked in black
\newcommand{\subscr}[2]{#1_{\textup{#2}}}
\newcommand{\supscr}[2]{#1^{\textup{#2}}}

\DeclareMathAlphabet{\mymathbb}{U}{BOONDOX-ds}{m}{n}
\makeatletter
\newlength{\dispup}
\newlength{\dispdn}
\newlength{\pskipcomp}
\newcommand{\pullup}[1]{\vspace{\dimexpr#1+\pskipcomp\relax}}
\g@addto@macro\normalsize{%
  \abovedisplayskip 3\p@ \@plus 2\p@ \@minus 2\p@
  \belowdisplayskip 3\p@ \@plus 2\p@ \@minus 2\p@
  \abovedisplayshortskip \z@ \@plus 2\p@
  \belowdisplayshortskip 2\p@ \@plus 2\p@ \@minus 1\p@}
\g@addto@macro\small{%
  \abovedisplayskip 3\p@ \@plus 2\p@ \@minus 2\p@
  \belowdisplayskip 3\p@ \@plus 2\p@ \@minus 2\p@
  \abovedisplayshortskip \z@ \@plus 2\p@
  \belowdisplayshortskip 2\p@ \@plus 2\p@ \@minus 1\p@}
\normalsize % re-select so the appended values take effect immediately
\allowdisplaybreaks[4]
\let\aut@thebibliography\@thebibliography
\def\@thebibliography#1{\aut@thebibliography{#1}%
  \setlength{\itemsep}{0.05\p@}\setlength{\parsep}{\z@}%
  \linespread{0.92}\selectfont}
\makeatother
\begin{document}

\begin{frontmatter}

\title{Baseline-improved Economic Model Predictive Control for Optimal Microgrid Dispatch} % Title, preferably not more 

%\thanks[footnoteinfo]{This paper was not presented at any IFAC 
%meeting. Corresponding author M.~T.~Cicero. Tel. +XXXIX-VI-mmmxxi. 
%Fax +XXXIX-VI-mmmxxv.}
\vspace{-1.5em}
\author[Paestum]{Avik Ghosh}\ead{avghosh@ucsd.edu},    % Add the 
\author[Paestum]{Adil Khurram}\ead{akhurram@ucsd.edu},               % e-mail address 
\author[Paestumb]{Cristian Cortes-Aguirre}\ead{cgcortes@uc.cl},    % Add the
\author[Paestum]{Jan Kleissl}\ead{jkleissl@ucsd.edu},  % (ead) as shown
\author[Paestum]{Sonia Mart{\'i}nez}\ead{soniamd@ucsd.edu} 

\address[Paestum]{Department of Mechanical and Aerospace Engineering, University of California, San Diego, California 92093-0411, USA}  % Please supply      
\address[Paestumb]{Solar Energy Research Center, SERC-Chile, Santiago, Chile}
% \address[Paestumb]{Department of Mechanical and Aerospace Engineering, University of California, San Diego, California 92093-0411, USA} 

                             % keyword list or with the 
                                          % help of the Automatica 
                                          % keyword wizard

\vspace{-1em}
\begin{abstract}                          % Abstract of not more than 200 words.
As opposed to stabilizing to a reference trajectory or state, Economic Model Predictive Control (EMPC) optimizes economic performance over a prediction horizon, making it particularly attractive for economic microgrid (MG) dispatch. However, as load and generation forecasts are only known $24-48$~h in advance, economically optimal steady states or periodic trajectories are not available and the EMPC-based works that rely on these signals are inadequate. In addition, demand charges, based on maximum monthly grid import power of the MG, cannot be easily cast as an additive cost, which prevents the application of the principle of optimality if introduced naively. Motivated by this, in this work we propose to close this mismatch between the EMPC prediction horizon and existing monthly timescales by means of an appropriately generated baseline reference trajectory. To do this, we first propose an EMPC formulation for a generic deterministic discrete non-linear time-varying system subject to hard state and input constraints. We then show that, under \new{appropriate terminal ingredients, like sequential control invariance of the terminal region, and existence of a terminal control law that causes a Lyapunov-like decrease of the terminal cost}, the asymptotic average economic cost of the proposed method is no worse than a baseline given by any arbitrary reference trajectory that is only known online. In particular, this results in a practical, finite-time upper bound on the average economic cost difference with the baseline that decreases linearly to zero as time goes to infinity. We then show how the proposed EMPC framework can be used to solve optimal MG dispatch problems, introducing various costs and constraints that conform to the required assumptions. By means of this framework, we conduct realistic simulations with data from the Port of San Diego MG, which demonstrate that the proposed method can reduce monthly electricity costs in closed-loop with respect to reference trajectories; which are either generated by the optimization of the electricity cost over the prediction horizon, or by tracking an ideal grid import curve.

\end{abstract}
%\clearpage
\end{frontmatter}

\section{Introduction}\label{intro}
Economic Model Predictive Control (EMPC) has recently gained popularity in the real-time optimal control of microgrids (MG) with high penetration of variable renewable energy (VRE)~\cite{hu2023economic}. Instead of stabilizing to a reference trajectory or state as in Tracking MPC, EMPC optimizes the actual economic performance over the prediction horizon. As lowering MG operating costs, i.e., electricity costs paid by the MG to the utility, is one of the primary control objectives of MG operators, EMPC becomes particularly attractive. 

There are two crucial bottlenecks to certifying and designing \emph{practically useful} controllers with provable performance guarantees for EMPC, especially for applications such as optimal MG dispatch.
\new{First, performance metrics may neither be quadratic nor positive definite, and are often not differentiable. Moreover, non-differentiability may also be inherent to the underlying nonlinear control system.}\footnote{\new{For optimal MG dispatch, the performance metric, i.e., electricity cost, is negative when an MG exports energy back to the main grid violating positivity. Moreover, the electricity cost is generally calculated over monthly windows composed of linear and max terms which violate quadratic behavior and differentiability. See more in Section~\ref{electricity_cost_structure}.}} This imposes significant difficulty in designing control invariant terminal regions, and Lyapunov-like terminal cost functions --- which are standard requirements for performance guarantees.

\new{Second, a standard requirement in EMPC formulations with performance guarantees is that, the overall performance of the system over its window of operation is encapsulated as the sum of stage costs that are a function of only the current state and control input~\cite{risbeck2019economic}. However, for many applications, this additivity is violated from the necessity of including demand charge type penalties~\cite{mo2026real}. The latter, which can constitute a substantial part of the performance metric, are levied based on the maximum value of a quantity (control input or a state) over a given period.}\footnote{For MG power dispatch, demand
charges are the cost levied for the peak load (control input) demanded by the MG
from the main grid (i.e., grid import) over a monthly window. For commercial and industrial customers, demand charge costs are typically $30-70\%$~\cite{Demand_charges}, and can get as high as $90\%$ of the overall performance metric, i.e., the monthly electricity cost~\cite{fitzgerald2017evgo,lee2020pricing}.} 
Demand charges can thus cause considerable discrepancy between the predicted open-loop MPC solutions and the actual closed-loop realizations, even under nominal (i.e., uncertainty-free) conditions, which hinders transparent controller tuning. This occurs because naively incorporating the demand charge term into the objective function violates the principle of optimality~\cite{jones2017solving}. 

\new{As there may be significant mismatches between EMPC prediction horizons and the window over which the overall performance metric is evaluated, there can be significant differences between %the performance of an 
 an infinite-horizon optimal control strategy %\footnote{That is, a one-shot optimization of monthly electricity cost using perfect forecast data over the entire month. Note that for discrete systems, as the month is finite,   For optimal MG dispatch, EMPC prediction horizons are typically between $24-48$~h, because load and VRE generation forecasts are only accurate one or two days ahead~\cite{cristian_empc}.} 
and a real-time receding horizon %closed-loop standard 
EMPC.}
This is because the former will adjust its prior demand charge type penalties to account for future demand peaks, which the finite horizon standard EMPC would be oblivious to, suggesting an avenue for improvement during real-time operation.  

\new{Thus, it is of particular importance to MG operators to devise EMPC algorithms with no-worse, and potentially better, performance guarantees over \emph{any} arbitrarily generated online reference trajectory or baseline,\footnote{An online reference trajectory
is one which is not known a-priori for all times under consideration, but continuously generated online at each time-step, while satisfying the dynamics and constraints.} where the performance metric includes non-positive, non-quadratic terms, and demand charge type penalties}.
These performance guarantees can be important because the mismatch of timescales can give rise to online reference trajectories with lower monthly electricity costs than the closed-loop standard EMPC. 
Note that notwithstanding our motivating application, we also aim to propose a generic framework suitable for analysis of EMPC applied to a wide class of discrete non-linear time varying systems subject to \new{appropriate assumptions on terminal ingredients; i.e., terminal region, control law, and cost}.

The recent literature on EMPC methods is extensive, and focuses both on asymptotic performance and stability guarantees, generally with respect to an optimal economic steady state~\cite{rawlings2012fundamentals,angeli10average,amrit2011economic}. Studies such as~\cite{rawlings2012fundamentals,angeli10average} have also looked into asymptotic stability with respect to an a-priori known optimal periodic trajectory. However, asymptotic stability guarantees require strict dissipativity assumptions with respect to an optimal economic steady state/trajectory~\cite{rawlings2012fundamentals,amrit2011economic,angeli10average,mpc_book},
which is hard to satisfy in practice, and is of less importance than economic performance~\cite{grune2013economic,mcallister2023suboptimal}, especially for MG operators. \new{Additionally, dissipativity assumptions ensuring asymptotic stability are ill-posed for our problem class where the reference trajectory is updated online and may itself be economically suboptimal.}

Thus, in this study we focus solely on studies with asymptotic performance guarantees for various EMPC formulations, as the assumptions required for asymptotic performance guarantees are milder than those required for stability, %\margin{I thought we had removed the 'mild'} Avik\s response - We just don't call our terminal ingredients mild, but in general performance guarantee requirements are milder than satbility requirements - that is true of everyone. So we can keep it here.
and are more useful for MG operators in practice. The asymptotic performance guarantees for EMPC with the existence of an optimal economic steady state or optimal periodic trajectory as explored in~\cite{rawlings2012fundamentals,amrit2011economic,angeli10average}, cannot be straightforwardly extended to EMPC for MGs because of: (i) difficulty of expressing the demand charge term in the electricity cost of MGs as an additive stage cost; %which is a prerequisite for the theoretical guarantee; 
(ii) %non-usefulness and moreover, 
the possible non-existence of an optimal economic steady state and optimal periodic trajectory respectively in MG dispatch, and, more generally, for generic time-varying non-linear systems. An optimal economic steady state is not useful for MG dispatch as directly minimizing economics usually leads to non-steady operating regimes. For example, in an MG, the battery energy storage system (BESS, where BESS state-of-charge (SOC) is the state) is expected to operate in a non-steady region, constantly charging and discharging to reduce costs. Having better economic performance with respect to a case where the state is steady~\cite{rawlings2012fundamentals,amrit2011economic,angeli10average} is trivial, and thus not \emph{practically useful}. The existence and a-priori knowledge of an optimal periodic trajectory for MG dispatch is impractical under high temporal (i.e., daily or seasonal) variability of load and VRE, and where accurate forecasts are typically known only $24-48$~h ahead~\cite{cristian_empc}. 

While the studies~\cite{rawlings2012fundamentals,amrit2011economic,angeli10average} considered stage costs that are time-invariant, the authors in~\cite{angeli2015average,angeli2016theoretical} extended the former works to prove asymptotic performance guarantees for stage costs that are time varying, but are either periodic or fulfill certain average behavior. Stage cost functions that are periodic or the existence and a-priori knowledge of an averaged stage cost function are likely to be violated for VRE dominated MGs, where real-time electricity tariffs are non-periodic, and anticipated only 24~h ahead~\cite{ambec2021real}. In addition, the studies~\cite{angeli2015average,angeli2016theoretical}, like~\cite{rawlings2012fundamentals,angeli10average}, still prove asymptotic performance guarantees with respect to an a-priori known optimal periodic reference trajectory, retaining its restrictiveness.

Gr{\"u}ne~\cite{grune2013economic} provided approximate optimal performance guarantees for both the infinite-horizon and transient (i.e., finite-time) closed-loop operation of a receding horizon EMPC. The result is powerful and is achieved without setting a terminal constraint to the EMPC, contrary to~\cite{rawlings2012fundamentals,amrit2011economic,angeli10average,angeli2015average,angeli2016theoretical}, which may increase the feasible region of operation of the controller. However, the sufficient conditions provided in~\cite{grune2013economic} %for ensuring the performance guarantees 
require strict dissipativity and controllability conditions that imply optimal steady state operation (which further implies the turnpike property), which %as discussed before, 
is unrealistic for MGs. %and economically oriented non-linear systems in general.      

None of the above works~\cite{rawlings2012fundamentals,amrit2011economic,grune2013economic, angeli10average,angeli2015average,angeli2016theoretical} consider the system dynamics to be time-varying making it unsuitable for handling demand charges in MGs, as certain demand charges are based on the time-of-day when peaks are reached.\footnote{See Section~\ref{augmeted_sys_NCDC_OPDC} for details.} The authors in~\cite{grune2017closed} extend the work in~\cite{grune2013economic} to encompass time-varying system dynamics. This work resorted to the time-varying turnpike property notion and the continuity of the value function to provide a bound on the deviation of the finite-horizon closed-loop EMPC cost from the infinite horizon optimal control trajectory, which is assumed to exist but unknown. The time-varying turnpike property assumption guarantees that the solutions to both problems remain close most of the time. However, in MGs with demand charges, an infinite-horizon optimal trajectory is significantly different from a finite-horizon EMPC one, violating the turnpike property.  

More recently, the authors in~\cite{risbeck2019economic} propose an EMPC framework for generic non-linear time-varying systems with asymptotic performance guarantees that are no worse than that of an arbitrary reference trajectory. The authors also showed the efficacy of their method for optimal BESS dispatch for demand charge reduction in an MG through a case study where an optimal periodic reference trajectory was assumed. However, the reference trajectory needs to be known a-priori for all future times, making the method impractical for most realistic systems involving forecasts. As an MG always needs to maintain a power balance with the main grid, it operates under a limited forecast of load and VRE generation (if applicable) of $24-48$~h ahead. 
In contrast, the authors in~\cite{risbeck2019economic} assume that forecasts are available months or weeks in advance to fix the reference trajectory a-priori. Thus, further realistic theoretical analyses, in addition to realistic case studies, are required to encourage the design and implementation of EMPC algorithms by MG operators.

The present work generalizes the asymptotic performance guarantees in~\cite{risbeck2019economic} with respect to an arbitrary reference trajectory which is known online only until the current time-step. The proposed EMPC framework is then leveraged to solve an optimal MG dispatch problem with both energy and demand charges incorporating BESS losses.

\section{Contributions}\label{contributions}

\new{The contributions of the present work are twofold. The stated contributions~1 and~2 are of a theoretical nature, and pertain to the proposed EMPC framework for a generic deterministic non-linear time-varying system; while the contributions~3 and~4 are application specific, and concern the particularization of this framework to the optimal MG dispatch problem. They are summarized as follows.}

\begin{enumerate}
    \item Under appropriate assumptions on terminal ingredients, i.e., terminal region, control law and cost, we first provide 
    an upper bound on the asymptotic average performance of the proposed scheme, which guarantees economic improvement over the baseline provided by a reference trajectory measured online.  
    We then relax this result to a practical finite-time guarantee, with an upper bound that decreases linearly to zero over time. %Our assumptions are milder because they only rely on information about the baseline until the current time, reducing dependency on long term accurate forecasts. 
    Our formulation allows the baseline to be continuously updated online at each time-step, thereby implicitly improving our scheme's performance.

    \item \new {We propose a novel constructive design of the terminal cost function and provide a checkable sufficient condition for ensuring the above asymptotic/practical performance guarantee.}
    
    \item For MG power dispatch, we convert the one-time demand charge component of the MG monthly electricity costs into an additive stage cost, \new{thereby restoring the principle of optimality. Then we provide several designs of terminal ingredients, certifying the above theoretical performance guarantee for our application.} %In particular, this method avoids unnecessary reductions of the admissible state-control input sequence for the problem, and can be used to generate other variations of terminal regions, control laws, and cost functions. This helps us further improve the finite-time performance of the scheme.
    
    \item By exploiting insights from a realistic demand charge rate structure in MGs, we identify online reference trajectories with superior performance in economic objectives than the standard EMPC. These references are then used in our proposed EMPC scheme to improve the system closed-loop performance. In this way, our work introduces an effective mechanism to reduce the MG monthly electricity costs, by considering prediction horizons that are much smaller than a month.

\end{enumerate}

The rest of the paper is organized as follows. Section~\ref{notations} presents the notations used in this study. Section~\ref{problem_form} introduces the receding horizon EMPC problem formulation, and provides the related assumptions, performance guarantee, and the design of the terminal cost function. Section~\ref{MG_dispatch} presents the optimal MG dispatch problem and its reformulation into the EMPC structure of Section~\ref{problem_form}. Section~\ref{case_study} presents the case study for a realistic grid-connected MG with PV, load, and BESS, with results and discussions in Section~\ref{results}. Section~\ref{conclusions} concludes the paper summarizing the takeaways of the study and laying down possible future directions of work.

\section{Notations}\label{notations}
%\subsection{Notations}\label{notations}

The following is common notation employed throughout the manuscript. The real vector space of dimension $n$ is denoted by $\mathbb{R}^{n}$. Similarly, the sets of nonnegative and non-positive real numbers are represented by $\mathbb{R}_{\geq 0}$ and $\mathbb{R}_{\leq 0}$, respectively. The set of natural numbers including 0 is denoted by $\mathbb{N}$, the set of consecutive natural numbers $\{i,i+1,\ldots,j\}$  by $\mathbb{N}_{i}^{j}$, and the set of all natural numbers greater or equal to $k\in \mathbb{N}$ by $\mathbb{N}_{\geq k}$. The $n$-dimensional vector of zeros and ones is represented by $\mymathbb{0}_n$ and $\mymathbb{1}_n$, respectively. 
For matrices $A$ and $B$ of equal dimensions, the operators $\{<,\leq,=,>,\geq\}$ are understood to hold component wise. The $\supscr{i}{th}$ row, and the element from the $\supscr{i}{th}$ row and $\supscr{j}{th}$ column of a matrix $A$ are denoted by $A_{i}$ and $A_{ij}$, respectively, while the $\supscr{i}{th}$ element of a vector $x$ is marked as $x_{i}$, unless mentioned otherwise. For a vector $x$, $|x|$ and $\left\|x\right\|_2$ denote its 1-norm and 2-norm, respectively. 

\section{Problem formulation}\label{problem_form}

This section introduces the problem setup consisting of the receding horizon EMPC problem we aim to solve, the baseline reference trajectory whose performance we seek to improve upon, the assumptions required to do so, and lastly the associated proof of performance.

\subsection{System description}\label{original_system}

We consider discrete time-varying systems governed by
\begin{equation} \label{dynamics_real}
\begin{aligned}
&x(t+1) = f(x(t),u(t),t), \qquad \forall t,
\end{aligned}
\end{equation}
where the state $x(t)$ at discrete time-step $t \in \mathcal{T} \subseteq \mathbb{N}$, 
belongs to a set $\mathcal{X}(t)$, 
that is, $x(t) \in \mathcal{X}(t)$. % $\mathcal{T}\subseteq\mathbb{N}$ is the set of discrete time-steps.  
Similarly, for control inputs, $u(t) \in \mathcal{U}(t)$. The composite sets $\mathcal{X}$ and $\mathcal{U}$ are defined by $\mathcal{X}:=\bigcup\limits_{t \in \mathcal{T}}\mathcal{X}(t)$ and $\mathcal{U}:=\bigcup\limits_{t \in \mathcal{T}}\mathcal{U}(t)$, which overall implies $f:\mathcal{X} \times \mathcal{U} \times \mathcal{T} \to \mathcal{X}$.

We denote the solution to~\eqref{dynamics_real} over a horizon $N$ starting from an initial state $x_0$ and time $t_0$\footnote{In the sequel, time and (discrete) time-step are used interchangeably for the ease of exposition.}; i.e.,~$x(t_0)=x_0\in \mathcal{X}(t_0)$, 
 under the control sequence parameterized by $x_0$ and $t_0$; i.e.,~$\mathbf {u}(x_0,t_0)\!=\!\bigl( u(t_0;(x_0,t_0)), u(t_0+1;(x_0,t_0)), \ldots, u(t_0+N-1;(x_0,t_0)) \bigr)^\top$, as the state sequence $\mathbf{x}(\mathbf{u},x_0,t_0)=\bigl( x_0, x(t_0+1;(x_0,t_0)), \ldots, x(t_0+N;(x_0,t_0)) \bigr)^\top$. %\footnote{Note that to avoid confusion, a state and control input sequence are denoted by boldface $\textbf{\text x}$ and $\textbf{\text u}$, respectively. The state and control input sequences are composed of a collection of individual state and control inputs varying over time. The individual state and control inputs at time $t$ are denoted by non-boldface $x(t)$ and $u(t)$ respectively.} 
Hereon, we simplify $u(t_0+k;(x_0,t_0))$ as $u(t_0+k), \; \forall k\in \mathbb{N}$, and, similarly, $x(t_0+k;(x_0,t_0))$ by $x(t_0+k),  \; \forall k\in \mathbb{N}$.

Let an arbitrary solution of~\eqref{dynamics_real} be a  reference trajectory, which evolves according to the dynamics
\begin{equation} \label{dynamics_reference}
\begin{aligned}
&x^{r}(t+1) = f(x^{r}(t),u^{r}(t),t), \qquad \forall t,
\end{aligned}
\end{equation}
where $x^{r}(t) \in \mathcal{X}(t)$ and $u^{r}(t) \in \mathcal{U}(t)$. This reference trajectory 
is  known from $x^r(t_0)$ up to and including $x^r(t+1)$ through~\eqref{dynamics_reference}; however, $u^{r}(t+1)$ is not available. %\footnote{Note that at time $t$, although $x^{r}(t+1)$ is known, $u^{r}(t+1)$ is unknown.}
This reference trajectory can be generated %offline/
online from a separate optimization or a rule-based method based on operator domain knowledge. 

\subsection{Receding horizon economic MPC}\label{empc}
Given a receding horizon $N$, at each $t$ we solve an economic MPC problem from state $x(t)$ given by
\small{\setlength{\abovedisplayskip}{-0em}
\setlength{\belowdisplayskip}{0pt}
\setlength{\abovedisplayshortskip}{0pt}
\setlength{\belowdisplayshortskip}{0pt}
\setlength{\jot}{0pt}
\begin{subequations}\label{empc_general}
\begin{align}
\label{objective_empc}
V_N^0(x(t),t) &= \mathop{\text{minimize}}\limits_{\mathbf{u}(x(t),t) \in \mathcal{U}^{N}}\, 
                 V_N\!\big(x(t),\mathbf{u}(x(t),t),t\big) \nonumber\\
&= \mathop{\text{minimize}}\limits_{\mathbf{u}(x(t),t) \in \mathcal{U}^{N}}\,
   \sum_{k=0}^{N-1} l\!\big(x(t\!+\!k|t),u(t\!+\!k|t),t\!+\!k\big) \nonumber \\
   &\qquad \qquad \qquad + V_f\!\big(x(t\!+\!N|t),t\!+\!N\big),\end{align}
\text{subject to} \\
\begin{align}
\label{dynamics_empc}
&x(t\!+\!k\!+\!1|t) = f\!\big(x(t\!+\!k|t), u(t+k|t), t\!+\!k\big),
\nonumber \\ & \qquad\qquad\qquad\qquad\qquad\quad\;\;\;\;\;\;\forall\, 
k \in \mathbb{N}_{0}^{N-1}, %\\ %\forall\, t \in \mathcal{T},
\end{align}%\vspace{-3em}
\begin{align}
\label{state_empc}
\quad x(t\!+\!k|t) &\in \mathcal{X}(t{+}k),
\quad \;\;\;\forall\, k \in \mathbb{N}_{0}^{N-1},\\ %\forall\, t \in \mathcal{T},\\
\label{input_empc}
\quad u(t\!+\!k|t) &\in \mathcal{U}(t{+}k),
\qquad\forall\, k \in \mathbb{N}_{0}^{N-1},\\ %\forall\, t \in \mathcal{T},\\
\label{terminal_state_empc}
\quad x(t\!+\!N|t) &\in \mathcal{X}_f(t{+}N)\subseteq \mathcal{X}(t{+}N),
\quad %\;\;\;\forall\, t \in \mathcal{T},
\end{align}
\end{subequations}}

\normalsize
where $V_N:\mathcal{X} \times \mathcal{U}^N \times \mathcal{T} \to \mathbb{R}$, 
consists of the additive economic stage cost function $l:\mathcal{X} \times \mathcal{U} \times \mathcal{T} \to \mathbb{R}$, and the terminal cost function $V_f:\mathcal{X}_f \times \mathcal{T} \to \mathbb{R}$. The composite terminal constraint set is defined by $\mathcal{X}_f:=\bigcup\limits_{t \in \mathcal{T}}\mathcal{X}_f(t)$.\footnote{Note that from the EMPC problem setup~\eqref{empc_general}, $\mathcal{X}_f(t)$ can be empty, if $t\in \mathbb{N}_{0}^{N-1}$.} 
The optimal control sequence from solving~\eqref{empc_general} is given by $\mathbf {u}^0(x(t),t)=\bigl( u^0(t|t), u^0(t\!+\!1|t), \ldots, u^0(t\!+\!N\!-\!1|t) \bigr)^\top$. Similarly, the optimal state sequence after applying the optimal control sequence is given by $\mathbf{x}(\mathbf {u}^0,x(t),t)=\bigl( x(t), x^0(t+1|t), \ldots, x^0(t+N-1|t), x^0(t+N|t) \bigr)^\top$. The closed-loop control law is given by the first element of $\mathbf{ u}^0(x(t),t)$, represented as $\kappa_N(x(t),t):=u^0(t|t)$ which is applied to the system at time $t$. The closed-loop system is thus represented by 
\begin{equation} \label{closed_loop_empc}
\begin{aligned}
x(t+1)=f(x(t),\kappa_N(x(t),t),t), \; \forall t \in \mathcal{T},
\end{aligned}
\end{equation}
where $\kappa_N:\mathcal{X}\times \mathcal{T} \to \mathcal{U}$, and after $x(t+1)$ is calculated from~\eqref{closed_loop_empc}, the optimization in~\eqref{empc_general} is repeated from the updated state $x(t+1)$, and time $t+1$.

\begin{assum}\label{assumption_continuity}\emph{(\textbf{Continuity of system and cost}).
{The functions $f$, $l$ and $V_f$ are continuous.}\oprocend} 
\end{assum}

\begin{assum}\label{assumption_sets}\emph{(\textbf{Properties of constraint sets}).
{For each $t\in \mathcal{T}$, $\mathcal{X}(t)$ is closed, and $\mathcal{U}(t)$ is compact. For each $t \in \mathcal{T}$, $\mathcal{X}_f(t)$ is compact. The reference is feasible; i.e.~$x^r(t) \in \mathcal{X}(t)$ and $u^r(t) \in \mathcal{U}(t)$. %For each $t\in \mathbb{T}$, $\mathbb{U}(t)$ is uniformly bounded.
The composite set $\mathcal{U}$ is bounded. }\oprocend} 
\end{assum}

\begin{rem}\label{remark_existence}\emph{(\textbf{Existence of a solution of the optimal control problem}). 
{Under Assumptions~\ref{assumption_continuity} and~\ref{assumption_sets}, a solution to~\eqref{empc_general} exists for $x(t) \in \mathcal{X}(t),\; \forall t\in \mathcal{T}$, if~\eqref{dynamics_empc},~\eqref{state_empc},~\eqref{input_empc} and~\eqref{terminal_state_empc} are satisfied over the prediction horizon $N$ \cite[Proposition 2.29]{mpc_book}.} \oprocend } 
\end{rem}

\begin{assum}\label{assumption_terminal_cost}\emph{(\textbf{Properties of the terminal cost, region, and control law}).
{For each $t \in \mathcal{T}$, if $x(t) \in \mathcal{X}_f(t)$, there exists a terminal control law $\kappa_f:\mathcal{X}_f\times \mathcal{T} \to \mathcal{U}$, such that the terminal cost follows a Lyapunov-like adjacent difference condition. Specifically,}
\small{
\setlength{\abovedisplayskip}{\dispup}
\setlength{\belowdisplayskip}{0pt}
\setlength{\abovedisplayshortskip}{0em}
\setlength{\belowdisplayshortskip}{0em}
\setlength{\jot}{0pt}
\begin{align}
&V_f\bigl(f(x(t),\kappa_f(x(t),t),t),t+1\bigr)-V_f(x(t),t) \nonumber
\\&\leq \!\!-l\bigl(x(t),\kappa_f(x(t),t),t\bigr)\!+\!l\bigl(x^{r}(t\!-\!N),u^{r}(t\!-\!N),t\!-\!N\bigr). \label{eq:adjacent_diff_lyapunov}
\end{align} 
}\oprocend}
\end{assum}
\normalsize
Note that Assumption~\ref{assumption_terminal_cost} implies that, applying the terminal control law $\kappa_f$ to $x(t) \in \mathcal{X}_f(t)$, leads to $f\bigl(x(t),\kappa_f(x(t),t),t\bigr) \in \mathcal{X}_f(t+1), \forall t\in \mathcal{T}$. In other words, the sequence of sets $\mathcal{X}_f(t), \forall t \in \mathcal{T}$ is sequentially control invariant for~\eqref{dynamics_real}~\cite[Definition 2.27]{mpc_book}.

\begin{assum}\label{assumption_lower_bound}\emph{(\textbf{Lower boundedness of stage and terminal cost}).
{The function $l$ is uniformly bounded from below for $\bigl(x(t),u(t),t\bigr) \in \mathcal{X} \times \mathcal{U} \times \mathcal{T}$. Similarly, the function  $V_f$ is uniformly bounded from below for $\bigl(x(t),t\bigr) \in \mathcal{X}_f \times \mathcal{T}$.}\oprocend}
\end{assum}

\begin{thm}\label{theorem_asymp_cost}(\textbf{Asymptotic average cost guarantee of the receding horizon EMPC}).
{Let Assumptions~\ref{assumption_continuity},~\ref{assumption_sets},~\ref{assumption_terminal_cost}, and~\ref{assumption_lower_bound} hold. The asymptotic average economic cost of the closed-loop system given by~\eqref{closed_loop_empc} is no worse than the asymptotic average economic cost of the reference trajectory given by~\eqref{dynamics_reference}. Specifically,
%\vspace{-1em}
\small{
\setlength{\abovedisplayskip}{\dispup}
\setlength{\belowdisplayskip}{0em}
\setlength{\abovedisplayshortskip}{\dispup}
\setlength{\belowdisplayshortskip}{\dispdn}
\setlength{\jot}{0pt}
\begin{align*}
\limsup_{T\to\infty}\!\!\sum_{t=t_0}^{t_0+T-1} \frac{1}{T}\biggl(l\bigl(x(t),u(t),t\bigr)\!-\!l\bigl(x^{r}(t),u^{r}(t),t\bigr)\biggr)\!\leq\!0,
\end{align*}}
\normalsize
where $u(t)=\kappa_N(x(t),t), \; \forall t \in \mathcal{T}$.}
\end{thm}
\begin{pf}
From Remark~\ref{remark_existence}, %and~Assumption~\ref{assumption_uniqueness}, 
we know that at time $t$, an optimal solution to~\eqref{empc_general} exists and is given by $\mathbf{u}^0(x(t),t)$, with $V_N^0(x(t),t) = V_N(x(t),{\mathbf{u}^0(x(t),t)},t)$. A feasible control sequence at time $t+1$ is given by the final $N-1$ control inputs of $\mathbf{u}^0(x(t),t)$, and appending a terminal control input by following Assumption~\ref{assumption_terminal_cost}, as $x(t\!+\!N|t) \in \mathcal{X}_f(t+N)$. Specifically, after  $\kappa_N(x(t),t)$ is applied to $x(t)$, which makes the system state evolve to $x(t+1)$, a feasible control sequence at $t+1$ for~\eqref{empc_general} is given by $\mathbf{u}(x(t\!+\!1),t\!+\!1)=\bigl(u^0(t\!+\!1|t), \ldots, u^0(t\!+\!N\!-\!1|t), \kappa_f(x^0(t+N|t),t+N) \bigr)^\top$. %For conciseness of presentation, henceforth denote $x^0(t+N|t; (x(t),t))=x^0(t+N|t)$. 
As the optimal cost is a lower bound of any feasible cost at any time, we conclude $V_N^0(x(t\!+\!1),t\!+\!1)\leq V_N(x(t\!+\!1),{\mathbf{u}(x(t\!+\!1),t\!+\!1)},t+1)$. Thus, it follows that,

\small{\setlength{\abovedisplayskip}{\dispup}
\setlength{\belowdisplayskip}{0pt}
\setlength{\abovedisplayshortskip}{0pt}
\setlength{\belowdisplayshortskip}{0pt}
\setlength{\jot}{0pt}
\begin{align} \label{succesive_substraction}
&V_N^0(x(t\!+\!1),t\!+\!1)-V_N^0(x(t),t) \nonumber \\ 
&\leq V_N(x(t\!+\!1),{\mathbf{u}(x(t\!+\!1),t\!+\!1)},t+1) - V_N^0(x(t),t) \nonumber \\
&=l(x^0(t+N|t), \kappa_f(x^0(t+N|t),t\!+\!N), t\!+\!N)\nonumber\\
& \quad+\!V_f\bigl(f(x^0(t\!+\!N|t),\kappa_f(x^0(t\!+\!N|t),t\!+\!N),t\!+\!N),t\!+\!N\!+\!1\bigr) \nonumber \\ 
& \quad-l(x(t),\kappa_N(x(t),t),t) - V_f(x^0(t+N|t),t+N) \nonumber \\ 
&\leq-l(x(t),\kappa_N(x(t),t),t) +{l(x^r(t),u^r(t),t)}
\end{align}}

\normalsize
The RHS in the last inequality, which only uses information of the reference trajectory until the current time $t$, follows from~\eqref{eq:adjacent_diff_lyapunov} in Assumption~\ref{assumption_terminal_cost}, on the properties of terminal cost, region and control law. Summing the inequality in~\eqref{succesive_substraction} from an initial time $t_0$ to $t_0+T-1$ and dividing by the total number of time-steps, $T$, gives
\small{
\setlength{\abovedisplayskip}{\dispup}
\setlength{\belowdisplayskip}{0pt}
\setlength{\abovedisplayshortskip}{-0em}
\setlength{\belowdisplayshortskip}{-0em}
\setlength{\jot}{0pt}
\begin{align} \label{avg_sum}
&\sum_{t=t_0}^{t_0+T-1} \frac{V_N^0(x(t\!+\!1),t\!+\!1)-V_N^0(x(t),t)}{T}  \\
&\leq \sum_{t=t_0}^{t_0+T-1} \frac{-l(x(t),\kappa_N(x(t),t),t) + l(x^r(t),u^r(t),t)}{T}. \nonumber
\end{align}}
\normalsize

Rearranging by interchanging the LHS and RHS of~\eqref{avg_sum} and taking the limit superior of both sides 
as $T \to \infty$,
\small{\setlength{\abovedisplayskip}{\dispup}
\setlength{\belowdisplayskip}{0pt}
\setlength{\abovedisplayshortskip}{0pt}
\setlength{\belowdisplayshortskip}{0pt}
\setlength{\jot}{0pt}
\begin{align}\label{final_form_cost}
&\limsup_{T\to\infty}\!\!\sum_{t=t_0}^{t_0+T-1} \frac{l\bigl(x(t),\kappa_N(x(t),t),t\bigr)\!-\!l\bigl(x^r(t),u^r(t),t\bigr)}{T} \nonumber \\
&\leq \limsup_{T\to\infty}\frac{V_N^0(x(t_0),t_0)\!-\!V_N^0(x(t_0+T),t_0+T)}{T}.
\end{align}}
\normalsize

\noindent We know from Assumption~\ref{assumption_lower_bound}, on the lower boundedness of stage and terminal cost, that $V_N^0(x(t),t)$ is lower bounded by some constant $C \in \mathbb{R},\; \forall t \in \mathcal{T}$, and $V_N^0(x(t_0),t_0)$ is also a constant. Thus, the RHS of~\eqref{final_form_cost} can be written as 
\small{\setlength{\abovedisplayskip}{\dispup}
\setlength{\belowdisplayskip}{0pt}
\setlength{\abovedisplayshortskip}{0pt}
\setlength{\belowdisplayshortskip}{0pt}
\setlength{\jot}{0pt}
\begin{align}\label{final_RHS}
&\limsup_{T\to\infty}\frac{1}{T}\biggl(V_N^0(x(t_0),t_0)-V_N^0(x(t_0+T),t_0+T)\biggr) \nonumber \\
&\leq \limsup_{T\to\infty}\frac{1}{T}\biggl(V_N^0(x(t_0),t_0)-C\biggr)= 0.
\end{align}}
\normalsize

Combining~\eqref{final_form_cost} and~\eqref{final_RHS} completes the proof with $u(t)=\kappa_N(x(t),t), \; \forall t$.%\qed
\end{pf}

\begin{rem}\label{remark_terminal_constraint}\emph{(\textbf{Practical applicability of Assumption~\ref{assumption_terminal_cost}}).
Assumption~\ref{assumption_terminal_cost}, on the properties of terminal cost, region and control law,  modifies \cite[Assumption 3]{risbeck2019economic} to make analysis and terminal cost function design possible (thereby increasing the practical viability of Assumption~\ref{assumption_terminal_cost}) when the reference trajectory is not fixed a-priori for all future times, but is only known until the current time. This design implies that the reference trajectory can be continuously updated online at each time-step. If \cite[Assumption 3]{risbeck2019economic} is used instead of Assumption~\ref{assumption_terminal_cost}, the last term in the RHS of~\eqref{succesive_substraction} would be $l(x^r(t+N),u^r(t+N),t+N)$. A fixed apriori reference trajectory cannot adapt to forecast updates for at least $N$ future time-steps (see previous sentence), or at all, which can significantly degrade system performance especially when maximum terms (like demand charges) are involved in the performance function (ultimately risking making the reference cost more), which weakens the performance guarantee. Also, fixing a reference trajectory a-priori for all future times is not always possible when long term forecasts are involved.
\oprocend}
\end{rem}

\begin{cor}\label{corollary_practical}(\textbf{Practical extension of Theorem~\ref{theorem_asymp_cost}}).
When $T$ is sufficiently (even though not infinitely) large, Theorem~\ref{theorem_asymp_cost} becomes
\small{
\setlength{\abovedisplayskip}{\dispup}
\setlength{\belowdisplayskip}{0pt}
\setlength{\abovedisplayshortskip}{0em}
\setlength{\belowdisplayshortskip}{0em}
\setlength{\jot}{0pt}
\begin{align*}
    \sum_{t=t_0}^{t_0+T-1} \frac{1}{T}\biggl(l\bigl(x(t),u(t),t\bigr)-l\bigl(x^r(t),u^r(t),t\bigr)\biggr)\leq \epsilon,
\end{align*}}
\normalsize

where $\epsilon>0$ is small. The implication is that the average economic cost difference of the proposed system~\eqref{closed_loop_empc} with the baseline given by the reference trajectory~\eqref{dynamics_reference} is upper bounded by a small positive constant, $\epsilon \propto \frac{1}{T}$.
\end{cor}
\begin{pf}
    The proof follows similarly to the proof of Theorem~\ref{theorem_asymp_cost} until~\eqref{avg_sum}, and then for $T$ being sufficiently large, and $V_N^0(x(t),t)$ being lower bounded by some constant $C \in \mathbb{R},\; \forall t \in \mathcal{T}$, using
    \small{
\setlength{\abovedisplayskip}{\dispup}
\setlength{\belowdisplayskip}{0pt}
\setlength{\abovedisplayshortskip}{0em}
\setlength{\belowdisplayshortskip}{0em}
\setlength{\jot}{0pt}
    \begin{align*}
    &\frac{1}{T}\biggl(V_N^0(x(t_0),t_0)-V_N^0(x(t_0+T),t_0+T)\biggr) \nonumber \\
&\leq \frac{1}{T}\biggl(V_N^0(x(t_0),t_0)-C\biggr)= \epsilon>0.
\end{align*}}
\end{pf}
%
%\vspace{-2em}
\normalsize
 \new{
\begin{prop}\label{prop_terminal_cost_design}(\textbf{Constructive design of the terminal cost function}).
%\margin{I've edited this so it is more clear}
{Let the terminal region sequence $\mathcal{X}_f(t),\, t\in\mathcal{T}$, and the terminal control law $\kappa_f:\mathcal{X}_f\times\mathcal{T}\to\mathcal{U}$ be such that $\{\mathcal{X}_f(t)\}_{t\in\mathcal{T}}$ is sequentially control invariant. Let $W:\mathcal{X}_f\times\mathcal{T}\to\mathbb{R}$ be a uniformly lower bounded continuous function, and define 
%and consider a terminal cost of the form
%\small{
%\setlength{\abovedisplayskip}{\dispup}
%\setlength{\belowdisplayskip}{0pt}
%\setlength{\abovedisplayshortskip}{0em}
%\setlength{\belowdisplayshortskip}{0em}
%\setlength{\jot}{0pt}
%\begin{align}\label{terminal_cost_decomposition}
%V_f\bigl(x(t),t\bigr)=W\bigl(x(t),t\bigr)-h(t),
%\end{align}}
%\normalsize

%where $h:\mathcal{T}\to\mathbb{R}$ depends on time only. 
%Define
\small{\setlength{\abovedisplayskip}{\dispup}
\setlength{\belowdisplayskip}{0pt}
\setlength{\abovedisplayshortskip}{0pt}
\setlength{\belowdisplayshortskip}{0pt}
\setlength{\jot}{0pt}
\begin{align}\label{psi_definition}
\psi_{\kappa_f}\bigl(x(t),t\bigr)&:=W\bigl(f(x(t),\kappa_f(x(t),t),t),t\!+\!1\bigr)-W\bigl(x(t),t\bigr)\nonumber\\
&\quad+l\bigl(x(t),\kappa_f(x(t),t),t\bigr)\nonumber\\
&\quad-l\bigl(x^{r}(t\!-\!N),u^{r}(t\!-\!N),t\!-\!N\bigr),
\end{align}}%
\normalsize
and $G(t)\geq \sup_{x(t)\in\mathcal{X}_f(t)}\psi_{\kappa_f}\bigl(x(t),t\bigr)$. Let
\small{
\setlength{\abovedisplayskip}{\dispup}
\setlength{\belowdisplayskip}{\dispdn}
\setlength{\abovedisplayshortskip}{0em}
\setlength{\belowdisplayshortskip}{0em}
\setlength{\jot}{0pt}
\begin{align}\label{h_construction}
h(t)=\sum_{k=N}^{t-1}G(k), \qquad t\geq N,
\end{align}}
\normalsize

be initialized at $h(N)=0$. Define
\small{
\setlength{\abovedisplayskip}{\dispup}
\setlength{\belowdisplayskip}{0pt}
\setlength{\abovedisplayshortskip}{0em}
\setlength{\belowdisplayshortskip}{0em}
\setlength{\jot}{0pt}
\begin{align}\label{terminal_cost_decomposition}
V_f\bigl(x(t),t\bigr)=W\bigl(x(t),t\bigr)-h(t).
\end{align}}
\normalsize

If \hspace{0.01em} $\sup_{t\geq N}h(t)<\infty$, then the terminal cost~\eqref{terminal_cost_decomposition} complies with both Assumption~\ref{assumption_terminal_cost}, on the properties of the terminal cost, and Assumption~\ref{assumption_lower_bound}, on the lower boundedness of the terminal cost. }
\end{prop}
\begin{pf}
Fix any $t\in\mathcal{T}$ and $x(t)\in\mathcal{X}_f(t)$, and denote the successor state under the terminal control law by $x^{+}:=f\bigl(x(t),\kappa_f(x(t),t),t\bigr)$. By sequential control invariance, $x^{+}\in\mathcal{X}_f(t+1)$, so both $W(x^{+},t\!+\!1)$ and $V_f(x^{+},t\!+\!1)$ are well-defined. Substituting the decomposition~\eqref{terminal_cost_decomposition} into the adjacent difference of the terminal cost yields
\small{\setlength{\abovedisplayskip}{\dispup}
\setlength{\belowdisplayskip}{\dispdn}
\setlength{\abovedisplayshortskip}{0pt}
\setlength{\belowdisplayshortskip}{0pt}
\setlength{\jot}{0pt}
\begin{align}\label{Vf_difference_expansion}
&V_f\bigl(x^{+},t\!+\!1\bigr)-V_f\bigl(x(t),t\bigr)\nonumber\\
&=W\bigl(x^{+},t\!+\!1\bigr)-W\bigl(x(t),t\bigr)-\bigl(h(t\!+\!1)-h(t)\bigr).
\end{align}}%
\normalsize

Hence, using~\eqref{Vf_difference_expansion} and Definition~\eqref{psi_definition}, the inequality~\eqref{eq:adjacent_diff_lyapunov} of Assumption~\ref{assumption_terminal_cost} holds if and only if
\small{\setlength{\abovedisplayskip}{\dispup}
\setlength{\belowdisplayskip}{\dispdn}
\setlength{\abovedisplayshortskip}{0pt}
\setlength{\belowdisplayshortskip}{0pt}
\setlength{\jot}{0pt}
\begin{align}\label{h_pointwise_condition}
h(t+1)-h(t)\geq \psi_{\kappa_f}\bigl(x(t),t\bigr).
\end{align}}
\normalsize

As $h(t+1)-h(t)$ is independent of the state, and~\eqref{h_pointwise_condition} must hold for every $x(t)\in\mathcal{X}_f(t)$, it is equivalent to the state-independent condition
\small{\setlength{\abovedisplayskip}{\dispup}
\setlength{\belowdisplayskip}{0em}
\setlength{\abovedisplayshortskip}{0pt}
\setlength{\belowdisplayshortskip}{0pt}
\setlength{\jot}{0pt}
\begin{align}\label{h_sup_condition}
h(t+1)-h(t)\geq \sup_{x(t)\in\mathcal{X}_f(t)}\psi_{\kappa_f}\bigl(x(t),t\bigr).
\end{align}}
\normalsize
%Note that if the terminal control law $\kappa_f(\cdot,t)$ is continuous, then $\psi_{\kappa_f}(\cdot,t)$ is continuous (as $W$, $l$, and $f$ are continuous); since $\mathcal{X}_f(t)$ is nonempty and compact, the supremum is attained, hence finite, at each $t\ge N$ 
The construction~\eqref{h_construction} yields
\small{\setlength{\abovedisplayskip}{\dispup}
\setlength{\belowdisplayskip}{0pt}
\setlength{\abovedisplayshortskip}{0pt}
\setlength{\belowdisplayshortskip}{0pt}
\setlength{\jot}{0pt}
\begin{align}\label{h_diff_G}
h(t+1)-h(t)&=\sum_{k=N}^{t}G(k)-\sum_{k=N}^{t-1}G(k)\nonumber\\
&=G(t)\geq\sup_{x(t)\in\mathcal{X}_f(t)}\psi_{\kappa_f}\bigl(x(t),t\bigr),
\end{align}}%
\normalsize
where the last inequality is the definition of $G(t)$. Hence,~\eqref{h_sup_condition}, and therefore~\eqref{h_pointwise_condition} and the inequality~\eqref{eq:adjacent_diff_lyapunov} of Assumption~\ref{assumption_terminal_cost}, hold for all $x(t)\in\mathcal{X}_f(t)$.%, provided the sequence $h$ is real-valued.

It remains to establish that Assumption~\ref{assumption_lower_bound} holds, which follows from $\sup_{t\geq N}h(t)<\infty$. Let $M:=\sup_{t\geq N}h(t)<\infty$, and %regard $h$, defined by~\eqref{h_construction}, as an a priori extended-real-valued sequence. We show by induction on $t\geq N$ that $G(t)<\infty$ and $h(t)$ is finite. The base case holds, as $h(N)=0$. Suppose $h(t)$ is finite for some $t\geq N$. Since $\mathcal{X}_f(t)\neq\emptyset$, $G(t)>-\infty$; and if $G(t)=+\infty$, then $h(t+1)=h(t)+G(t)=+\infty>M$, contradicting the definition of $M$. Hence, $G(t)<\infty$ and $h(t+1)=h(t)+G(t)$ is finite, completing the induction. 
from~\eqref{terminal_cost_decomposition}, for all $x(t)\in\mathcal{X}_f(t)$ and $t\geq N$,
\small{\setlength{\abovedisplayskip}{\dispup}
\setlength{\belowdisplayskip}{0pt}
\setlength{\abovedisplayshortskip}{0pt}
\setlength{\belowdisplayshortskip}{0pt}
\setlength{\jot}{0pt}
\begin{align}\label{Vf_lower_bound}
V_f\bigl(x(t),t\bigr)&=W\bigl(x(t),t\bigr)-h(t)\nonumber\\
&\geq \underline{W}-\sup_{t\geq N}h(t)=\underline{W}-M,
\end{align}}%
\normalsize
where $\underline{W}:=\inf_{(x,t)\in\mathcal{X}_f\times\mathcal{T}}W(x,t)>-\infty$, as $W$ is uniformly bounded from below. Since $\underline{W}-M$ is a finite constant, $V_f$ is uniformly bounded from below; i.e., Assumption~\ref{assumption_lower_bound} holds which completes the proof.
\end{pf}}
%
\begin{comment}
\begin{rem}\label{remark_terminal_design}\emph{(\textbf{Design freedom in the terminal cost decomposition}).
The decomposition~\eqref{terminal_cost_decomposition} separates the terminal cost into a designer-chosen function $W$ of the state and a purely time-dependent offset $h$. Once $W$, $\mathcal{X}_f$, and $\kappa_f$ are fixed, the offset $h$ is determined by~\eqref{h_construction}, so these are the only design degrees of freedom. Continuity of $V_f=W-h$ (Assumption~\ref{assumption_continuity}) follows directly from the continuity of $W$. The remaining requirement, $\sup_{t\geq N}h(t)<\infty$, couples $W$, $\mathcal{X}_f$, and $\kappa_f$ through~\eqref{psi_definition}--\eqref{h_construction}: it caps the accumulated growth of the offset $h$ and, together with the lower boundedness of $W$, secures Assumption~\ref{assumption_lower_bound}. \oprocend}
\end{rem}
\end{comment}
%

\section{Application of EMPC to optimal MG dispatch}\label{MG_dispatch} 

In this section, we apply the EMPC structure from Section~\ref{empc} to the optimal MG dispatch problem. To do this, as explained in Section~\ref{sec:MG_model}, we consider a grid-connected MG with local gross load demand, PV generation, with an installed BESS. Our proposed EMPC formulation will aim to solve the MG dispatch problem as stated below.

\begin{prob}\label{problem}
    {At each time $t$, under a given perfect forecast of load and VRE (PV in this case) generation over a prediction horizon $N$, compute the optimal control inputs, i.e., BESS dispatch profile and grid import, considering electricity costs over the prediction horizon, and implement the first time-step computed control inputs in a receding horizon fashion in closed-loop, with the idealized goal of minimizing monthly electricity costs.} \oprocend
    
\end{prob}

\subsection{Microgrid (MG) model}\label{sec:MG_model}
Here, we first consider a standard MG model from the literature (see~\cite{ghosh2023adaptive,ghosh2025adaptive,cristian_empc,mcclone2023hybrid}), which is a linear-time-invariant (LTI) system consisting of one constituent state $x_{1}(t)$, and two constituent control inputs represented by $\setlength\arraycolsep{3pt} u(t)\!=\!{\begin{bmatrix} u_{1}(t) &  u_{2}(t)\end{bmatrix}}^{\top}$. Later, we will augment this original MG model to add additional states $x_{2}(t)$, $x_{3}(t)$, with a motivation and dynamics for $x_{2}(t)$ and $x_{3}(t)$ that will become clear in Section~\ref{electricity_cost_structure} and formulations presented in Sections~\ref{augmeted_sys_NCDC_OPDC} and~\ref{candidate_NCDC_OPDC_choice_1}; respectively. The original MG model is represented in the structure of the receding horizon EMPC as in Section~\ref{empc} as
\small{\setlength{\abovedisplayskip}{\dispup}
\setlength{\belowdisplayskip}{0em}
\setlength{\abovedisplayshortskip}{0pt}
\setlength{\belowdisplayshortskip}{0pt}
\setlength{\jot}{0pt}
\begin{subequations}\label{MG_equations}
\begin{align}
\label{dynamics_MG}
x_{1}(t{+}1) &= Ax_{1}(t)+Bu(t), && \forall\, t,\\
\label{inputs_limit_MG}
Su(t) &\leq s, && \forall\, t,\\
\label{inputs_coupling_MG}
Mu(t) &= c(t), && \forall\, t,\\
\label{state_limit_MG}
Gx_{1}(t) &\leq g, && \forall\, t,
\end{align}
\end{subequations}}

\normalsize
where $x_{1}(t) \in \mathbb{R}_{\geq0}$ is the BESS SOC. The control inputs are $\setlength\arraycolsep{3pt} u(t)\!=\!{\begin{bmatrix} u_{1}(t) &  u_{2}(t)\end{bmatrix}}^{\top}$, with $u_{1}(t) \in \mathbb{R}$ being the BESS dispatch power, and $u_{2}(t)\in \mathbb{R}$, the grid import power. When $u_{1}(t)>0$, the BESS is charging, while $u_{2}(t)>0$ denotes power import from the main grid to the MG. The EMPC prediction horizon is subdivided into $N$ equal time-steps of $\Delta t$ duration (sampling period) each. 

The system matrices are $A=[1]$, $B={\begin{bmatrix} \frac{\Delta t}{\subscr{\rm BESS}{en}} & \!0 \end{bmatrix}}$, %
where $\subscr{\rm BESS}{en}$ is the energy capacity of the BESS. The system matrices handle the SOC update of the BESS due to charging/discharging by~\eqref{dynamics_MG}. 
For the hard control input constraints~\eqref{inputs_limit_MG},{
\setlength{\abovedisplayskip}{0pt}
\setlength{\belowdisplayskip}{0pt}
\setlength{\abovedisplayshortskip}{0pt}
\setlength{\belowdisplayshortskip}{0pt}
\setlength{\jot}{0pt}
\begin{align*}
    S=\begin{bmatrix}
    1 & 0 \\
    -1 & 0 \\
    0 & 1 \\
    0 & -1 \\
    \end{bmatrix}, \text{ and } \setlength\arraycolsep{3pt} s={\begin{bmatrix} \subscr{\rm BESS}{max} \\ \subscr{\rm BESS}{max} \\ \hat{b} \\ -\hat{a} \end{bmatrix}},
\end{align*}}

which constrain the maximum charging/discharging power of the BESS and MG power exchange with the main grid. Practically, $|\hat {a}|$ and $|\hat{b}|$ are chosen to be very large so that the MG power exchange with the grid constraints are never active, {negating} the risk of violating~\eqref{inputs_coupling_MG}, {which is possible} as the main grid can be conceptualized as an infinite bus. The PV generation and gross load are denoted by $\text {PV}(t)$ and $L(t)$; respectively, and are used as forecast inputs to the MPC and assumed to be  exactly known for the prediction horizon. For the time-varying equality constraints~\eqref{inputs_coupling_MG} coupling the control inputs, we take $\setlength\arraycolsep{3pt} M={\begin{bmatrix} 1 & -1 \end{bmatrix}}$, and ${c(t)}=[{\rm PV}(t)-L(t)]$, which ensures the power balance of the MG with the main grid. The hard state constraints represented by~\eqref{state_limit_MG} limit the SOC of the BESS between upper ($\subscr {\rm SOC}{max}$) and lower bounds ($\subscr {\rm SOC}{min}$), resulting in $G={\begin{bmatrix} 1 & \!-1 \end{bmatrix}}^{\top}$, and $\setlength\arraycolsep{3pt} g={\begin{bmatrix} \subscr {\rm SOC}{max} & -\subscr {\rm SOC}{min} \end{bmatrix}}^{\top}$.

\subsection{Electricity cost structure}\label{electricity_cost_structure} 
In general, the monthly electricity costs for an MG are composed of volumetric energy charges (\$/kWh) and demand charges (\$/kW). Two distinct time-of-use demand charges are used: one based on maximum grid imports for the whole month called non-coincident demand peak (NCDP), and another one which is based on the maximum grid import between the on-peak (OP) hours 16:00--21:00~h 
of all days of the month, called on-peak demand peak (OPDP). The demand charges associated with NCDP and OPDP are called non-coincident demand charges (NCDC), and on-peak demand charges (OPDC); respectively. The pricing structure can be extended to other forms of demand charges or be adjusted to account for NCDC only by setting the OPDC terms to 0.

As demand charges are calculated over a monthly window, consider a month being equally subdivided into $T$ equal time-steps of $\Delta t$ duration each. For each month, let $\mathcal{T}_w\subseteq \mathcal{T}$ be the set of all time-steps corresponding to the monthly window. Specifically, $\mathcal{T}_w=\{0,1,2, \ldots,T-1\}$. $\mathcal{T}_w'\subset \mathcal{T}_w$ 
denotes the set of time-steps of the month over which OPDC is calculated.

The monthly electricity costs are formulated as~\cite{ghosh2022effects,ghosh2023adaptive,ghosh2025adaptive}
\small{\setlength{\abovedisplayskip}{\dispup}
\setlength{\belowdisplayskip}{0em}
\setlength{\abovedisplayshortskip}{0pt}
\setlength{\belowdisplayshortskip}{0pt}
\setlength{\jot}{0pt}
\begin{align} \label{monthly_cost_NCDC_OPDC}
%\begin{aligned}
\sum_{t=0}^{T-1}\Bigl[&R_{\rm EC}(t)\Delta t\Bigl(u_{2}(t)+ \frac{(1\!-\!\eta)}{2}|u_{1}(t)|\Bigr)\Bigr]
\nonumber\\
&+ R_{\rm NC}\max_{k \in \mathcal{T}_w}{u_{2}(k)} + R_{\rm OP}\max_{k \in \mathcal{T}_w'}{u_{2}(k)},
%\end{aligned}
\end{align}}
\normalsize

where the first, second, third, and fourth terms cover the energy charges, BESS losses, NCDC, and OPDC; respectively. The time-of-use energy charge rate, NCDC rate, OPDC rate, and BESS round-trip efficiency are denoted by $R_{\rm EC}(t)$, $R_{\rm NC}$, $R_{\rm OP}$, and $\eta$; respectively, and are all non-negative. Note that demand charges can never be negative. %\footnote{In~\eqref{monthly_cost_NCDC_OPDC}, it is implicitly assumed that $\max_{k \in \mathcal{T}_w}{u_{2}(k)} \geq 0$, implying that, at some time within the month, the grid import is positive, which is a reasonable assumption for an MG with a BESS installation. The third term in~\eqref{monthly_cost_NCDC_OPDC} can be generalized further to $R_{\rm NC}\max\{0,\max_{k \in \mathcal{T}_w}{u_{2}(k)\}}$ to incorporate the non-negativity of demand charges explicitly; however, this is avoided here for brevity. A similar implicit assumption entails for the OPDC term, where it is implicitly assumed that $\max_{k \in \mathcal{T}_w'}{u_{2}(k)} \geq 0$.} 
The first two terms in~\eqref{monthly_cost_NCDC_OPDC}; i.e., the terms inside the summation, can be encapsulated as an additive economic stage cost, depending only on the current input and time. However, because of the presence of demand charges which involve a maximization over multiple time-steps, it is challenging to encapsulate the demand charges, i.e., the third and fourth terms in~\eqref{monthly_cost_NCDC_OPDC} as an additive economic stage cost depending only on the current state, input, and time as in Section~\ref{empc}. Thus, to mitigate this issue, inspired by~\cite{jones2017solving}, 
we formulate the dynamics of the second and third constituent states, $x_{2}(t)$ and $x_{3}(t)$ respectively by augmenting the original MG model in Section~\ref{augmeted_sys_NCDC_OPDC} to represent the demand charge term as an additive economic stage cost as in Section~\ref{empc}.

\subsection{Augmented system}\label{augmeted_sys_NCDC_OPDC}

An additional constituent state $x_{2}(t) \in \mathbb{R}_{\geq 0}, \; \forall t$, is defined for the augmented system, which tracks the `running' NCDP (i.e., the maximum grid import until the current time $t \in \mathcal{T}_w$) as
\begin{equation} \label{running_demand_peak_NCDC}
\begin{aligned}
&x_{2}(t+1) = \max\bigl(x_{2}(t),u_{2}(t)\bigr), \qquad \forall t.
\end{aligned}
\end{equation}
To incorporate the OPDC, we define another state $x_{3}(t) \in \mathbb{R}_{\geq 0}, \; \forall t$, which tracks the `running' OPDP (i.e., maximum grid import during OP hours until the current time $t\in\mathcal{T}_w$) as
%\vspace{-1em}
\begin{equation} \label{running_demand_peak_NCDC_OPDC}
\begin{aligned}
x_{3}(t+1) &= \max\bigl(x_{3}(t),\beta(t)u_{2}(t)\bigr), \qquad \forall t,
\end{aligned}
\end{equation}
where 
\pullup{-1em}
\small{\setlength{\abovedisplayskip}{\dispup}
\setlength{\belowdisplayskip}{\dispdn}
\setlength{\abovedisplayshortskip}{0pt}
\setlength{\belowdisplayshortskip}{0pt}
\setlength{\jot}{0pt}
\begin{align*}
\beta(t) & = \begin{cases}
    1, & t \in \mathcal{T}_w',\\
    0, & \text{otherwise.}
\end{cases}
\end{align*}}
\normalsize

The augmented state is then denoted as $\setlength\arraycolsep{3pt} x(t)={\begin{bmatrix} x_{1}(t) & x_{2}(t) & x_{3}(t)\end{bmatrix}}^{\top}$ with the augmented MG dynamics and additive economic stage cost being rewritten in~\eqref{augmented_dynamics_NCDC_OPDC} and~\eqref{augmented_stage_cost_NCDC_OPDC}; respectively. In all, we have
\small{\setlength{\abovedisplayskip}{\dispup}
\setlength{\belowdisplayskip}{0em}
\setlength{\abovedisplayshortskip}{0pt}
\setlength{\belowdisplayshortskip}{0pt}
\setlength{\jot}{0pt}
\begin{subequations}\label{augmented_system_NCDC_OPDC}
\begin{align} \label{augmented_dynamics_NCDC_OPDC}
%\begin{aligned}
x(t\!+\!1)=&f(x(t),u(t),t) \nonumber\\ :=&\begin{pmatrix}
           Ax_{1}(t)+Bu(t) \\
           \max\bigl(x_{2}(t),u_{2}(t)\bigr) \\
           \max\bigl(x_{3}(t),\beta(t)u_{2}(t)\bigr) \\
         \end{pmatrix}, \quad \forall t,\\
%\end{aligned}
%\end{equation}
%\begin{equation} 
\label{augmented_stage_cost_NCDC_OPDC}
%\begin{aligned}
l(x(t),u(t),t):=& R_{\rm EC}(t)\Delta t\Bigl(u_{2}(t)+ \frac{(1\!-\!\eta)}{2}|u_{1}(t)|\Bigr) \nonumber \\
&+R_{\rm NC}\Bigl(a(t\!+\!1)x_{2}(t\!+\!1)-a(t)x_{2}(t)\Bigr) \nonumber \\
&+R_{\rm OP}\Bigl(b(t\!+\!1)x_{3}(t\!+\!1)-b(t)x_{3}(t)\Bigr),  \forall t, 
\end{align}
%\end{equation}
\end{subequations}}
\normalsize
%\vspace{-0.5em}

where, $a:\mathcal{T}\to\mathbb{R}$ is a function defining coefficients for $x_{2}(t), \; \forall t\in{\mathcal{T}_{w}}$ 
with $a(T)=1$, and $b:\mathcal{T}\to\mathbb{R}$ is a function defining coefficients for $x_{3}(t), \; \forall t\in {\mathcal{T}_{w}}$ 
with $b(T)=1$.\footnote{Note that $x(t+1)$ is a function of $\{x(t),u(t),t\}$ by~\eqref{dynamics_real}. Thus, the terms multiplying $R_{\rm NC}$ and $R_{\rm OP}$ in the RHS of~\eqref{augmented_stage_cost_NCDC_OPDC} are still implicitly dependent only on the current state, input, and time.} Equation~\eqref{augmented_stage_cost_NCDC_OPDC} has similar form to~\cite[Remark 8]{risbeck2019economic}, but relaxes the more restrictive requirement that the coefficients $a(0)$ and $b(0)$ have to be $0$.
%\vspace{-1em}
\begin{prop}\label{prop_monthly_cost_NCDC}{(\textbf{Representing monthly electricity costs via augmented, additive-economic stage costs})}.
{The monthly electricity costs in~\eqref{monthly_cost_NCDC_OPDC}} are exactly represented by the summation of the augmented additive stage cost in~\eqref{augmented_stage_cost_NCDC_OPDC}, over the demand charge window; i.e., a month.
\end{prop}

\pullup{-2em}
\begin{pf}
Considering the demand charge window as $\mathcal{T}_w=\{0,1,2, \ldots,T-1\}$, $\mathcal{T}_w'\subset \mathcal{T}_w$, and summing~\eqref{augmented_stage_cost_NCDC_OPDC} over all the time-steps in $\mathcal{T}_{w}$ gives
\small{\setlength{\abovedisplayskip}{\dispup}
\setlength{\belowdisplayskip}{\dispdn}
\setlength{\abovedisplayshortskip}{0pt}
\setlength{\belowdisplayshortskip}{0pt}
\setlength{\jot}{0pt}
\begin{align} \label{prop_eqn_NCDC_OPDC}
&\sum_{t=0}^{T-1}l(x(t),u(t),t) \nonumber \\
& =\sum_{t=0}^{T-1}\biggl[R_{\rm EC}(t)\Delta t\Bigl(u_{2}(t) +\frac{(1-\eta)}{2}|u_{1}(t)|\Bigr)\nonumber \\
&\qquad\quad+R_{\rm NC}\Bigl(a(t+1)x_{2}(t+1)-a(t)x_{2}(t)\Bigr) \nonumber \\
&\qquad\quad+R_{\rm OP}\Bigl(b(t+1)x_{3}(t+1)-b(t)x_{3}(t)\Bigr)\biggr] \nonumber\\
&=\sum_{t=0}^{T-1}\Bigl[R_{\rm EC}(t)\Delta t\Bigl(u_{2}(t)+ \frac{(1\!-\!\eta)}{2}|u_{1}(t)|\Bigr)\Bigr] \nonumber\\
&\quad +R_{\rm NC}\biggl(a(T)x_{2}(T)-a(0)x_{2}(0)\biggr) \nonumber \\
&\quad + R_{\rm OP}\biggl(b(T)x_{3}(T)-b(0)x_{3}(0)\biggr).
%\end{aligned}
\end{align}}

\normalsize
Note, $x_{2}(0)\!=\!0$, $a(T)\!=\!1$, and $x_{2}(T)\!=\!\max_{k \in \mathcal{T}_w}{u_{2}(k)}$. Also, $x_{3}(0)\!=\!0$, $b(T)\!=\!1$, and $x_{3}(T)\!=\!\max_{k \in \mathcal{T}_w'}{u_{2}(k)}$. Substituting the above in~\eqref{prop_eqn_NCDC_OPDC}, makes the last equality in~\eqref{prop_eqn_NCDC_OPDC} equal to~\eqref{monthly_cost_NCDC_OPDC} which completes the proof. 
\end{pf}

\begin{rem}\label{remark_cost}\emph{(\textbf{Application of Theorem~\ref{theorem_asymp_cost}/Corollary~\ref{corollary_practical} to Problem~\ref{problem}}).
To be able to apply Theorem~\ref{theorem_asymp_cost} to Problem~\ref{problem}, where the MG is modeled as~\eqref{MG_equations},~\eqref{augmented_dynamics_NCDC_OPDC}, and the stage cost is modeled as~\eqref{augmented_stage_cost_NCDC_OPDC}, the $T$ in the demand charge window (i.e., a month) $\mathcal{T}_w$  %Theorem~\ref{theorem_asymp_cost} 
has to correspond to $\infty$. That is, the granularity of the control inputs should be infinitely large, implying $\Delta t \to 0$. \new{For a discrete time system, $\Delta t \to 0$ is not possible, and costs computed over a finite window (a month) implies $T<\infty$. Thus, the finite window performance guarantee, i.e., Corollary~\ref{corollary_practical} is the main operative guarantee we seek to fulfill, which bears importance due to the linear decrease of worst-case average cost difference with increase of granularity (i.e., increase of $T$). \new{Note that fulfilling the conditions of Corollary~\ref{corollary_practical} entails no weakening of the performance guarantee: Corollary~\ref{corollary_practical} is precisely the form Theorem~\ref{theorem_asymp_cost} takes over the finite window on which Problem~\ref{problem} is defined.}} \oprocend}
\end{rem}

%We emphasize that satisfying the conditions of Corollary~\ref{corollary_practical} for a discrete-time system is equivalent to satisfying those of Theorem~\ref{theorem_asymp_cost} for the corresponding infinite-horizon problem: both rest on the same assumptions --- in particular, on the time-dependent offset $h$ in the terminal cost admitting a time-independent upper bound, $\sup_{t\geq N}h(t)=\bar{C}<\infty$ (cf.~Proposition~\ref{prop_terminal_cost_design}) --- and differ only in the length of the window over which the average cost is evaluated. Owing to the monthly nature of demand charges, an infinite-horizon formulation of Problem~\ref{problem} does not exist ($T<\infty$ for any $\Delta t>0$). 

\subsection{A $\supscr{1}{\rm st}$-choice problem specification of terminal cost, terminal region, and control law}\label{candidate_NCDC_OPDC_choice_1} 
Having set up the optimal MG dispatch problem in the standard EMPC form with additive stage costs in the objective function, in this section and Section~\ref{candidate_NCDC_OPDC_choice_2}, we will investigate terminal cost functions, regions, and control laws which satisfy Assumptions~\ref{assumption_continuity},~\ref{assumption_sets},~\ref{assumption_terminal_cost}, and~\ref{assumption_lower_bound}. Satisfaction of the above assumptions ensures that the optimal MG dispatch problem has the same structure as the EMPC of Section~\ref{empc}, and, as a consequence, the results of Corollary~\ref{corollary_practical}, on the practical economic cost guarantee, hold. To determine a candidate terminal cost function, region, and control law, we introduce Assumption~\ref{assumption_length_prediction_horizon}. 

\begin{assum}\label{assumption_length_prediction_horizon}\emph{(\textbf{Length of the prediction horizon}).
{For the optimal MG dispatch problem (Problem~\ref{problem}), the prediction horizon $N$ is at least equal to the minimum time required to traverse %
the entire admissible SOC set.}\oprocend}
\end{assum}

Assumption~\ref{assumption_length_prediction_horizon}, though theoretically limiting, is generally true for MG dispatch problems with prediction horizons generally being $24-48$~h, which is enough to traverse the entire SOC set with the BESS dispatch power. From the MG model in Sections~\ref{sec:MG_model} and~\ref{augmeted_sys_NCDC_OPDC}, the admissible state set can thus be represented as $\mathcal{X}(t)=[\subscr{\rm SOC}{min}, \subscr{\rm SOC}{max}]$$\times[0,\hat{b}]\times[0,\hat{b}], \; \forall t\in \mathcal{T}$, where $\hat{b}<\infty$ in any practical scenario. The admissible control input set is given by $\mathcal{U}(t)=[-\subscr {\rm BESS}{max}, \subscr{\rm BESS}{max}]$$\times [\hat{a},\hat{b}],\; \forall t \in \mathcal{T}$, where $-\infty<\hat{a}<0<\hat{b}<\infty$, in practical scenarios.

With Assumption~\ref{assumption_length_prediction_horizon} active, obvious choices of the terminal cost and region are $V_f(t):=0$ and $\mathcal{X}_f(t):= \{x^{r}(t-N)\}$; respectively, $\forall t \in \mathcal{T}$.\footnote{Here, we implicitly assume $\mathcal{X}_f(t)=\emptyset$, $\forall t\in \mathbb{N}_{0}^{N-1}$.} 
\new{However, such a choice would render a highly restrictive terminal control law because of the structure of the iterated max operation in~\eqref{augmented_dynamics_NCDC_OPDC}, which may lead to infeasibility, or suboptimality in case of a poor reference. %the above terminal region makes $x^r_{2}(t-N)$ and $x^r_{3}(t-N)$ a hard upper bound on $x_{2}(t-N+k\,|\,t-N)$ and $x_{3}(t-N+k\,|\,t-N)$; respectively, $\forall k\in \mathbb{N}, \forall t \in \mathcal{T}$, within the optimization in~\eqref{empc_general}, which may lead to infeasibility. In addition, the above terminal region gives no incentive to choose $x_{2}(t-N+k\,|\,t-N)<x^r_{2}(t-N)$ or $x_{3}(t-N+k\,|\,t-N)<x^r_{3}(t-N)$, $\forall k\in \mathbb{N}, \forall t \in \mathcal{T}$, which could be advantageous if the reference trajectory is economically suboptimal.

To mitigate the above restrictiveness, we design the following terminal cost function, region and control law, which allows for the reference bound to be exceeded or not realized depending on the economic objective. The terminal region is defined as
%
%\begin{subequations} \label{candidate_terminal} 
\small{\setlength{\abovedisplayskip}{\dispup}
\setlength{\belowdisplayskip}{\dispdn}
\setlength{\abovedisplayshortskip}{0pt}
\setlength{\belowdisplayshortskip}{0pt}
\setlength{\jot}{0pt}
\begin{equation} \label{candidate_terminal_region_NCDC_OPDC}
\begin{aligned}
\mathcal{X}_f(t):=\Biggl\{\begin{pmatrix}
           x^r_{1}(t-N) \\
           x_{2}(t) \\
           x_{3}(t)\\ 
         \end{pmatrix} \Big| \; x_2(t), x_3(t) \in [0,\hat{b}] \Biggr\}, 
\end{aligned}
\end{equation}}
\normalsize

$\forall t \in \mathcal{T}$, which implies that the BESS SOC in the optimization~\eqref{empc_general} 
must terminate (at the end of the prediction horizon) exactly at the current reference SOC, while the running NCDP, $x_{2}(t)$, and the running OPDP, $x_{3}(t)$ can terminate at any value depending on the economic objective. 
The terminal control law is defined as
\small{\setlength{\abovedisplayskip}{\dispup}
\setlength{\belowdisplayskip}{\dispdn}
\setlength{\abovedisplayshortskip}{0pt}
\setlength{\belowdisplayshortskip}{0pt}
\setlength{\jot}{0pt}
\begin{equation} \label{candidate_terminal_control_law}
\begin{aligned}
\kappa_f(x(t),t):=\begin{pmatrix}
           u^r_{1}(t-N) \\
           u^r_{1}(t-N)-c(t) \\
         \end{pmatrix}, \forall t \in \mathcal{T}. 
\end{aligned}
\end{equation}}
\normalsize

Note that the sequential control invariance of $\mathcal{X}_f(t), \; \forall t \in \mathcal{T}$ defined in~\eqref{candidate_terminal_region_NCDC_OPDC} holds under the terminal control law~\eqref{candidate_terminal_control_law}.

The $\supscr{1}{\rm st}$-choice terminal cost function is defined by~\eqref{candidate_terminal_cost_1_NCDC_OPDC}, in which the time-dependent offset $h$ is \emph{constructed}, following Proposition~\ref{prop_terminal_cost_design}, so as to satisfy Assumption~\ref{assumption_terminal_cost}, on the properties of the terminal cost, region, and control law, as well as Assumption~\ref{assumption_lower_bound}, on its lower boundedness. Our $\supscr{1}{\rm st}$ choice for the terminal cost function is given by %the following: %\vspace{-1em}
\small{\setlength{\abovedisplayskip}{\dispup}
\setlength{\belowdisplayskip}{0pt}
\setlength{\abovedisplayshortskip}{0pt}
\setlength{\belowdisplayshortskip}{0pt}
\setlength{\jot}{0pt}
\begin{align} \label{candidate_terminal_cost_1_NCDC_OPDC}
&V_f(x(t),t)\nonumber \\
&:= R_{\rm NC}\Bigl[\max\Bigl(a(t)x_{2}(t),
a(t-N+1)x^r_{2}(t-N+1)\Bigr)\Bigr] \nonumber \\
%&\qquad\qquad-a(t-N+1)x^r_{2}(t-N+1)\Bigr] \nonumber \\
&\quad\;\;+R_{\rm OP}\Bigl[\max\Bigl(b(t)x_{3}(t),
b(t-N+1)x^r_{3}(t-N+1)\Bigr)\Bigr] \nonumber \\
%&\qquad\qquad\;\;\;-b(t-N+1)x^r_{3}(t-N+1)\Bigr]
&\quad\;\;-h(t), 
\end{align}}
\normalsize

$\forall t \in \mathcal{T}$, where $h:\mathcal{T}\to \mathbb{R}$ is a purely time-dependent offset. Equation~\eqref{candidate_terminal_cost_1_NCDC_OPDC} is exactly of the form $V_f(x(t),t)=W(x(t),t)-h(t)$ of the decomposition~\eqref{terminal_cost_decomposition}, with $W(x(t),t)$ given by its two $\max$-terms. As $a(t),b(t)\in(0,1]$ and $x_{2},x_{3},x^r_{2},x^r_{3}\in[0,\hat b]$, $W$ is continuous and $0\leq W(x(t),t)\leq (R_{\rm NC}+R_{\rm OP})\hat b$, hence uniformly lower bounded, as required by Proposition~\ref{prop_terminal_cost_design}. Substituting the terminal region~\eqref{candidate_terminal_region_NCDC_OPDC}, control law~\eqref{candidate_terminal_control_law}, dynamics~\eqref{augmented_dynamics_NCDC_OPDC}, and stage cost~\eqref{augmented_stage_cost_NCDC_OPDC} into the definition~\eqref{psi_definition} of $\psi_{\kappa_f}$, and denoting $A_1(t)=a(t)x_{2}(t)$, $B_1(t)=b(t)x_{3}(t)$, $A_2(t)=a(t)x^r_{2}(t)$, $B_2(t)=b(t)x^r_{3}(t)$, $C(t)=u^r_{1}(t-N)$, and $D(t)=u^r_{1}(t)$, the pointwise condition $h(t{+}1)-h(t)\geq\psi_{\kappa_f}(x(t),t)$ of Proposition~\ref{prop_terminal_cost_design} reads, $\forall t \in \mathcal{T}$,
{\small \setlength{\abovedisplayskip}{\dispup}
\setlength{\belowdisplayskip}{0pt}
\setlength{\abovedisplayshortskip}{0pt}
\setlength{\belowdisplayshortskip}{0pt}
\setlength{\jot}{0pt}
\begin{align}\label{big_eqn_1}
& h(t+1)-h(t)   \nonumber \\ &\geq R_{\rm EC}(t)\Delta t\Bigl(C(t)-c(t) + \frac{(1\!-\!\eta)}{2}|C(t)|\Bigr) \nonumber \\
&\quad-R_{\rm EC}(t\!-\!N)\Delta t\Bigl(D(t\!-\!N)-c(t\!-\!N)
+\frac{(1\!-\!\eta)}{2}|D(t\!-\!N)|\Bigr) \nonumber \\ 
&\quad+R_{\rm NC}\Bigl[A_1(t\!+\!1)-A_1(t)
+\max\Bigl(A_1(t\!+\!1),A_2(t\!-\!N\!+\!2)\Bigr) \nonumber \\
&\qquad\qquad-\max\Bigl(A_1(t),A_2(t\!-\!N\!+\!1)\Bigr)\nonumber\\
&\qquad\qquad+A_2(t\!-\!N)-A_2(t\!-\!N\!+\!1)\Bigr] \nonumber \\
&\quad+R_{\rm OP}\Bigl[B_1(t\!+\!1)-B_1(t)
+\max\Bigl(B_1(t\!+\!1),B_2(t\!-\!N\!+\!2)\Bigr) \nonumber \\
&\qquad\qquad-\max\Bigl(B_1(t),B_2(t\!-\!N\!+\!1)\Bigr) \nonumber\\&\qquad\qquad+B_2(t\!-\!N)-B_2(t\!-\!N\!+\!1)\Bigr].
\end{align}}

\normalsize
\noindent The RHS of~\eqref{big_eqn_1} is precisely $\psi_{\kappa_f}(x(t),t)$ of Proposition~\ref{prop_terminal_cost_design}, particularized to the $\supscr{1}{\rm st}$-choice specification. Let us assume $a(t)$ and $b(t)$ to be nondecreasing in $t.$ We denote by $\bar{\psi}_{\kappa_f}(x(t),t)$ the function obtained by dropping from the RHS of~\eqref{big_eqn_1} its non-positive state-dependent and reference difference terms, ensuring $\bar{\psi}_{\kappa_f}(x(t),t)\geq{\psi}_{\kappa_f}(x(t),t)$. Hence, %i.e., $-R_{\rm NC}\bigl[A_1(t)+\max\bigl(A_1(t),A_2(t\!-\!N\!+\!1)\bigr)+A_2(t\!-\!N\!+\!1)-A_2(t\!-\!N)\bigr]$ and $-R_{\rm OP}\bigl[B_1(t)+\max\bigl(B_1(t),B_2(t\!-\!N\!+\!1)\bigr)+B_2(t\!-\!N\!+\!1)-B_2(t\!-\!N)\bigr]$. Note that $A_2(t\!-\!N\!+\!1)-A_2(t\!-\!N) \geq 0$, as $A_2(t)=a(t)x^r_{2}(t)$ is the product of two non-negative nondecreasing sequences: $a(t)$ is nondecreasing by assumption, and the reference running peak $x^r_{2}(t)$ is nondecreasing by the max-dynamics~\eqref{running_demand_peak_NCDC}; i.e., $x^r_{2}(t\!+\!1)=\max\bigl(x^r_{2}(t),u^r_{2}(t)\bigr)\geq x^r_{2}(t)$. By the identical argument, $B_2(t\!-\!N\!+\!1)-B_2(t\!-\!N)\geq0$. Hence,
{\small \setlength{\abovedisplayskip}{\dispup}
\setlength{\belowdisplayskip}{0pt}
\setlength{\abovedisplayshortskip}{0pt}
\setlength{\belowdisplayshortskip}{0pt}
\setlength{\jot}{0pt}
\begin{align}\label{kappa_bar}
& \bar{\psi}_{\kappa_f}(x(t),t) = R_{\rm EC}(t)\Delta t\Bigl(C(t)-c(t) + \frac{(1\!-\!\eta)}{2}|C(t)|\Bigr) \nonumber \\
&\quad-R_{\rm EC}(t\!-\!N)\Delta t\Bigl(D(t\!-\!N)-c(t\!-\!N)
+\frac{(1\!-\!\eta)}{2}|D(t\!-\!N)|\Bigr) \nonumber \\ 
&\quad+R_{\rm NC}\Bigl[A_1(t\!+\!1)
+\max\Bigl(A_1(t\!+\!1),A_2(t\!-\!N\!+\!2)\Bigr)\Bigr] \nonumber \\
&\quad+R_{\rm OP}\Bigl[B_1(t\!+\!1)
+\max\Bigl(B_1(t\!+\!1),B_2(t\!-\!N\!+\!2)\Bigr)\Bigr].
\end{align}}

Following the construction~\eqref{h_construction}, we accordingly set
{\small\setlength{\abovedisplayskip}{\dispup}
\setlength{\belowdisplayskip}{0pt}
\setlength{\abovedisplayshortskip}{0pt}
\setlength{\belowdisplayshortskip}{0pt}
\setlength{\jot}{0pt}
\begin{align}\label{G_h_choice_1}
G(t):=\!\!\!\sup_{x_{2}(t),x_{3}(t)\in[0,\hat b]}\!\!\!\bar{\psi}_{\kappa_f}(x(t),t),\qquad h(t):=\!\sum_{k=N}^{t-1}\!G(k),
\end{align}}%
\normalsize
for $t\geq N$. Note that $G(t)\geq \sup_{x_{2}(t),x_{3}(t)\in[0,\hat b]}\!{\psi}_{\kappa_f}(x(t),t)$, and the supremum is taken over $(x_{2}(t),x_{3}(t))$ as ${\psi}_{\kappa_f}$ is independent of $x_{1}(t)$.}
\new{
\begin{prop}\label{prop_satisfy_assum_4_NCDC_OPDC}(\textbf{Satisfaction of Assumptions~\ref{assumption_terminal_cost} and~\ref{assumption_lower_bound} for the $\supscr{1}{\rm st}$-choice terminal cost}).
{Consider the $\supscr{1}{\rm st}$-choice terminal region~\eqref{candidate_terminal_region_NCDC_OPDC}, control law~\eqref{candidate_terminal_control_law}, and terminal cost~\eqref{candidate_terminal_cost_1_NCDC_OPDC}, subject to the dynamics~\eqref{augmented_dynamics_NCDC_OPDC} and the additive stage cost~\eqref{augmented_stage_cost_NCDC_OPDC}, with $G$ and $h$ constructed as in~\eqref{G_h_choice_1}, with $\bar{\psi}_{\kappa_f}(x(t),t)$ obtained from~\eqref{kappa_bar}. Suppose $|c(t)|\leq\bar c<\infty$; $R_{\rm EC}$ is $N$-periodic with $|R_{\rm EC}(t)|\leq\bar R_{\rm EC}<\infty$; $a(t)=b(t)=\lambda^{T-t}$ for $t\in[0,T]$ with $\lambda\in(0,1)$; the terminal grid import $w(t)=u^r_{1}(t-N)-c(t)\in[0,\hat b]$. Then,
{\small\setlength{\abovedisplayskip}{\dispup}\setlength{\belowdisplayskip}{0pt}
\begin{align}\label{h_sup_bound_choice_1}
\sup_{t\geq N}h(t)\leq\bar R_{\rm EC}\Delta t[2N\bar c]\!+\!R_{\rm NC}\big[\tfrac{2\hat b}{1-\lambda}\big]\!+\!R_{\rm OP}\big[\tfrac{2\hat b}{1-\lambda}\big]=\bar{C}<\infty,
\end{align}}%
\normalsize
and hence, by Proposition~\ref{prop_terminal_cost_design}, the $\supscr{1}{\rm st}$-choice terminal cost~\eqref{candidate_terminal_cost_1_NCDC_OPDC} satisfies Assumptions~\ref{assumption_terminal_cost} and~\ref{assumption_lower_bound}. }
\end{prop}
\begin{pf}
%As the terms dropped from the RHS of~\eqref{big_eqn_1} to form $\bar{\psi}_{\kappa_f}$ are non-positive ($a(\cdot),b(\cdot)>0$, and $x_{2},x_{3},x^r_{2},x^r_{3}\geq0$), $\bar{\psi}_{\kappa_f}\geq\psi_{\kappa_f}$ pointwise, and hence $G$ in~\eqref{G_h_choice_1} satisfies $G(t)\geq\sup_{x_{2}(t),x_{3}(t)\in[0,\hat b]}\psi_{\kappa_f}(x(t),t)$, while $h$ obeys the construction~\eqref{h_construction}. 
As $W$ is continuous and uniformly lower bounded, Proposition~\ref{prop_terminal_cost_design} yields Assumptions~\ref{assumption_terminal_cost} and~\ref{assumption_lower_bound} once $\sup_{t\geq N}h(t)<\infty$ is shown. It thus suffices to prove~\eqref{h_sup_bound_choice_1}.

We first evaluate the supremum in~\eqref{G_h_choice_1} in closed form. Split $\bar{\psi}_{\kappa_f}(x(t),t)=\bar{\psi}_{\rm EC}(x(t),t)+\bar{\psi}_{\rm NC}(x(t),t)+\bar{\psi}_{\rm OP}(x(t),t)$ into its energy charge, NCDC, and OPDC terms in~\eqref{kappa_bar}. 

\emph{Energy charge terms:} Under~\eqref{candidate_terminal_control_law}, $C(t)=D(t-N)=u^r_{1}(t-N)$, so with $R_{\rm EC}$ $N$-periodic the $|C(t)|$ and $|D(t-N)|$ terms in~\eqref{kappa_bar} cancel, giving $\bar{\psi}_{\rm EC}(x(t),t)=R_{\rm EC}(t)\Delta t\,\bigl(c(t-N)-c(t)\bigr)$, independent of $\bigl(x_{2}(t),x_{3}(t)\bigr)$. Thus, $\sup_{x_2(t),x_{3}(t)\in[0,\hat b]}\bar{\psi}_{\rm EC}(x(t),t)=\bar{\psi}_{\rm EC}(x(t),t)$.

\emph{NCDC terms:} The NCDC terms are independent of $x_3(t)$. With $x_{2}^{+}=\max\bigl(x_{2}(t),w(t)\bigr)$,
{\small\setlength{\abovedisplayskip}{\dispup}\setlength{\belowdisplayskip}{0pt}\setlength{\belowdisplayskip}{0pt}
\begin{align*}
\bar{\psi}_{\rm NC}(x(t),t)=&{R_{\rm NC}}\Bigl[a(t{+}1)x_{2}^{+}+\max\!\big(a(t{+}1)x_{2}^{+},A_2(t{-}N{+}2)\big)\Bigr].
\end{align*}}%
\normalsize
Both $x_{2}$-dependent terms are nondecreasing in $x_{2}(t)$, so their supremum over $x_{2}(t)\in[0,\hat b]$ is attained at $x_{2}(t)=\hat b$, where $x_{2}^{+}=\hat b$, as $w(t)\in[0,\hat b]$. Moreover, $a(t)=\lambda^{T-t}$ is nondecreasing in $t$, so $A_2(t{-}N{+}2)\leq a(t{-}N{+}2)\hat b\leq a(t{+}1)\hat b$, and $\max\bigl(a(t{+}1)\hat b,A_2(t{-}N{+}2)\bigr)=a(t{+}1)\hat b$. Hence,
{\small\setlength{\abovedisplayskip}{\dispup}\setlength{\belowdisplayskip}{0pt}
\begin{align*}
\sup_{x_{2}(t),x_3(t)\in[0,\hat b]}\bar{\psi}_{\rm NC}(x(t),t)=R_{\rm NC}\,2a(t{+}1)\hat b.
\end{align*}}%
\normalsize

\emph{OPDC terms:} The OPDC terms are independent of $x_2$. The OPDC terms analysis is similar to the NCDC analysis above, with $b$, $x_{3}$, and $B_2$ in place of $a$, $x_{2}$, and $A_2$; respectively. Thus,

{\small\setlength{\abovedisplayskip}{\dispup}\setlength{\belowdisplayskip}{0pt}
\begin{align*}
\sup_{x_2(t),x_{3}(t)\in[0,\hat b]}\bar{\psi}_{\rm OP}(x(t),t)=R_{\rm OP}\,2b(t{+}1)\hat b.
\end{align*}}%

Therefore,
{\small\setlength{\abovedisplayskip}{\dispup}\setlength{\belowdisplayskip}{0pt}\setlength{\jot}{0pt}
\begin{align}\label{G_closed_form}
G(k)=&\;R_{\rm EC}(k)\Delta t\,\bigl(c(k{-}N)-c(k)\bigr)\nonumber\\
&+R_{\rm NC}\,2a(k{+}1)\hat b+R_{\rm OP}\,2b(k{+}1)\hat b.
\end{align}}%
\normalsize

Now, sum~\eqref{G_closed_form} over $k=N,\ldots,t-1$ to get $h(t)$ as shown below. 

\emph{Sum over energy charge terms:} Writing $g(k):=R_{\rm EC}(k)c(k)$ and using $R_{\rm EC}(k)c(k-N)=R_{\rm EC}(k-N)c(k-N)=g(k-N)$ due to $N$-periodicity of $R_{\rm EC}$, the sum telescopes in $N$-blocks

{\small\setlength{\abovedisplayskip}{\dispup}\setlength{\belowdisplayskip}{0pt}\setlength{\jot}{0pt}
\begin{align*}
\sum_{k=N}^{t-1}\!R_{\rm EC}(k)\Delta t\,\bigl(c(k{-}N)\!-\!c(k)&\bigr)=\Delta t\!\sum_{k=N}^{t-1}\!\bigl(g(k\!-\!N)-g(k)\bigr)\\
&=\Delta t\Big(\sum_{k=0}^{t-N-1}\!\!g(k)-\!\sum_{k=N}^{t-1}\!g(k)\Big)\\
&=\Delta t\Big(\sum_{k=0}^{N-1}\!g(k)-\!\!\sum_{k=t-N}^{t-1}\!\!g(k)\Big) \\
\leq\Delta &t\Big(\Big|\sum_{k=0}^{N-1}\!g(k)\Big|+\Big|\sum_{k=t-N}^{t-1}\!\!g(k)\Big|\Big) \\
&\leq 2N\bar R_{\rm EC}\bar c\,\Delta t,
\end{align*}}%
\normalsize
where the second equality follows from the change of the summation index $k\to k+N$ in the first sum, and the third equality follows from cancelling the terms common to both sums. The final inequality holds as at most $2N$ terms remain, each satisfying $|g(k)|\leq\bar R_{\rm EC}\bar c$, as $|R_{\rm EC}(k)|\leq\bar R_{\rm EC}$ and $|c(k)|\leq\bar c$. This is the first term in the RHS of the inequality in~\eqref{h_sup_bound_choice_1}.

\emph{Sum over NCDC terms:} As $a(k{+}1)=\lambda^{T-k-1}$, $\sum_{k=N}^{t-1}2a(k{+}1)\hat b\leq2\hat b\sum_{j=0}^{\infty}\lambda^{j}=\tfrac{2\hat b}{1-\lambda}$; i.e., the $R_{\rm NC}$ part of the sum is at most $R_{\rm NC}\big[\tfrac{2\hat b}{1-\lambda}\big]$, the second term in the RHS of the inequality in~\eqref{h_sup_bound_choice_1}. 

\emph{Sum over OPDC terms:} The analysis is identical to the sum over NCDC terms, with the $R_{\rm OP}$ part of the sum being at most $R_{\rm OP}\big[\tfrac{2\hat b}{1-\lambda}\big]$, the third term in the RHS of the inequality in~\eqref{h_sup_bound_choice_1}. 

\emph{Supremum of $h(t)$:} Adding the three bounds gives $h(t)=\sum_{k=N}^{t-1}G(k)\leq\bar R_{\rm EC}\Delta t[2N\bar c]+R_{\rm NC}[\tfrac{2\hat b}{1-\lambda}]+R_{\rm OP}[\tfrac{2\hat b}{1-\lambda}]$, with $h$ initialized by $h(N)=0$. As the RHS of the last inequality is independent of $t$,~\eqref{h_sup_bound_choice_1} follows, which completes the proof.   %As the running peaks, and the construction~\eqref{G_h_choice_1} restart with each successive demand charge window, the same bound holds within every window, and thus for all $t\geq N$, which is~\eqref{h_sup_bound_choice_1} with any finite constant $\bar C$ no smaller than its middle expression. Proposition~\ref{prop_terminal_cost_design} then gives Assumptions~\ref{assumption_terminal_cost} and~\ref{assumption_lower_bound}.
\end{pf}}
%
\begin{comment}
  
\begin{rem}\label{possible_failure}\emph{(\textbf{Guaranteed satisfaction of Assumptions~\ref{assumption_terminal_cost} and~\ref{assumption_lower_bound} for the $\supscr{1}{\rm st}$-choice specification}).
%Unlike postulating an operational condition on the terminal cost, the offset $h$ in~\eqref{G_h_choice_1} is \emph{constructed} from the problem data through Proposition~\ref{prop_terminal_cost_design}. 
The bound~\eqref{h_sup_bound_choice_1} certifies $\sup_{N\leq t\leq T}h(t)\leq\bar{C}<\infty$ over the entire monthly window under mild, site-verifiable conditions: a bounded net load ($|c(t)|\leq\bar c$), an $N$-periodic and bounded energy-charge rate, and geometric weights $a(t)=b(t)=\lambda^{T-t}$. Consequently, Assumptions~\ref{assumption_terminal_cost} and~\ref{assumption_lower_bound} hold for all finite times, so that the practical cost guarantee of Corollary~\ref{corollary_practical} always applies to the $\supscr{1}{\rm st}$-choice specification. %; in the idealized limit $T\to\infty$, where, $\Delta t\to0$, the asymptotic guarantee of Theorem~\ref{theorem_asymp_cost} applies as well. %The bound scales linearly with the peak-import ceiling $\hat b$ and the effective discount horizon $\tfrac{1}{1-\lambda}$, and with $N\bar c$ for the energy component. 
\oprocend}
\end{rem}
\end{comment}
\new{
\begin{rem}\label{lambda_tradeoff}\emph{(\textbf{Effect of the discount parameter $\lambda$}).
The choice of $\lambda\in(0,1)$ in the weights $a(t)=b(t)=\lambda^{T-t}$ trades off the tightness of the bound~\eqref{h_sup_bound_choice_1} against the intra-month quality of demand charge penalization. As $\lambda\to1$, the demand charge terms in~\eqref{h_sup_bound_choice_1}, scaling with $\tfrac{2\hat b}{1-\lambda}$, grow unbounded, weakening the performance guarantee in Corollary~\ref{corollary_practical}, as the lower bound on $V_N^0$ varies inversely with the upper bound on $h$ (see~\eqref{monthly_upper_bound}). However, as the weights $a(t),b(t)$ remain close to $1$ throughout the demand charge window, peaks reached early in the month are penalized in the stage cost~\eqref{augmented_stage_cost_NCDC_OPDC} and the terminal cost~\eqref{candidate_terminal_cost_1_NCDC_OPDC} nearly as strongly as those reached near the end of the month. Such uniform penalization is desirable in practice: by the max-dynamics~\eqref{running_demand_peak_NCDC}, the running peaks are non-decreasing, so an under-penalized early peak irreversibly locks in demand charges for the remainder of the month. Conversely, as $\lambda$ decreases, the bound~\eqref{h_sup_bound_choice_1} tightens, approaching $\bar R_{\rm EC}\Delta t[2N\bar c]+2\hat b(R_{\rm NC}+R_{\rm OP})$ as $\lambda\to0$; however, $a(t)=\lambda^{T-t}\approx0$ for $t\ll T$, so early peaks are severely under-penalized, which may permit large grid imports early in the month, and is thus undesirable from a practical standpoint. The choice of $\lambda$ should hence balance the strength of the theoretical guarantee against uniform peak penalization over the demand charge window, as is desirable in practice. \oprocend}
\end{rem}
}
  
Substituting~\eqref{augmented_stage_cost_NCDC_OPDC} and~\eqref{candidate_terminal_cost_1_NCDC_OPDC} in the objective function of the EMPC~\eqref{objective_empc} and simplifying yields
\new{
%\vspace*{-0.8cm}
{\small\setlength{\abovedisplayskip}{\dispup}\setlength{\belowdisplayskip}{0pt}\setlength{\jot}{-1pt}
\begin{align}\label{substitute_objective_1b_NCDC_OPDC}
&{V}_N(x(t),{\mathbf{u}(x(t),t)},t) \nonumber\\
&= \sum_{k=0}^{N-1}\Bigl[R_{\rm EC}(t\!+\!k)\Delta t\Bigl(u_{2}(t\!+\!k|t) 
+\frac{(1-\eta)}{2}|u_{1}(t\!+\!k|t)|\Bigr)\Bigr] \nonumber \\
&\quad+\Biggl[R_{\rm NC}\Biggl(a(t\!+\!N)x_{2}(t\!+\!N|t) - a(t)x_{2}(t) \nonumber \\
&\qquad\qquad\quad+\max\Bigl(a(t\!+\!N)x_{2}(t\!+\!N|t),
a(t\!+\!1)x^r_{2}(t\!+\!1)\Bigr)\Biggr)\Biggr] \nonumber \\
&\quad+\Biggl[R_{\rm OP}\Biggl(b(t+N)x_{3}(t+N|t) -b(t)x_{3}(t) \nonumber \\
&\qquad\qquad\quad+\max\Bigl(b(t+N)x_{3}(t+N|t),b(t+1)x^r_{3}(t+1)\Bigr)\Biggr)\Biggr] \nonumber \\
&\quad -h(t\!+\!N).
\end{align}}}
\normalsize
%\end{subequations}

The objective function and constraints of Problem~\ref{problem} are now set up in the EMPC structure of Section~\ref{empc}. \new{Note that the terms $a(t)x_{2}(t)$, $b(t)x_{3}(t)$, and $
h(t+N)$ of~\eqref{substitute_objective_1b_NCDC_OPDC} are independent of ${\mathbf{u}(x(t),t)}$ for a given initial state $x(t)$, reference states $x^{r}(t)$, $x^{r}(t+1)$ and time $t$. Thus, the values of $a(t)x_{2}(t)$, $b(t)x_{3}(t)$, and $
h(t+N)$ do not affect the optimal control sequence; i.e., the solution $\mathbf{u}^0$, on solving the EMPC. Thus, while an $h$ can be constructed according to~\eqref{G_h_choice_1} to guarantee satisfaction of Corollary~\ref{corollary_practical},\footnote{For Problem~\ref{problem}, by inspection we can additionally conclude that in practical scenarios Assumptions~\ref{assumption_continuity} and~\ref{assumption_sets} are also satisfied.} in practical implementation and computation, $h$ need not be designed at all, and the $h(t+N)$ term along with $a(t)x_{2}(t)$, $b(t)x_{3}(t)$ can be dropped from the optimization. For a concise representation of the $\supscr{1}{\rm st}$-choice problem specification, see Table~\ref{table_choice_summary}.}

\subsection{A $\supscr{2}{\rm nd}$ and a $\supscr{3}{\rm rd}$-choice problem specification of terminal cost, terminal region, and control law}\label{candidate_NCDC_OPDC_choice_2}

Although the $\supscr{1}{\rm st}$-choice problem specification in Section~\ref{candidate_NCDC_OPDC_choice_1} ensures the practical cost guarantee of Corollary~\ref{corollary_practical}, the terminal region (specifically the requirement that the SOC at the end of the optimization ends at the current reference SOC due to terminal constraint~\eqref{candidate_terminal_region_NCDC_OPDC}) in the $\supscr{1}{\rm st}$-choice may lead to suboptimal finite-time economic performance. We now seek to take advantage of Assumption~\ref{assumption_length_prediction_horizon}, on the length of the prediction horizon, to further relax the conditions for constructing alternative problem specifications.%terminal region, control law, and cost function.

\subsubsection{Elements of the $\supscr{2}{\rm nd}$-choice problem specification}\label{subsec:second_choice}
As Assumption~\ref{assumption_length_prediction_horizon}, on the length of the prediction horizon, holds, we can define $\mathcal{X}_f(t):=\mathcal{X}(t),\; \forall t \in \mathcal{T}$; i.e., each of the terminal SOC and running NCDP and OPDP are free variables, as in~\eqref{candidate_terminal_region_NCDC_OPDC_2}. %In this case, any admissible control input, $\setlength\arraycolsep{3pt} u(t)\!=\!{\begin{bmatrix} u_{1}(t) &  u_{1}(t)-c(t)\end{bmatrix}}^{\top}$ and successor state $x(t+1)$, satisfying~\eqref{inputs_limit_MG}, and~\eqref{augmented_dynamics_NCDC_OPDC}, \eqref{state_limit_MG}; respectively, can be a terminal control law (an obvious example is $\setlength\arraycolsep{3pt} u(t)\!=\!{\begin{bmatrix} 0 &  -c(t)\end{bmatrix}}^{\top}$). Specifically, it means $\forall t\in \mathcal{T}$,
{\small\setlength{\abovedisplayskip}{\dispup}\setlength{\belowdisplayskip}{0pt}\setlength{\jot}{-0pt}
\begin{equation} \label{candidate_terminal_region_NCDC_OPDC_2}
\begin{aligned}
\mathcal{X}_f(t)
  := \Biggl\{
      \begin{pmatrix}
        x_{1}(t) \\
        x_{2}(t) \\
        x_{3}(t)
      \end{pmatrix}
      \;\Bigg|\
      \begin{aligned}
        &x_1(t) \in [\subscr{\rm SOC}{min}, \subscr{\rm SOC}{max}];\\
        &x_2(t),\, x_3(t) \in [0,\, \hat{b}]
      \end{aligned}
    \Biggr\}, 
\end{aligned}
\end{equation}}
\normalsize
\new{
Similarly, the candidate terminal control law can be defined as,
% \pskipcomp keeps the full original gap here: the tall \Biggl\{ descends far
% enough below the display baseline to touch the next paragraph without it.
{\small\setlength{\abovedisplayskip}{\dispup}\setlength{\belowdisplayskip}{\pskipcomp}\setlength{\jot}{-0pt}
\begin{equation} \label{candidate_terminal_control_law_relaxed}
\begin{aligned}
\kappa_f(x(t),t)\!:=\!\Biggl\{\!\!\begin{pmatrix}
           0 \\
           -c(t) \\
         \end{pmatrix}
        \!\! \Bigg|\
      \begin{aligned}
        &-c(t) \in [\hat{a}, \hat{b}]
      \end{aligned}\Biggr\}. 
\end{aligned}
\end{equation}}
\normalsize

The $\supscr{2}{\rm nd}$-choice terminal cost function is defined by~\eqref{candidate_terminal_cost_1_NCDC_OPDC}. Denoting $A_1(t)=a(t)x_{2}(t)$, $B_1(t)=b(t)x_{3}(t)$, $A_2(t)=a(t)x^r_{2}(t)$, $B_2(t)=b(t)x^r_{3}(t)$, $C(t)=0$, and $D(t)=u^r_{1}(t)$, the pointwise condition $h(t{+}1)-h(t)\geq\psi_{\kappa_f}(x(t),t)$ of Proposition~\ref{prop_terminal_cost_design} reads, as~\eqref{big_eqn_1} $\forall t \in \mathcal{T}$. Note the change of $C(t)$ in the $\supscr{2}{\rm nd}$-choice problem specification as opposed to the $\supscr{1}{\rm st}$-choice. Here too, the sequential control invariance of $\mathcal{X}_f(t), \; \forall t \in \mathcal{T}$ defined in~\eqref{candidate_terminal_region_NCDC_OPDC_2} holds under the terminal control law~\eqref{candidate_terminal_control_law_relaxed}. Similar to the $\supscr{1}{\rm st}$-choice, in the $\supscr{2}{\rm nd}$-choice problem specification, we denote by $\bar{\psi}_{\kappa_f}(x(t),t)$ the function obtained by dropping from the RHS of~\eqref{big_eqn_1} its non-positive reference loss term, i.e., $-R_{\rm EC}(t\!-\!N)\Delta t\frac{(1\!-\!\eta)}{2}|D(t\!-\!N)|$, in addition to the non-positive state-dependent and reference difference terms as in~\eqref{kappa_bar}, ensuring $\bar{\psi}_{\kappa_f}(x(t),t)\geq{\psi}_{\kappa_f}(x(t),t)$. $G(t)$
and $h(t)$ are then set according to~\eqref{G_h_choice_1} with the updated $\bar{\psi}_{\kappa_f}(x(t),t)$.

\begin{prop}\label{prop_satisfy_assum_4_NCDC_OPDC_b}(\textbf{Satisfaction of Assumptions~\ref{assumption_terminal_cost} and~\ref{assumption_lower_bound}
for the $\supscr{2}{\rm nd}$-choice terminal cost}).
{Consider the $\supscr{2}{\rm nd}$-choice terminal region~\eqref{candidate_terminal_region_NCDC_OPDC_2}, control law~\eqref{candidate_terminal_control_law_relaxed}, and terminal cost~\eqref{candidate_terminal_cost_1_NCDC_OPDC}, subject to the dynamics~\eqref{augmented_dynamics_NCDC_OPDC} and the additive stage cost~\eqref{augmented_stage_cost_NCDC_OPDC}, with $G$ and $h$ constructed as in~\eqref{G_h_choice_1}, with $\bar{\psi}_{\kappa_f}(x(t),t)$ obtained by dropping from the RHS of~\eqref{kappa_bar} the $-R_{\rm EC}(t\!-\!N)\Delta t\frac{(1\!-\!\eta)}{2}|D(t\!-\!N)|$ term. Suppose $|c(t)|\leq\bar c<\infty$; $|R_{\rm EC}(t)|\leq\bar R_{\rm EC}<\infty$; $\sup_{t}\sum_{k=0}^{t}\Big|R_{\rm EC}(k+1)-R_{\rm EC}(k)\Big|=\bar{D}<\infty$ which implies bounded total variation of $R_{\rm EC}$; $a(t)=b(t)=\lambda^{T-t}$ for $t\in[0,T]$ with $\lambda\in(0,1)$; the terminal grid import $w(t)=-c(t)\in[0,\hat b]$. Then,
{\small\setlength{\abovedisplayskip}{\dispup}\setlength{\belowdisplayskip}{0pt}
\begin{align}\label{h_sup_bound_choice_2}
\sup_{t\geq N}h(t)&\leq\bar R_{\rm EC}\Delta t[2N\bar c] + {\rm BESS}_{\rm en}\Delta {\rm SOC}[\bar R_{\rm EC}+\bar{D}]\!\nonumber\\
&\quad +R_{\rm NC}\big[\tfrac{2\hat b}{1-\lambda}\big]\!+\!R_{\rm OP}\big[\tfrac{2\hat b}{1-\lambda}\big]=\bar{C}<\infty,
\end{align}}%
\normalsize
where $\Delta {\rm SOC}=\rm{SOC}_{\rm max}-\rm{SOC}_{\rm min}$, and hence, by Proposition~\ref{prop_terminal_cost_design}, the $\supscr{2}{\rm nd}$-choice terminal cost~\eqref{candidate_terminal_cost_1_NCDC_OPDC} satisfies Assumptions~\ref{assumption_terminal_cost} and~\ref{assumption_lower_bound}. }
\end{prop}

\begin{pf}
The proof follows similarly to the proof of Proposition~\ref{prop_satisfy_assum_4_NCDC_OPDC}, with the only difference being in the energy charge terms, which we expand below. Under~\eqref{candidate_terminal_control_law_relaxed}, $C(t)=0$, and writing $g(k):=R_{\rm EC}(k)c(k)$ yields
{\small\setlength{\abovedisplayskip}{\dispup}\setlength{\belowdisplayskip}{0pt}
\begin{align*}
\bar{\psi}_{\rm EC}(x(t),t)=&\;\Delta t\bigl(g(t{-}N)-g(t)\bigr)\\
&-R_{\rm EC}(t{-}N)\Delta t\,u^r_{1}(t{-}N).
\end{align*}}%
\normalsize
%Contrary to the $\supscr{1}{\rm st}$ choice, the difference form $g(t{-}N)-g(t)$ arises directly, without requiring the $N$-periodicity of $R_{\rm EC}$.

\emph{Sum over energy charge terms:} The sum of $g(t{-}N)-g(t)$ over $k=N,\ldots,t-1$ is at most $2N\bar R_{\rm EC}\bar c\,\Delta t$, exactly as in the proof of Proposition~\ref{prop_satisfy_assum_4_NCDC_OPDC}, and is the first term in the RHS of the inequality in~\eqref{h_sup_bound_choice_2}. For the reference dispatch term, from the reference BESS dynamics~\eqref{dynamics_MG}, $\Delta t\,u^r_{1}(j)={\rm BESS}_{\rm en}\bigl(\tilde x^r_{1}(j{+}1)-\tilde x^r_{1}(j)\bigr)$, where $\tilde x^r_{1}(j):=x^r_{1}(j)-{\rm SOC}_{\rm min}\in[0,\Delta{\rm SOC}]$, by the feasibility of the reference trajectory (Assumption~\ref{assumption_sets}). The change of the summation index $k\to k+N$, with $M:=t-N$, and summation by parts then give
{\small\setlength{\abovedisplayskip}{\dispup}\setlength{\belowdisplayskip}{0pt}\setlength{\jot}{-1pt}
\begin{align*}
&-\Delta t\!\sum_{k=N}^{t-1}\!R_{\rm EC}(k{-}N)u^r_{1}(k{-}N)\\
&=-{\rm BESS}_{\rm en}\!\sum_{j=0}^{M-1}\!R_{\rm EC}(j)\bigl(\tilde x^r_{1}(j{+}1)-\tilde x^r_{1}(j)\bigr)\\
&={\rm BESS}_{\rm en}\Bigl[R_{\rm EC}(0)\tilde x^r_{1}(0)-R_{\rm EC}(M{-}1)\tilde x^r_{1}(M)\\
&\qquad\qquad+\sum_{j=1}^{M-1}\bigl(R_{\rm EC}(j)-R_{\rm EC}(j{-}1)\bigr)\tilde x^r_{1}(j)\Bigr]\\
&\leq{\rm BESS}_{\rm en}\Delta{\rm SOC}\bigl[\bar R_{\rm EC}+\bar D\bigr],
\end{align*}}%
\normalsize
where the last inequality holds as: (i) $R_{\rm EC}(0)\tilde x^r_{1}(0)\leq\bar R_{\rm EC}\Delta{\rm SOC}$; (ii) $-R_{\rm EC}(M{-}1)\tilde x^r_{1}(M)\leq0$, as energy charge rates are non-negative; and (iii) the last sum is at most $\bar D\,\Delta{\rm SOC}$, by the bounded total variation of $R_{\rm EC}$ and $\tilde x^r_{1}(j)\in[0,\Delta{\rm SOC}]$. These are the second and third terms in the RHS of the inequality in~\eqref{h_sup_bound_choice_2}; the fourth and fifth terms in the RHS of the inequality in~\eqref{h_sup_bound_choice_2} are the same as those of Proposition~\ref{prop_satisfy_assum_4_NCDC_OPDC}, which completes the proof.
\end{pf}
For a concise representation of the $\supscr{2}{\rm nd}$-choice problem specification, see Table~\ref{table_choice_summary}.}

\subsubsection{Elements of the $\supscr{3}{\rm rd}$-choice problem specification}\label{subsec:third_choice}
The candidate terminal region and control law of the $\supscr{3}{\rm rd}$ choice are the same as those of the $\supscr{2}{\rm nd}$ choice (i.e.,~\eqref{candidate_terminal_region_NCDC_OPDC_2} and~\eqref{candidate_terminal_control_law_relaxed}; respectively), with the main difference being given by the terminal cost function.
\new{In the objective function~\eqref{substitute_objective_1b_NCDC_OPDC} used in both the $\supscr{1}{\rm st}$ and $\supscr{2}{\rm nd}$ choices, based on the terminal cost~\eqref{candidate_terminal_cost_1_NCDC_OPDC}, the objective penalizes the terminal running NCDP, $x_{2}(t\!+\!N|t)$, and the terminal running OPDP, $x_{3}(t\!+\!N|t)$ twice: once from the additive stage cost and a second time from the terminal cost.} This can make the second penalization of $x_{2}(t\!+\!N|t)$ and $x_{3}(t\!+\!N|t)$ 
coming from the terminal cost~\eqref{candidate_terminal_cost_1_NCDC_OPDC} practically redundant, when peak demand charges dominate over BESS losses and energy cost within a prediction horizon --- e.g., when the peak demand penalty is orders of magnitude higher than the penalty on grid imports or BESS losses. %\footnote{We refer to the redundancy as `practical' since double penalizing already dominant terms in the objective function would generally have negligible practical effect on the control law, but there might be slight numerical differences.} 
\new{Thus, it might be helpful to remove this redundancy in favor of penalizing the increase of next-step predicted peaks, i.e., $x_{2}(t\!+\!1|t)$ and $x_{3}(t\!+\!1|t)$, from the reference peaks, in the terminal cost function (see Appendix~\ref{choice_3_motivation} for a motivation for such a
penalty). 
The $\supscr{3}{\rm rd}$ choice terminal cost function is thus formulated as 
\small{\setlength{\abovedisplayskip}{\dispup}
\setlength{\belowdisplayskip}{0pt}
\setlength{\abovedisplayshortskip}{0pt}
\setlength{\belowdisplayshortskip}{0pt}
\setlength{\jot}{0pt}
\begin{align} \label{candidate_terminal_cost_2_NCDC_OPDC}
%\begin{aligned}
&V_f(x(t),t)\nonumber \\
&:= R_{\rm NC}\Bigl[\max\Bigl(a(t\!-\!N\!+\!1)x_{2}(t\!-\!N\!+\!1),\nonumber \\
&\qquad \qquad \qquad \;\; a(t\!-\!N\!+\!1)x^r_{2}(t\!-\!N\!+\!1)\Bigr)\Bigr] \nonumber\\
&\quad\;\;+R_{\rm OP}\Bigl[\max\Bigl(b(t\!-\!N\!+\!1)x_{3}(t\!-\!N\!+\!1),\nonumber \\
&\qquad \qquad \qquad \quad\;\;\; b(t\!-\!N\!+\!1)x^r_{3}(t\!-\!N\!+\!1)\Bigr)\Bigr] -h(t),
\end{align}}

\normalsize
$\forall t \in \mathcal{T}$, where $h:\mathcal{T}\to \mathbb{R}$ is a purely time-dependent offset.\footnote{Note that $V_f$ in~\eqref{candidate_terminal_cost_2_NCDC_OPDC} is implicitly a function of only $t$, as $x(t)$ does not appear in the RHS.} Substituting the terminal region~\eqref{candidate_terminal_region_NCDC_OPDC_2}, control law~\eqref{candidate_terminal_control_law_relaxed}, dynamics~\eqref{augmented_dynamics_NCDC_OPDC}, and stage cost~\eqref{augmented_stage_cost_NCDC_OPDC} into the definition~\eqref{psi_definition} of $\psi_{\kappa_f}$, and denoting $A_1(t)=a(t)x_{2}(t)$, $B_1(t)=b(t)x_{3}(t)$, $A_2(t)=a(t)x^r_{2}(t)$, $B_2(t)=b(t)x^r_{3}(t)$, $C(t)=0$, and $D(t)=u^r_{1}(t)$, the pointwise condition $h(t{+}1)-h(t)\geq\psi_{\kappa_f}(x(t),t)$ of Proposition~\ref{prop_terminal_cost_design} reads, $\forall t \in \mathcal{T}$,
{\small \setlength{\abovedisplayskip}{\dispup}
\setlength{\belowdisplayskip}{0pt}
\setlength{\abovedisplayshortskip}{0pt}
\setlength{\belowdisplayshortskip}{0pt}
\setlength{\jot}{0pt}
\begin{align}\label{big_eqn_2}
& h(t+1)-h(t)   \nonumber \\ &\geq R_{\rm EC}(t)\Delta t\Bigl(C(t)-c(t) + \frac{(1\!-\!\eta)}{2}|C(t)|\Bigr) \nonumber \\
&\quad-R_{\rm EC}(t\!-\!N)\Delta t\Bigl(D(t\!-\!N)-c(t\!-\!N)
+\frac{(1\!-\!\eta)}{2}|D(t\!-\!N)|\Bigr) \nonumber \\ 
&\quad+R_{\rm NC}\Bigl[A_1(t\!+\!1)-A_1(t)
+\max\Bigl(A_1(t\!-\!N\!+\!2),A_2(t\!-\!N\!+\!2)\Bigr) \nonumber \\
&\qquad\qquad-\max\Bigl(A_1(t\!-\!N\!+\!1),A_2(t\!-\!N\!+\!1)\Bigr)\nonumber\\
&\qquad\qquad+A_2(t\!-\!N)-A_2(t\!-\!N\!+\!1)\Bigr] \nonumber \\
&\quad+R_{\rm OP}\Bigl[B_1(t\!+\!1)-B_1(t)
+\max\Bigl(B_1(t\!-\!N\!+\!2),B_2(t\!-\!N\!+\!2)\Bigr) \nonumber \\
&\qquad\qquad-\max\Bigl(B_1(t\!-\!N\!+\!1),B_2(t\!-\!N\!+\!1)\Bigr) \nonumber\\&\qquad\qquad+B_2(t\!-\!N)-B_2(t\!-\!N\!+\!1)\Bigr].
\end{align}}

\normalsize
\noindent The RHS of~\eqref{big_eqn_2} is precisely $\psi_{\kappa_f}(x(t),t)$ of Proposition~\ref{prop_terminal_cost_design}, particularized to the $\supscr{3}{\rm rd}$-choice specification. We denote by $\bar{\psi}_{\kappa_f}(x(t),t)$ the function obtained by dropping from the RHS of~\eqref{big_eqn_2} its non-positive state-dependent, reference difference, reference loss terms like the $\supscr{2}{\rm nd}$-choice specification, in addition to the non-positive reference relative penalization terms, ensuring $\bar{\psi}_{\kappa_f}(x(t),t)\geq{\psi}_{\kappa_f}(x(t),t)$. Hence, %i.e., $-R_{\rm NC}\bigl[\max\bigl(A_1(t\!-\!N\!+\!1),A_2(t\!-\!N\!+\!1)\bigr)\bigr]$ and $-R_{\rm OP}\bigl[\max\bigl(B_1(t\!-\!N\!+\!1),B_2(t\!-\!N\!+\!1)\bigr)\bigr]$.

{\small \setlength{\abovedisplayskip}{\dispup}
\setlength{\belowdisplayskip}{0pt}
\setlength{\abovedisplayshortskip}{0pt}
\setlength{\belowdisplayshortskip}{0pt}
\setlength{\jot}{0pt}
\begin{align}\label{kappa_bar_2}
& \bar{\psi}_{\kappa_f}(x(t),t) = R_{\rm EC}(t)\Delta t\Bigl(C(t)-c(t) + \frac{(1\!-\!\eta)}{2}|C(t)|\Bigr) \nonumber \\
&\quad-R_{\rm EC}(t\!-\!N)\Delta t\Bigl(D(t\!-\!N)-c(t\!-\!N)\Bigr) \nonumber \\ 
&\quad+R_{\rm NC}\Bigl[A_1(t\!+\!1)
+\max\Bigl(A_1(t\!-\!N\!+\!2),A_2(t\!-\!N\!+\!2)\Bigr)\Bigr] \nonumber \\
&\quad+R_{\rm OP}\Bigl[B_1(t\!+\!1)
+\max\Bigl(B_1(t\!-\!N\!+\!2),B_2(t\!-\!N\!+\!2)\Bigr)\Bigr].
\end{align}}

\begin{prop}\label{prop_satisfy_assum_4_NCDC_OPDC_2}(\textbf{Satisfaction of Assumptions~\ref{assumption_terminal_cost} and~\ref{assumption_lower_bound} for the $\supscr{3}{\rm rd}$-choice terminal cost}).
{Consider the $\supscr{3}{\rm rd}$-choice terminal region~\eqref{candidate_terminal_region_NCDC_OPDC_2}, control law~\eqref{candidate_terminal_control_law_relaxed}, and terminal cost~\eqref{candidate_terminal_cost_2_NCDC_OPDC}, subject to the dynamics~\eqref{augmented_dynamics_NCDC_OPDC} and the additive stage cost~\eqref{augmented_stage_cost_NCDC_OPDC}, with $G$ and $h$ constructed as in~\eqref{G_h_choice_1}, with $\bar{\psi}_{\kappa_f}(x(t),t)$ obtained from~\eqref{kappa_bar_2}. Suppose $|c(t)|\leq\bar c<\infty$; $|R_{\rm EC}(t)|\leq\bar R_{\rm EC}<\infty$; $\sup_{t}\sum_{k=0}^{t}\Big|R_{\rm EC}(k+1)-R_{\rm EC}(k)\Big|=\bar{D}<\infty$ which implies bounded total variation of $R_{\rm EC}$; $a(t)=b(t)=\lambda^{T-t}$ for $t\in[0,T]$ with $\lambda\in(0,1)$; the terminal grid import $w(t)=-c(t)\in[0,\hat b]$. Then,
{\small\setlength{\abovedisplayskip}{\dispup}\setlength{\belowdisplayskip}{0pt}
\begin{align}\label{h_sup_bound_choice_3}
\sup_{t\geq N}h(t)&\leq\bar R_{\rm EC}\Delta t[2N\bar c] + {\rm BESS}_{\rm en}\Delta {\rm SOC}[\bar R_{\rm EC}+\bar{D}]\!\nonumber\\
&\quad +R_{\rm NC}\big[\tfrac{2\hat b}{1-\lambda}\big]\!+\!R_{\rm OP}\big[\tfrac{2\hat b}{1-\lambda}\big]=\bar{C}<\infty,
\end{align}}%
\normalsize
where $\Delta {\rm SOC}=\rm{SOC}_{\rm max}-\rm{SOC}_{\rm min}$, and hence, by Proposition~\ref{prop_terminal_cost_design}, the $\supscr{3}{\rm rd}$-choice terminal cost~\eqref{candidate_terminal_cost_2_NCDC_OPDC} satisfies Assumptions~\ref{assumption_terminal_cost} and~\ref{assumption_lower_bound}. }
\end{prop}

\begin{pf}
Similar to the proof of Propositions~\ref{prop_satisfy_assum_4_NCDC_OPDC} and~\ref{prop_satisfy_assum_4_NCDC_OPDC_b}.
\end{pf}

Table~\ref{table_choice_summary} summarizes the three problem specifications and the conditions under which each satisfies the practical cost guarantee of Corollary~\ref{corollary_practical}.}

\begin{table*}[!t]
\centering
\footnotesize
\setlength{\aboverulesep}{0.15ex}\setlength{\belowrulesep}{0.25ex}
\renewcommand{\arraystretch}{1.1}%
\setlength{\belowcaptionskip}{2pt}\new{
\caption{Summary of the three problem specifications for optimal MG dispatch (Problem~\ref{problem}).}
\label{table_choice_summary}
\begin{tabularx}{\textwidth}{@{}>{\raggedright\arraybackslash}p{3.2cm}>{\raggedright\arraybackslash}X>{\raggedright\arraybackslash}X>{\raggedright\arraybackslash}X@{}}
\toprule
 & $\supscr{1}{\rm st}$ choice & $\supscr{2}{\rm nd}$ choice & $\supscr{3}{\rm rd}$ choice \\
\midrule
Terminal region & \eqref{candidate_terminal_region_NCDC_OPDC}: terminal SOC fixed at $x^r_{1}(t{-}N)$; running peaks free & \eqref{candidate_terminal_region_NCDC_OPDC_2}: free; i.e., $\mathcal{X}_f(t)=\mathcal{X}(t)$ & \eqref{candidate_terminal_region_NCDC_OPDC_2}: free; i.e., $\mathcal{X}_f(t)=\mathcal{X}(t)$ \\
Terminal control law & \eqref{candidate_terminal_control_law}: $u_{1}(t)=u^r_{1}(t{-}N)$, $u_{2}(t)=u^r_{1}(t{-}N)-c(t)$ & \eqref{candidate_terminal_control_law_relaxed}: $u_{1}(t)=0$, $u_{2}(t)=-c(t)$ & \eqref{candidate_terminal_control_law_relaxed}: $u_{1}(t)=0$, $u_{2}(t)=-c(t)$ \\
Terminal cost & \eqref{candidate_terminal_cost_1_NCDC_OPDC}: penalizes weighted terminal running peaks; constructed offset $h$ from~\eqref{G_h_choice_1} & \eqref{candidate_terminal_cost_1_NCDC_OPDC}: same as the $\supscr{1}{\rm st}$ choice & \eqref{candidate_terminal_cost_2_NCDC_OPDC}: penalizes weighted running peaks at $t{-}N{+}1$ relative to the reference peaks; constructed offset $h$ from~\eqref{G_h_choice_1} \\
Assumptions on $R_{\rm EC}$ & Bounded rate, $|R_{\rm EC}(t)|\leq\bar R_{\rm EC}$; $N$-periodic, $R_{\rm EC}(t)=R_{\rm EC}(t+N)$ & Bounded rate, $|R_{\rm EC}(t)|\leq\bar R_{\rm EC}$; bounded total variation, $\sup_{t}\sum_{k=0}^{t}\Big|R_{\rm EC}(k+1)-R_{\rm EC}(k)\Big|<\infty$ & Bounded rate, $|R_{\rm EC}(t)|\leq\bar R_{\rm EC}$; bounded total variation, $\sup_{t}\sum_{k=0}^{t}\Big|R_{\rm EC}(k+1)-R_{\rm EC}(k)\Big|<\infty$ \\
Additional operational assumptions & Bounded net load, $|c(t)|\leq\bar c$; geometric weights, $a(t)=b(t)=\lambda^{T-t}$, $\lambda\in(0,1)$; sufficiently long prediction horizon (Assumption~\ref{assumption_length_prediction_horizon}) & Bounded net load, $|c(t)|\leq\bar c$; geometric weights, $a(t)=b(t)=\lambda^{T-t}$, $\lambda\in(0,1)$; sufficiently long prediction horizon (Assumption~\ref{assumption_length_prediction_horizon}) & Bounded net load, $|c(t)|\leq\bar c$; geometric weights, $a(t)=b(t)=\lambda^{T-t}$, $\lambda\in(0,1)$; sufficiently long prediction horizon (Assumption~\ref{assumption_length_prediction_horizon}) \\
Corollary~\ref{corollary_practical} holds & Yes (Proposition~\ref{prop_satisfy_assum_4_NCDC_OPDC}) & Yes (Proposition~\ref{prop_satisfy_assum_4_NCDC_OPDC_b}) & Yes (Proposition~\ref{prop_satisfy_assum_4_NCDC_OPDC_2}) \\
\bottomrule
\end{tabularx}}
\end{table*}

\section{Case study}\label{case_study}
In this section, we implement our proposed method characterized by $\supscr{1}{\rm st}$, $\supscr{2}{\rm nd}$ and $\supscr{3}{\rm rd}$ choices, for simulating the optimal BESS dispatch strategy for a grid-connected MG with PV and load. The MG setup refers to the real-life MG at the Port of San Diego, described in~\cite{ghosh2023adaptive,cristian_empc,ghosh2025adaptive}. This model incorporates electricity prices with NCDC, OPDC, and energy charges. First, we assume a perfect day-ahead load and PV forecast for the analyses consistent with our MG model. 

\new{\emph{Generation of reference trajectories:}} It is by definition that the optimal electricity cost for the month can only be achieved if~\eqref{monthly_cost_NCDC_OPDC} is used as the objective function in an optimization that spans the entire month, which necessitates the availability of a perfect forecast for the entire month. As obtaining this information is unrealistic, the standard method in the literature~\cite{cristian_empc} carries out a receding horizon optimization with an economic objective function that is similar to~\eqref{monthly_cost_NCDC_OPDC}, while employing a day-ahead prediction horizon and information of previous running NCDP and OPDP (see~\eqref{substitute_reference_1a_NCDC_OPDC}), 
which is more realistic. The trajectory generated by this standard method is henceforth named the \emph{standard reference method}. \new{We motivate a \emph{tracking reference method} to generate reference trajectories in Appendix~\ref{choice_3_motivation} and use~\eqref{tracking_objective} as its objective function. Note that the reference trajectories are generated online in real-time with the same information (i.e., available measurements, forecast horizon and its predictions) and constraints that are available to the proposed method to closely reflect reality (i.e., MG model~\eqref{MG_equations},~\eqref{augmented_dynamics_NCDC_OPDC}). The predefined design parameters of the MG are shown in Table~\ref{table_parameters} in Appendix~\ref{appendix_table}. Terminal constraints are sometimes added to the reference methods to reflect varied scenarios of the MG operator's operational choices and conservativeness (from rule-of-thumb or domain knowledge expertise) to compensate for the aforementioned time-scale mismatch, and are described in Section~\ref{ref_terminal_case_study_settings}.}

\new{
\begin{rem}\label{remark_surrogate}\emph{(\textbf{Interpretation of the receding horizon implementation as a
surrogate for one-shot monthly cost optimization}).
Note that the proposed receding horizon scheme is not claimed to recover the monthly
optimum. Instead, its monthly-cost interpretation is inherited by
summing the guarantee of Corollary~\ref{corollary_practical} over the demand charge window
and applying Proposition~\ref{prop_monthly_cost_NCDC} yielding
$\sum_{t=0}^{T-1}l\bigl(x(t),u(t),t\bigr)\leq\sum_{t=0}^{T-1}l\bigl(x^{r}(t),u^{r}(t),t\bigr)
+T\epsilon$. This implies that the realized monthly bill under the proposed controller~\eqref{closed_loop_empc} is
no larger than the monthly bill of the reference trajectory plus the constant
$T\epsilon = V_N^0\bigl(x(0),0\bigr)-C\leq N\bar R_{\rm EC}\Delta t\Bigl(\hat b-\hat a+\tfrac{(1-\eta)}{2}{\rm BESS}_{\rm max}\Bigr)+2(R_{\rm NC}+R_{\rm OP})\hat b+\bar{C}$, whose upper bound (RHS of the inequality above) is independent of the window length $T$ and is computable at design time. $\bar{C}$ is given by~\eqref{h_sup_bound_choice_1},~\eqref{h_sup_bound_choice_2}, and~\eqref{h_sup_bound_choice_3} for the $\supscr{1}{\rm st}$, $\supscr{2}{\rm nd}$ and $\supscr{3}{\rm rd}$ choices; respectively. (see Appendix~\ref{appendix_upper_bound} for the proof).\oprocend}
\end{rem}

While we theoretically cannot guarantee a better finite-time economic performance in general, this case study provides an example that our proposed method, over and above the practical economic guarantees from Corollary~\ref{corollary_practical},  produces better finite-time economic performance, when compared to a variety of reference methods of varying conservativeness, which might be of great importance to MG operators.
}

\section{Results and discussion}\label{results}
\subsection{Reference method terminal constraints and case study settings} \label{ref_terminal_case_study_settings}
We analyze the month-wise results for the entire year 2019 at the Port of San Diego MG by comparing our proposed method to both the (\emph{standard} and \emph{tracking}) reference methods, each reference method being further subdivided into three test cases based on terminal constraints. 

Case~(i), where the terminal constraint set is the entire state space (i.e., no terminal constraints).

Case~(ii), where the reference methods have their terminal SOC ending at the starting SOC for the MPC given by \vspace{-1em}
{\small\setlength{\abovedisplayskip}{\dispup}\setlength{\belowdisplayskip}{\dispdn}
\begin{align}\label{eq:ref_final_initial}
x^{r}(t\!+\!N|t) = 
\begin{pmatrix}
x^r_{1}(t) \\
x^r_{2}(t+N|t) \\
x^r_{3}(t+N|t) \\
\end{pmatrix}, \; \forall t \in \mathcal{T}.
\end{align}}
\normalsize

This symbolizes a case where the MG operator wants to ensure that the BESS does not have any net energy export or import from the grid over the prediction horizon. 

Finally, we consider a Case~(iii), where the reference methods have a terminal SOC above the 50\% SOC for the MPC given by \vspace{-1em}
{\small\setlength{\abovedisplayskip}{\dispup}\setlength{\belowdisplayskip}{\dispdn}
\begin{align}\label{eq:ref_final_more_than_50}
x^{r}(t\!+\!N|t) \geq 
\begin{pmatrix}
0.5 \\
x^r_{2}(t+N|t) \\
x^r_{3}(t+N|t) \\
\end{pmatrix}, \; \forall t \in \mathcal{T}. 
\end{align}}
\normalsize

This captures the situation where the MG operator
is anxious about future demand peaks and wants to have enough storage in the BESS. 
\new{For simplicity of design, all of our simulations are run with $a(t)=b(t)=1$; i.e., the limiting case
$\lambda\to1^{-}$ of the geometric weights, to penalize the demand peaks uniformly in
time.\footnote{\new{Although
Propositions~\ref{prop_satisfy_assum_4_NCDC_OPDC},~\ref{prop_satisfy_assum_4_NCDC_OPDC_b},
and~\ref{prop_satisfy_assum_4_NCDC_OPDC_2} require $\lambda\in(0,1)$, re-running the
simulations with $\lambda<1$ is unnecessary to guarantee application of Corollary~\ref{corollary_practical}. For $\lambda$ sufficiently close to $1$
(e.g., $\lambda=1-10^{-9}$), the weights $a(t)$, $b(t)$ differ from $1$ by at most
$T(1-\lambda)<3\times10^{-6}$, below the optimality tolerance of numerical solvers, so
the computed dispatch is identical.}}}

Note that, as arbitrage is not applied for energy charges (see Table~\ref{table_parameters}), the energy cost differences between these cases are minor, and are only the result of different ending conditions for the SOC of the BESS at the end of the month. 
In Sections~\ref{yearly_results} and~\ref{stochastic_results}, we abbreviate the standard and tracking reference methods by `Std Ref' and `Track Ref'; respectively. 

\subsection{\new{Month-wise yearly analysis under perfect forecasts}}\label{yearly_results}

\begin{figure}[!t]
\centering
\includegraphics[width=0.9\columnwidth]{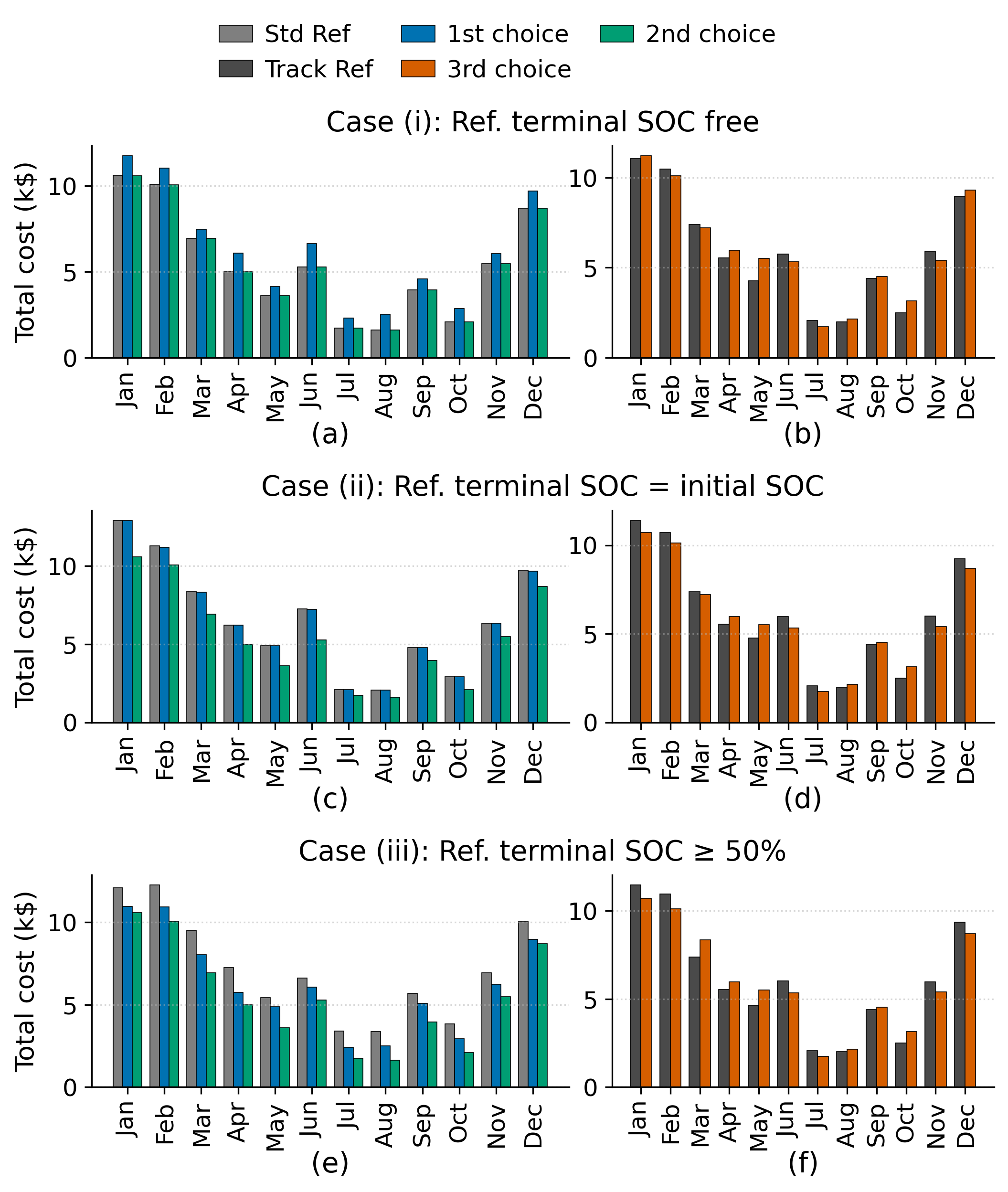}
\caption{\new{Month-wise total electricity costs for 2019 at the Port of San Diego MG under perfect forecasts. 
The left column ((a), (c), (e)) compares the Std Ref method with the $\supscr{1}{\rm st}$ and $\supscr{2}{\rm nd}$-choice methods, while the right column ((b), (d), (f)) compares the Track Ref method with the $\supscr{3}{\rm rd}$-choice method.}}
\label{fig_yearly_total}
\end{figure}

\begin{table}[!htpb]
\centering
\footnotesize
\caption{\new{Yearly (2019) total electricity costs of the reference and the proposed methods under \emph{perfect} forecasts, with the relative cost change with respect to the respective reference method in parentheses. %(negative percentages denote savings).
}}
\label{table_yearly_savings}
% \color declared outside \resizebox so the box metrics are untouched by it.
{\color{black}%
\resizebox{\columnwidth}{!}{%
\begin{tabular}{c|rrr|rr}
\toprule
 & Std Ref & $\supscr{1}{\rm st}$-choice & $\supscr{2}{\rm nd}$-choice & Track Ref & $\supscr{3}{\rm rd}$-choice \\
\midrule
Case (i)   & \$65{,}046 & \$75{,}107 ($+15.5\%$) & \$65{,}040 ($\approx0\%$)  & \$70{,}311 & \$71{,}571 ($+1.8\%$) \\
Case (ii)  & \$78{,}925 & \$78{,}719 ($-0.3\%$)  & \$65{,}041 ($-17.6\%$) & \$71{,}912 & \$70{,}472 ($-2.0\%$) \\
Case (iii) & \$86{,}455 & \$74{,}742 ($-13.5\%$) & \$65{,}041 ($-24.8\%$) & \$72{,}289 & \$71{,}626 ($-0.9\%$) \\
\bottomrule
\end{tabular}}}
\end{table}

\new{Fig.~\ref{fig_yearly_total} reports the month-wise total electricity costs (energy charges, NCDC, OPDC, and BESS usage losses) over the entire year 2019, and Table~\ref{table_yearly_savings} aggregates the corresponding yearly totals and the savings of each proposed method with respect to its reference. We benchmark the proposed $\supscr{1}{\rm st}$- and $\supscr{2}{\rm nd}$-choice methods against the Std Ref method, and the $\supscr{3}{\rm rd}$-choice method against the Track Ref method, as the $\supscr{3}{\rm rd}$-choice method is designed around the reference peaks laid out by the Track Ref method (cf.~Appendix~\ref{choice_3_motivation}). Refer to Appendix~\ref{appendix_huge_tables_deterministic} for a month-wise and electricity cost component-wise breakdown of these different methods.

Additionally, Fig.~\ref{fig_closed_loop_jan} shows the closed-loop mechanism behind the cost differences for January---a representative month, as the sign of its cost difference agrees with that of the yearly aggregate for all nine reference-method pairings in Fig.~\ref{fig_yearly_total}. The peak-setting events that fix the monthly demand charges concentrate within the first few days, so the SOC trajectory over that window largely determines the January monthly bill.

\begin{figure}[!htpb]
\centering
\includegraphics[width=0.9\columnwidth]{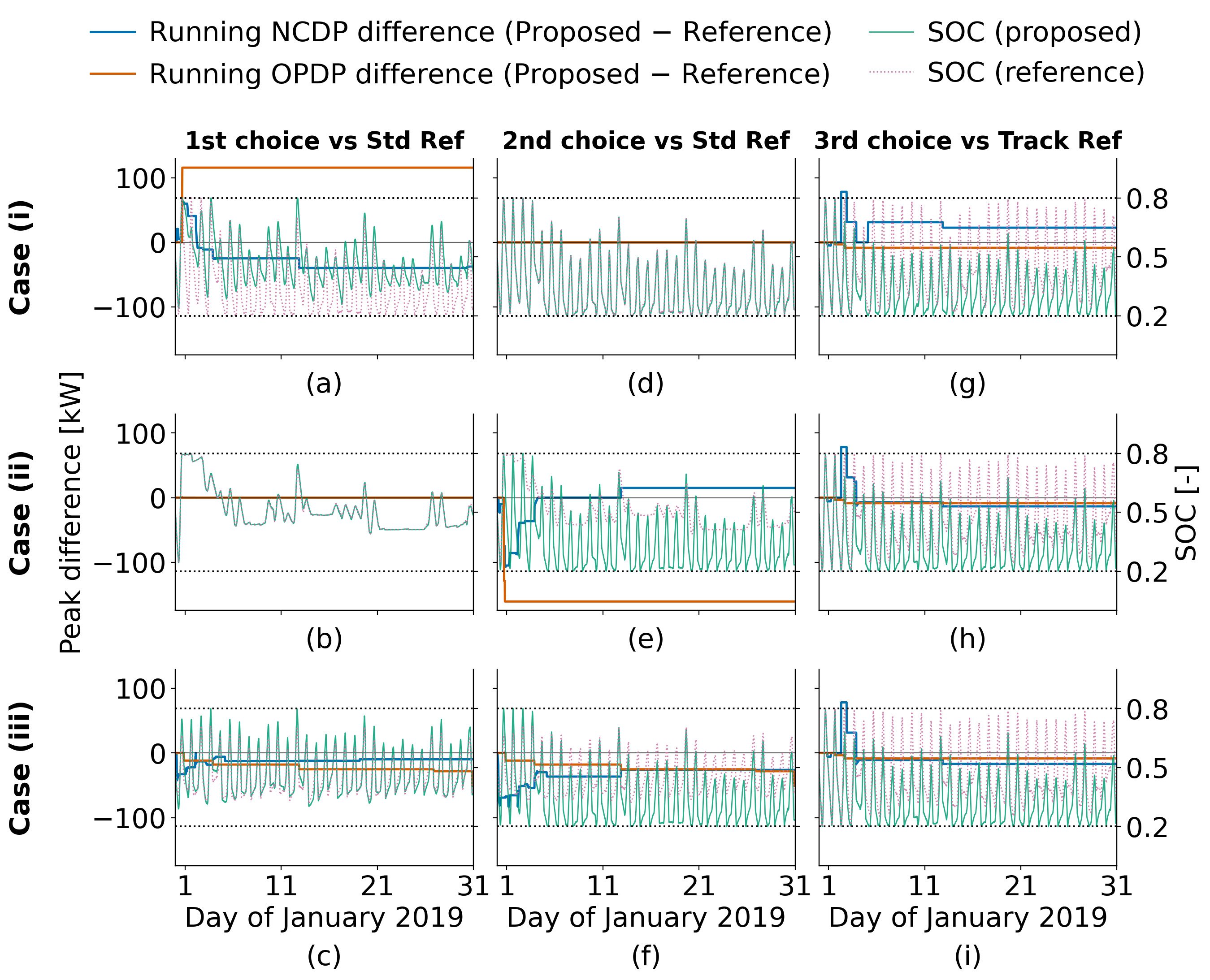}
\caption{\new{Closed-loop trajectories for Cases (i)--(iii) (rows) and the three reference-method pairings (columns) under perfect forecasts. Left axes: running NCDP and OPDP differences. Right axes: closed-loop SOC of the proposed and reference methods; the two horizontal black dotted lines mark the hard SOC limits $\subscr{\rm SOC}{min}$ and $\subscr{\rm SOC}{max}$.}}
\label{fig_closed_loop_jan}
\end{figure}

\emph{Comparison of $\supscr{1}{\rm st}$-choice method with Std Ref:} In Case (i), the $\supscr{1}{\rm st}$-choice method is outperformed by the Std Ref method in every month (a $15.5\%$ yearly cost increase -- see Fig.~\ref{fig_yearly_total}(a)), majorly because of higher OPDC. The terminal region~\eqref{candidate_terminal_region_NCDC_OPDC} in the $\supscr{1}{\rm st}$-choice method forces the BESS to return, at the end of every prediction horizon, to the SOC the reference realized $N$ steps earlier, while the Std Ref method is itself unconstrained. This leaves the $\supscr{1}{\rm st}$-choice method with little flexibility in BESS discharge to cover the on-peak window: resulting in it paying substantially more OPDC yearly than the Std Ref method, which discharges freely across the on-peak windows (cf. Table~\ref{table_breakdown_opdc}). For January, this behavior can be seen in Fig.~\ref{fig_closed_loop_jan}(a), with the SOC time-series of the $\supscr{1}{\rm st}$-choice method remaining consistently above the Std Ref as the terminal anchor holds charge through the
on-peak windows that the Std Ref method is able to spend. We also observe from Fig.~\ref{fig_closed_loop_jan}(a) that the OPDP penalty of the $\supscr{1}{\rm st}$-choice method is locked in within the first day. Taking advantage of this already accrued OPDP, it shifts energy from the off-peak to on-peak period for the rest of the month, leading to a lower NCDP than Std Ref.

In Case (ii), the $\supscr{1}{\rm st}$-choice method has approximately the same costs as the Std Ref method. As both methods start from the same initial SOC, the Case (ii) terminal constraint of the Std Ref~\eqref{eq:ref_final_initial} renders the two formulations equivalent (same terminal constraints and equivalent objective function because of redundant double penalization of terminal peaks in~\eqref{substitute_objective_1b_NCDC_OPDC}; see also Section~\ref{subsec:third_choice}). The closed-loop plots in Fig.~\ref{fig_closed_loop_jan}(b) are likewise identical.

In Case (iii), the $\supscr{1}{\rm st}$-choice method outperforms the Std Ref method in every month of the year (Fig.~\ref{fig_yearly_total}(e)), saving $13.5\%$ annually majorly because of lower NCDC and OPDC, as the $50\%$ terminal SOC requirement in~\eqref{eq:ref_final_more_than_50} makes the Std Ref too conservative.  This can be observed in Fig.~\ref{fig_closed_loop_jan}(c), where the Std Ref SOC keeps off the lower limit, whereas the $\supscr{1}{\rm st}$-choice method sweeps more of the entire $0.2$--$0.8$ band, depletes more ahead of peaks, and avoids the higher NCDP and OPDP that the Std Ref method incurs.

\emph{Comparison of $\supscr{2}{\rm nd}$-choice method with Std Ref:} The $\supscr{2}{\rm nd}$-choice method, in contrast, is never worse than the Std Ref method in any month of any case (Fig.~\ref{fig_yearly_total}(a), (c), (e)), and its yearly cost is nearly identical across the three cases. Note that the electricity costs of the $\supscr{2}{\rm nd}$-choice method recover the performance of the Std Ref in Case (i), which is a direct consequence of its relaxed terminal region~\eqref{candidate_terminal_region_NCDC_OPDC_2}, which can also be seen in the identical closed-loop plot of Fig.~\ref{fig_closed_loop_jan}(d). The resulting yearly savings of the $\supscr{2}{\rm nd}$-choice method therefore grow with the conservatism of the reference: $\approx0\%$ for Case (i), $17.6\%$ for Case (ii), and $24.8\%$ for Case (iii); respectively. Analogous behavior can be seen in Figs.~\ref{fig_closed_loop_jan}(e), (f), where the relaxed terminal constraint of the $\supscr{2}{\rm nd}$-choice method as compared to the Std Ref in Cases (ii) and (iii) respectively, makes the BESS SOC sweep more of the $0.2$--$0.8$ band, leading to deeper discharges and overall demand charge savings.

\emph{Comparison of $\supscr{3}{\rm rd}$-choice method with Track Ref:} The Track Ref baseline is markedly less sensitive to its terminal constraints, with yearly totals between $\$70.3$k and $\$72.3$k across the three cases, versus $\$65.0$k--$\$86.5$k for the Std Ref method, and is a stronger baseline for Cases (ii) and (iii). The $\supscr{3}{\rm rd}$-choice method (Fig.~\ref{fig_yearly_total}(b), (d), (f)) leads to a modest $1.8\%$ yearly cost increase for Case (i), and modest yearly savings of $2.0\%$ and $0.9\%$ for Cases (ii) and (iii); respectively, with the monthly comparisons split roughly evenly in each case. This behavior, also seen in Figs.~\ref{fig_closed_loop_jan}(g)--(i), and analogous to the $\supscr{2}{\rm nd}$-choice method comparisons, is explained by the conservatism of the Track Ref (Case (i) is less conservative than Cases (ii) and (iii)).

Overall, the proposed methods deliver their largest gains when operator-imposed terminal constraints render the reference conservative (Cases (ii) and (iii)), and match an already strong reference otherwise (Case (i), $\supscr{2}{\rm nd}$-choice). We recall that Corollary~\ref{corollary_practical} certifies no-worse performance up to a constant which, at the $\lambda\to1^{-}$ operating point used here, is large (cf.~Remark~\ref{lambda_tradeoff}); finite-time deviations in the worse direction, as observed for the $\supscr{1}{\rm st}$ and $\supscr{3}{\rm rd}$-choice methods in Case (i), are therefore consistent with the theory, and the reported improvements are empirical gains obtained over and above the baseline-relative guarantee in Appendix~\ref{appendix_upper_bound}.

\subsection{\new{Empirical extension under imperfect forecasts}}\label{stochastic_results}

We now repeat the study of Section~\ref{yearly_results} with the imperfect load and PV forecasts of~\cite{ghosh2025adaptive}. Note that the performance guarantee of Corollary~\ref{corollary_practical} is
established for a deterministic system, so the results of this
section are not theorem certified, and are empirical.

The only change required in the algorithm is in the terminal cost penalty:~\eqref{candidate_terminal_cost_1_NCDC_OPDC} for the $\supscr{1}{\rm st}$- and
$\supscr{2}{\rm nd}$-choice, and~\eqref{candidate_terminal_cost_2_NCDC_OPDC} for the $\supscr{3}{\rm rd}$-choice terminal
cost; respectively. Instead of penalizing against $x^r_{2}(t\!+\!1)$ and
$x^r_{3}(t\!+\!1)$ which are not known when the optimization at $t$
is solved, we penalize with respect to the reference's \emph{predicted} peaks
$\hat{x}^r_{2}(t\!+\!1|t)$ and $\hat{x}^r_{3}(t\!+\!1|t)$. See Appendix~\ref{appendix_huge_tables_stochastic} for additional discussions on the recursive feasibility and the feasibility of the initial state.

Table~\ref{table_yearly_savings_imperfect} reports the yearly totals, with their month-wise and cost component-wise breakdown in Appendix~\ref{appendix_huge_tables_stochastic}, obtained under
imperfect forecasts. The differences with the perfect forecast results are traceable almost entirely to the demand charges (cf.~Tables~\ref{table_breakdown_opdc} and~\ref{table_breakdown_opdc_stochastic}). Here, the mismatch between the realized and forecast net load must be absorbed in real-time, and whenever the planned BESS dispatch leaves no SOC margin, the shortfall is imported from the grid, irreversibly increasing the demand charges. An economic dispatch without terminal constraints plans the BESS discharge tightly around the \emph{forecast} net load and carries no such margin, whereas the operator imposed terminal SOC constraints act as a robustness margin, so the case-wise ordering of Table~\ref{table_yearly_savings} inverts: the Std Ref yearly total now \emph{falls} from Case (i) to Case (iii), due to increasing conservativeness. The proposed methods inherit exactly the margin, or lack thereof, of their terminal ingredients: the $\supscr{1}{\rm st}$-choice terminal region~\eqref{candidate_terminal_region_NCDC_OPDC} holds charge through the on-peak windows and thus outperforms the Std Ref method substantially in Case (i), modestly in Case (ii), while performing on par with it in Case (iii). The $\supscr{2}{\rm nd}$- and $\supscr{3}{\rm rd}$-choice methods, with the free terminal region~\eqref{candidate_terminal_region_NCDC_OPDC_2}, discard this margin, and their yearly totals are correspondingly high and nearly insensitive to the case. Finally, the Track Ref method is now the strongest baseline, as its tracking objective~\eqref{tracking_objective} discharges the BESS through every on-peak window irrespective of the forecast accuracy, keeping its OPDC low. %at the price of the highest BESS usage losses (Table~\ref{table_breakdown_bess_stochastic}) --- a protection the $\supscr{3}{\rm rd}$-choice method forgoes by reinstating the economic objective without terminal constraints.

}

\begin{table}[!t]
\centering
\footnotesize
\caption{\new{Yearly (2019) total electricity costs of the reference and the proposed
methods under \emph{imperfect} forecasts, with the relative cost change with
respect to the respective reference method in parentheses.}}
\label{table_yearly_savings_imperfect}
% \color declared outside \resizebox so the box metrics are untouched by it.
{\color{black}%
\resizebox{\columnwidth}{!}{%
\begin{tabular}{c|rrr|rr}
\toprule
 & Std Ref & $\supscr{1}{\rm st}$-choice & $\supscr{2}{\rm nd}$-choice & Track Ref & $\supscr{3}{\rm rd}$-choice \\
\midrule
Case (i)   & \$142{,}978 & \$123{,}752 ($-13.4\%$) & \$143{,}834 ($+0.6\%$)  & \$91{,}009 & \$147{,}084 ($+61.6\%$) \\
Case (ii)  & \$94{,}291  & \$93{,}132 ($-1.2\%$)   & \$144{,}054 ($+52.8\%$) & \$91{,}266 & \$147{,}083 ($+61.2\%$) \\
Case (iii) & \$89{,}974  & \$90{,}132 ($+0.2\%$)   & \$143{,}568 ($+59.6\%$) & \$89{,}589 & \$146{,}880 ($+63.9\%$) \\
\bottomrule
\end{tabular}}}
\end{table}

\section{Conclusions and future work}\label{conclusions}
%\vspace{-4cm}
The paper presents an economic MPC (EMPC) formulation for a generic deterministic discrete non-linear time-varying system. We first prove that under appropriate terminal ingredients, the asymptotic average closed-loop cost of the proposed EMPC is no worse than the asymptotic average closed-loop cost of an arbitrary reference trajectory for the same system.
A practical finite-time guarantee is also derived which quantifies the upper bound as a function of time that decreases linearly to zero as time goes to infinity.
Unlike previous works, we do not assume that the reference trajectory is known a-priori for all future times, which is critical in applications.

We then solve an optimal MG dispatch problem, which can be reformulated to the conditions for our proposed EMPC formulation ensuring practical performance guarantees. A realistic case study using PV and load forecast data from the microgrid at the Port of San Diego shows that our method is capable of reducing monthly electricity costs as compared to reference trajectories generated by directly optimizing the electricity cost function over the prediction horizon or by tracking an ideal grid import curve. Future work will focus on the design of techniques which can extend similar guarantees as this work for stochastic systems.

% \section*{Acknowledgements}
% CC acknowledges financial support from CONICYT PFCHA/DOCTORADO BECAS CHILE/2017 and the project ‘‘Solar Energy Research Center’’ - SERC-Chile (ANID/CIN250043).

\enlargethispage{3\baselineskip}
\newcounter{bibtrigger}
\let\origbibitem\bibitem
\renewcommand{\bibitem}{%
  \stepcounter{bibtrigger}%
  \ifnum\value{bibtrigger}=7\enlargethispage{2\baselineskip}\fi
  \origbibitem}

\bibliographystyle{plain}        % Include this if you use bibtex 
\bibliography{automatica}

\newpage
\appendix

\section{Motivation for the $\supscr{3}{\rm rd}$-choice problem specification for a type of reference trajectories}\label{choice_3_motivation}

The \emph{standard reference method} previously outlined, optimizes the actual economic cost over a prediction horizon subject to input and state constraints. As we perform a finite-horizon optimization over a much smaller time scale than a month, the standard reference method might provide a solution that performs poorly in closed-loop despite its \emph{a-priori}, open-loop, optimality. %\footnote{Additionally, sometimes, terminal constraints are imposed on the standard reference method by the MG operator from rule-of-thumb or domain knowledge expertise to compensate for this time scale mismatch (see Section~\ref{ref_terminal_case_study_settings} later). These terminal constraints, although intended to limit the aggressive discharge of BESS during a prediction horizon in anticipation of future peak loads, can often worsen system performance by being overtly conservative.} %\change{high quality open-loop solutions}. 
For example, once the standard reference method optimizer concludes that a predicted peak will be reached within the prediction horizon (howsoever far from the present it might be), the optimizer does not have any incentive to aggressively discharge the BESS at the first time-step to beyond what is necessary to equalize the grid import with the future predicted peak. However, the future predicted peak might never be realized in closed-loop as more forecast data becomes available while moving ahead in time. And once a poor decision has been made in the present time-step, as MPC applies the first time-step computed control in closed-loop, the solution cannot be corrected later, resulting in higher NCDP and OPDP overall. 

It is discussed in~\cite{ghosh2022effects} that if $\frac{R_{\rm NC}}{R_{\rm OP}}<\frac{\mu_{\rm off}}{\mu_{\rm on}}$, where $\mu_{\rm off}$ hours, and $\mu_{\rm on}$ hours, are the number of off-peak,\footnote{Off-peak hours are all hours of the day except the on-peak hours, i.e., all hours except the on-peak hours 16:00--21:00~h.} and on-peak hours of the day; respectively, it is economical to spread the grid import over the off-peak hours and keep the on-peak load at 0. The intuition behind this condition lies in the analysis of an increase of the NCDP, $P_{\rm NCDP}$ kW, and an increase of the OPDP, $P_{\rm OPDP}$ kW,  
from importing $\omega_{\rm GI}$ kWh 
from the grid in either off-peak or on-peak hours. If $R_{\rm NC}P_{\rm NCDP}<R_{\rm OP}P_{\rm OPDP}$, then adding the grid import over the off-peak hours is more economical than adding it over the on-peak hours. As the uniform addition of grid import over the available hours, by definition, leads to least peak load, we substitute $P_{\rm NCDP}=\frac{\omega_{\rm GI}}{\mu_{\rm off}}$ and $P_{\rm OPDP}=\frac{\omega_{\rm GI}}{\mu_{\rm on}}$ in the above inequality, 
which gives the condition $\frac{R_{\rm NC}}{R_{\rm OP}}<\frac{\mu_{\rm off}}{\mu_{\rm on}}$. 

Thus, drawing inspiration from the above discussion, as $\frac{R_{\rm NC}}{R_{\rm OP}}<\frac{\mu_{\rm off}}{\mu_{\rm on}}$ holds for most realistic electricity tariff structures~\cite{ghosh2022effects}, we calculate another reference trajectory by solving the receding horizon optimization with an %economic 
objective function that measures the deviation from a reference trajectory corresponding to an ideal grid import. This ideal grid import just spreads the entire forecast net load over the off-peak hours. The objective function for such a reference trajectory is given in~\eqref{tracking_objective} in Appendix~\ref{appendix_tertiary}. %with a detailed explanation of the procedure and logic behind calculating the ideal trajectory. 
The reference trajectory created from the method in this section is referred to as the \emph{tracking reference method}, as opposed to the \emph{standard reference method}.

\new{As the tracking reference method tries to minimize the deviation from an ideal grid import trajectory, it typically dispatches the BESS more aggressively than the standard reference method, suppressing the next-step predicted running peaks (lower $x^r_{2}(t+1|t)$ and $x^r_{3}(t+1|t)$) at the expense of higher predicted terminal peaks ($x^r_{2}(t+N|t)$ and $x^r_{3}(t+N|t)$). Therefore, in open-loop, the tracking reference method might be worse (or at best no better) than the standard reference method; in closed-loop, however, the distinction between next-step and terminal predicted peaks becomes crucial. In a receding horizon implementation, only the first control input is applied, so the next-step predicted peaks $x_{2}(t+1|t)$ and $x_{3}(t+1|t)$ are the only predicted peaks that are realized in closed-loop (exactly so under perfect forecasts). A peak predicted later in the horizon, in contrast, may never be realized at all: the controller re-optimizes at every time-step with fresh forecast data, and can revise the dispatch before the predicted peak occurs. Moreover, by the max-dynamics~\eqref{augmented_dynamics_NCDC_OPDC}, a realized peak is irreversible: once $x_{2}(t+1)$ or $x_{3}(t+1)$ increases, the correspondingly incurred demand charge cannot be undone for the remainder of the month. Suppressing the increase of the next-step peaks is thus disproportionately valuable compared to suppressing peaks predicted later in the horizon, which is precisely the closed-loop behavior that makes the tracking reference method empirically superior. Note, however, that there is no theoretical guarantee that the closed-loop economic costs of the tracking reference method (which explicitly does not consider economics) are always lower than the standard reference method -- and its superiority is based only on empirical findings.

The $\supscr{3}{\rm rd}$-choice method incorporates this insight while including the economic cost explicitly in the objective function (from~\eqref{augmented_stage_cost_NCDC_OPDC}), giving it the ability to relatively weigh the penalty on the predicted peaks in the prediction horizon temporally which is practically useful, and often leads to better performance (see Section~\ref{yearly_results}).}

\section{Objective function for the optimal MG dispatch problem for the reference methods}\label{appendix_tertiary}
{
\renewcommand{\theequation}{\thesection.\arabic{equation}}
  % redefine the command that creates the equation no.
  \setcounter{equation}{0}  % reset counter

The objective function for the standard reference method is given by
\small{\setlength{\abovedisplayskip}{\dispup}
\setlength{\belowdisplayskip}{0pt}
\setlength{\abovedisplayshortskip}{0pt}
\setlength{\belowdisplayshortskip}{0pt}
\setlength{\jot}{-2pt}
\begin{align}\label{substitute_reference_1a_NCDC_OPDC}
&\tilde{V}_{N}(x^{r}(t),{\mathbf{u}^r(x^{r}(t),t)},t)\!\nonumber \\
&=\! \sum_{k=0}^{N-1}\!\Bigl[R_{\rm EC}(t+k)\Delta t\Bigl(u^{r}_{2}(t+k|t) +\frac{(1-\eta)}{2}|u^r_{1}(t+k|t)|\Bigr)\Bigr] \nonumber \\
&\quad+ R_{\rm NC}\Bigl(a(t+N)x^r_{2}(t+N|t)\Bigr) \nonumber\\
&\quad+ R_{\rm OP}\Bigl(b(t+N)x^r_{3}(t+N|t)\Bigr).
\end{align}}
\normalsize

\new{The objective function for the tracking reference method is given by
\small{\setlength{\abovedisplayskip}{\dispup}
\setlength{\belowdisplayskip}{\dispdn}
\setlength{\abovedisplayshortskip}{0pt}
\setlength{\belowdisplayshortskip}{0pt}
\setlength{\jot}{0pt}
\begin{align}\label{tracking_objective}
&\tilde{V}_{N}(x^{r}(t),{\mathbf{u}^r(x^{r}(t),t)},t) \nonumber \\
&=\left\|P(t)\max \bigl[ \bigl(\mathbf{u}^r_{2}(x^{r}(t),t)-\bar{\mathbf{u}}^r_{2}(t)\bigr), \mymathbb{0}_N \bigr]\right\|_2,
\end{align}
}}}% 3rd brace closes the \theequation scoping group opened after \section (a \new{...} absorbed its old closer)
\normalsize

\noindent where $\bar{\mathbf{u}}^r_{2}(t)$ is the ideal grid import trajectory over the prediction horizon and $P(t)$ is the penalty matrix. $\bar{\mathbf{u}}^r_{2}(t)$ is composed of $0$ for the OP period time-steps of the prediction horizon, with the off-peak period elements being $\max\Bigl(x^r_{2}(t),-\frac{\sum_{k=0}^{N-1}c(t+k|t)}{N_{\rm off}}\Bigr)$, where $N_{\rm off}$ is the length of the prediction horizon coinciding with the off-peak period. $P(t)$ is a diagonal matrix that penalizes the positive deviation of the predicted grid import from the ideal, with more weighting for on-peak periods, as OPDC are charged on top of NCDC. $P(t)$ is composed of $1$ for off-peak period time-steps of the prediction horizon, and $\frac{R_{\rm NC}+R_{\rm OP}}{R_{\rm NC}}$ for the OP period time-steps of the prediction horizon, symbolizing that increasing demand during OP periods can increase both NCDC and OPDC.

%We do not use the running OPDP, $x^r_{3}(t)$, in $\bar{\mathbf{u}}^r_{2}(t)$, as increasing the ideal grid import trajectory in the OP period time-steps further risks aggravating future OP period peaks. However, in some situations, not increasing the ideal grid import trajectory over the OP period time-steps when a prior OPDP has been reached could increase the NCDP beyond necessary, especially if the NCDP happens after the OPDP. If the running OPDP is included in the ideal grid import trajectory, the time-steps of $\bar{\mathbf{u}}^r_{2}(t)$ corresponding to the OP period would be modified as $\max\bigl(0,x^r_{3}(t)\bigr)$. 

\section{Design parameters of the MG}\label{appendix_table}

\new{The design parameters of the MG, and the prediction horizon and sampling period common to the proposed and the reference methods, are listed in Table~\ref{table_parameters}.}

\begin{table}[ht]
% \normalsize
\centering
\caption{Design parameters of the MG at the Port of San Diego~\cite{ghosh2025adaptive}.}
\setlength{\belowcaptionskip}{2pt}%
\renewcommand{\arraystretch}{0.9}%
\resizebox{\columnwidth}{!}{%
{\begin{tabular}{lcc}
\cline{1-3}
Parameter            & Symbol              &  Value  \\ 
\cline{1-3}
NCDC  rate           & $R_{\rm NC}$       & \$24.48/kW        \\
OPDC  rate           & $R_{\rm OP}$       & \$19.19/kW       \\
Energy rate          & $R_{\rm EC}(t), \forall t$       & \$0.1/kWh \\
BESS round-trip efficiency      & $\eta$                &  0.8  \\
BESS energy capacity       & $\subscr{\rm BESS}{en}$     & 2,500 kWh \\
BESS power capacity  & $\subscr{\rm BESS}{max}$        &700 kW     \\
Upper bound of SOC   & $\subscr{\rm SOC}{max}$          & 0.8    \\
Lower bound of SOC   & $\subscr{\rm SOC}{min}$          & 0.2    \\
Initial SOC  & $x_{1}(0)$ and $x^r_{1}(0)$        &       0.5     \\
\new{Sampling period}      & \new{$\Delta t$}          & \new{0.25 h} \\
\new{Prediction horizon}   & \new{$N$}                 & \new{96 (one day ahead)} \\
\cline{1-3}
\end{tabular}}
}
\label{table_parameters}
%\vspace{-2em}
\end{table}

\new{
\section{Upper bound of the monthly cost difference with the reference}\label{appendix_upper_bound}
The constant $V_N^0\bigl(x(0),0\bigr)-C$ in Corollary~\ref{corollary_practical} can be upper bounded explicitly for each problem specification through the respective bound $\sup_{t\geq N}h(t)\leq\bar{C}$ of Propositions~\ref{prop_satisfy_assum_4_NCDC_OPDC},~\ref{prop_satisfy_assum_4_NCDC_OPDC_b}, and~\ref{prop_satisfy_assum_4_NCDC_OPDC_2}, and the bounds on the control inputs. Consider the stage cost~\eqref{augmented_stage_cost_NCDC_OPDC}. As $0\leq R_{\rm EC}(t)\leq\bar R_{\rm EC}$, $u_{2}(t)\in[\hat a,\hat b]$ with $\hat a<0$, and $|u_{1}(t)|\leq{\rm BESS}_{\rm max}$, the energy charge and BESS loss terms lie, per time-step, between $\bar R_{\rm EC}\Delta t\,\hat a$ and $\bar R_{\rm EC}\Delta t\bigl(\hat b+\frac{(1-\eta)}{2}{\rm BESS}_{\rm max}\bigr)$. The demand charge increments are non-negative, as $a(t)$, $b(t)$ are nondecreasing and the running peaks are non-negative and nondecreasing by~\eqref{augmented_dynamics_NCDC_OPDC}, and they telescope over the prediction horizon to at most $R_{\rm NC}\,a(N)x_{2}(N|0)+R_{\rm OP}\,b(N)x_{3}(N|0)\leq(R_{\rm NC}+R_{\rm OP})\hat b$. Finally, using $h(N)=0$ from the construction~\eqref{h_construction}, the terminal cost at time $0$ contributes at most $\sup_{x}W(x,N)\leq(R_{\rm NC}+R_{\rm OP})\hat b$, while, as $W\geq0$, the terminal cost satisfies $V_f=W-h\geq-\sup_{t\geq N}h(t)\geq-\bar{C}$ at all times. Hence, for all three problem specifications, $V_N^0\bigl(x(0),0\bigr)\leq N\bar R_{\rm EC}\Delta t\bigl(\hat b+\frac{(1-\eta)}{2}{\rm BESS}_{\rm max}\bigr)+2(R_{\rm NC}+R_{\rm OP})\hat b$, and the lower bound of $V_N^0$ in Corollary~\ref{corollary_practical} can be taken as $C=N\bar R_{\rm EC}\Delta t\,\hat a-\bar{C}$. Combining the two,
{\small\setlength{\abovedisplayskip}{\dispup}\setlength{\belowdisplayskip}{\dispdn}
\begin{align}\label{monthly_upper_bound}
T\epsilon&= V_N^0\bigl(x(0),0\bigr)-C\nonumber\\
&\leq N\bar R_{\rm EC}\Delta t\Bigl(\hat b-\hat a+\tfrac{(1-\eta)}{2}{\rm BESS}_{\rm max}\Bigr)\nonumber\\
&\quad+2(R_{\rm NC}+R_{\rm OP})\hat b+\bar{C},
\end{align}}%
\normalsize

where the only specification-dependent term is $\bar{C}$, given by~\eqref{h_sup_bound_choice_1} for the $\supscr{1}{\rm st}$ choice, and by~\eqref{h_sup_bound_choice_2} and~\eqref{h_sup_bound_choice_3} for the $\supscr{2}{\rm nd}$ and $\supscr{3}{\rm rd}$ choices; respectively. That is, the looser the certified bound on the terminal cost offset $\sup_{t\geq N}h(t)$, the looser the monthly-bill guarantee. %, with the dominant contribution $(R_{\rm NC}+R_{\rm OP})\tfrac{2\hat b}{1-\lambda}$ common to all three specifications (cf.~Remark~\ref{lambda_tradeoff}). 
All the quantities in the RHS of~\eqref{monthly_upper_bound} are known to the MG operator a-priori, rendering the bound computable at design time.}

\new{\section{Month-wise results under perfect forecasts}\label{appendix_huge_tables_deterministic}
Tables~\ref{table_breakdown_ncdc}--\ref{table_breakdown_total} give the month-wise breakdown of the year-2019 electricity costs behind Fig.~\ref{fig_yearly_total} and Table~\ref{table_yearly_savings}, one table per cost component, under perfect forecasts. The \emph{Yearly} row of Table~\ref{table_breakdown_total} reproduces Table~\ref{table_yearly_savings}. Note that the energy charges are negative in the summer months, when the MG exports enough PV generation to be a net seller of energy over the month---the violation of positive definiteness of the performance metric discussed in Section~\ref{intro}.}

\begin{table*}[!htpb]
\centering
\footnotesize
\caption{\new{Month-wise NCDC for the year 2019 under \emph{perfect} load and PV forecasts for each case and method. `Std' and `Track' abbreviate the Std Ref and Track Ref methods; `1st', `2nd' and `3rd' the correspondingly numbered proposed choices.}}
\label{table_breakdown_ncdc}
% Fixed font rather than \resizebox: scaling each table to \textwidth would
% set the narrow ones (OPDC) far larger than the wide ones (NCDC).
{\color{black}\scriptsize\setlength{\tabcolsep}{3.3pt}%
\begin{tabular}{l|rrrrr|rrrrr|rrrrr}
\toprule
 & \multicolumn{5}{c}{Case (i)} & \multicolumn{5}{c}{Case (ii)} & \multicolumn{5}{c}{Case (iii)} \\
\cmidrule(lr){2-6}\cmidrule(lr){7-11}\cmidrule(lr){12-16}
 & Std & 1st & 2nd & Track & 3rd & Std & 1st & 2nd & Track & 3rd & Std & 1st & 2nd & Track & 3rd \\
\midrule
January & 4{,}358 & 3{,}438 & 4{,}358 & 4{,}460 & 5{,}015 & 3{,}989 & 3{,}978 & 4{,}357 & 4{,}821 & 4{,}492 & 4{,}999 & 4{,}760 & 4{,}357 & 4{,}906 & 4{,}492 \\
February & 4{,}913 & 5{,}529 & 4{,}913 & 5{,}100 & 4{,}954 & 5{,}667 & 5{,}586 & 4{,}913 & 5{,}366 & 4{,}957 & 5{,}979 & 5{,}653 & 4{,}913 & 5{,}571 & 4{,}963 \\
March & 4{,}313 & 3{,}945 & 4{,}313 & 4{,}331 & 4{,}705 & 5{,}117 & 5{,}069 & 4{,}313 & 4{,}333 & 4{,}705 & 6{,}392 & 5{,}531 & 4{,}313 & 4{,}334 & 5{,}706 \\
April & 4{,}063 & 3{,}936 & 4{,}063 & 4{,}006 & 5{,}107 & 4{,}858 & 4{,}858 & 4{,}063 & 4{,}006 & 5{,}107 & 6{,}003 & 4{,}931 & 4{,}063 & 4{,}006 & 5{,}107 \\
May & 2{,}968 & 2{,}916 & 2{,}968 & 3{,}022 & 5{,}062 & 4{,}167 & 4{,}167 & 2{,}968 & 3{,}544 & 5{,}062 & 4{,}995 & 4{,}378 & 2{,}968 & 3{,}430 & 5{,}062 \\
June & 3{,}947 & 4{,}049 & 3{,}947 & 4{,}125 & 3{,}999 & 4{,}118 & 4{,}092 & 3{,}947 & 4{,}353 & 3{,}998 & 5{,}344 & 4{,}320 & 3{,}947 & 4{,}411 & 3{,}997 \\
July & 3{,}740 & 3{,}521 & 3{,}740 & 3{,}704 & 3{,}827 & 4{,}316 & 4{,}316 & 3{,}740 & 3{,}704 & 3{,}827 & 5{,}736 & 4{,}680 & 3{,}740 & 3{,}704 & 3{,}827 \\
August & 3{,}445 & 3{,}433 & 3{,}445 & 3{,}416 & 4{,}083 & 3{,}524 & 3{,}524 & 3{,}445 & 3{,}416 & 4{,}083 & 5{,}388 & 4{,}451 & 3{,}445 & 3{,}416 & 4{,}083 \\
September & 3{,}579 & 3{,}644 & 3{,}579 & 3{,}562 & 4{,}264 & 4{,}514 & 4{,}514 & 3{,}579 & 3{,}562 & 4{,}264 & 5{,}515 & 4{,}849 & 3{,}579 & 3{,}562 & 4{,}264 \\
October & 2{,}593 & 2{,}336 & 2{,}593 & 2{,}446 & 3{,}762 & 3{,}466 & 3{,}466 & 2{,}593 & 2{,}446 & 3{,}762 & 4{,}497 & 3{,}514 & 2{,}593 & 2{,}446 & 3{,}762 \\
November & 2{,}964 & 2{,}749 & 2{,}964 & 3{,}004 & 3{,}004 & 3{,}942 & 3{,}942 & 2{,}964 & 3{,}101 & 3{,}004 & 4{,}163 & 3{,}858 & 2{,}964 & 3{,}065 & 3{,}004 \\
December & 3{,}977 & 3{,}503 & 3{,}977 & 3{,}904 & 4{,}509 & 3{,}916 & 3{,}867 & 3{,}977 & 4{,}235 & 3{,}984 & 4{,}762 & 4{,}020 & 3{,}977 & 4{,}359 & 3{,}977 \\
\midrule
\textbf{Yearly} & 44{,}861 & 42{,}999 & 44{,}860 & 45{,}080 & 52{,}292 & 51{,}595 & 51{,}380 & 44{,}860 & 46{,}887 & 51{,}245 & 63{,}774 & 54{,}944 & 44{,}860 & 47{,}210 & 52{,}244 \\
\bottomrule
\end{tabular}}
\end{table*}

\begin{table*}[!htpb]
\centering
\footnotesize
\caption{\new{Month-wise OPDC for the year 2019 under \emph{perfect} load and PV forecasts for each case and method. `Std' and `Track' abbreviate the Std Ref and Track Ref methods; `1st', `2nd' and `3rd' the correspondingly numbered proposed choices.}}
\label{table_breakdown_opdc}
% Fixed font rather than \resizebox: scaling each table to \textwidth would
% set the narrow ones (OPDC) far larger than the wide ones (NCDC).
{\color{black}\scriptsize\setlength{\tabcolsep}{3.3pt}%
\begin{tabular}{l|rrrrr|rrrrr|rrrrr}
\toprule
 & \multicolumn{5}{c}{Case (i)} & \multicolumn{5}{c}{Case (ii)} & \multicolumn{5}{c}{Case (iii)} \\
\cmidrule(lr){2-6}\cmidrule(lr){7-11}\cmidrule(lr){12-16}
 & Std & 1st & 2nd & Track & 3rd & Std & 1st & 2nd & Track & 3rd & Std & 1st & 2nd & Track & 3rd \\
\midrule
January & 0 & 2{,}217 & 0 & 163 & 0 & 3{,}082 & 3{,}079 & 0 & 163 & 0 & 972 & 0 & 0 & 163 & 0 \\
February & 0 & 412 & 0 & 0 & 0 & 547 & 546 & 0 & 0 & 0 & 1{,}284 & 157 & 0 & 48 & 0 \\
March & 100 & 1{,}043 & 100 & 232 & 0 & 857 & 857 & 100 & 232 & 0 & 740 & 0 & 100 & 232 & 218 \\
April & 0 & 1{,}345 & 0 & 302 & 0 & 559 & 559 & 0 & 302 & 0 & 472 & 0 & 0 & 302 & 0 \\
May & 0 & 599 & 0 & 341 & 0 & 238 & 237 & 0 & 341 & 0 & 0 & 0 & 0 & 341 & 0 \\
June & 0 & 1{,}374 & 0 & 0 & 0 & 1{,}991 & 1{,}991 & 0 & 0 & 0 & 0 & 439 & 0 & 0 & 0 \\
July & 0 & 820 & 0 & 168 & 0 & 0 & 0 & 0 & 168 & 0 & 0 & 0 & 0 & 168 & 0 \\
August & 0 & 994 & 0 & 97 & 0 & 393 & 393 & 0 & 97 & 0 & 0 & 0 & 0 & 97 & 0 \\
September & 0 & 634 & 0 & 193 & 0 & 0 & 0 & 0 & 193 & 0 & 0 & 0 & 0 & 193 & 0 \\
October & 0 & 1{,}051 & 0 & 241 & 0 & 0 & 0 & 0 & 241 & 0 & 0 & 0 & 0 & 241 & 0 \\
November & 0 & 865 & 0 & 174 & 0 & 0 & 0 & 0 & 174 & 0 & 446 & 0 & 0 & 174 & 0 \\
December & 0 & 1{,}685 & 0 & 86 & 86 & 1{,}239 & 1{,}230 & 0 & 86 & 0 & 730 & 265 & 0 & 86 & 0 \\
\midrule
\textbf{Yearly} & 100 & 13{,}039 & 100 & 1{,}997 & 86 & 8{,}904 & 8{,}891 & 100 & 1{,}997 & 0 & 4{,}645 & 861 & 100 & 2{,}045 & 218 \\
\bottomrule
\end{tabular}}
\end{table*}

\begin{table*}[!htpb]
\centering
\footnotesize
\caption{\new{Month-wise BESS usage losses for the year 2019 under \emph{perfect} load and PV forecasts for each case and method. `Std' and `Track' abbreviate the Std Ref and Track Ref methods; `1st', `2nd' and `3rd' the correspondingly numbered proposed choices.}}
\label{table_breakdown_bess}
% Fixed font rather than \resizebox: scaling each table to \textwidth would
% set the narrow ones (OPDC) far larger than the wide ones (NCDC).
{\color{black}\scriptsize\setlength{\tabcolsep}{3.3pt}%
\begin{tabular}{l|rrrrr|rrrrr|rrrrr}
\toprule
 & \multicolumn{5}{c}{Case (i)} & \multicolumn{5}{c}{Case (ii)} & \multicolumn{5}{c}{Case (iii)} \\
\cmidrule(lr){2-6}\cmidrule(lr){7-11}\cmidrule(lr){12-16}
 & Std & 1st & 2nd & Track & 3rd & Std & 1st & 2nd & Track & 3rd & Std & 1st & 2nd & Track & 3rd \\
\midrule
January & 584 & 381 & 584 & 751 & 535 & 137 & 138 & 584 & 728 & 563 & 431 & 518 & 584 & 724 & 563 \\
February & 476 & 382 & 476 & 661 & 476 & 370 & 371 & 476 & 639 & 476 & 278 & 418 & 475 & 631 & 476 \\
March & 489 & 401 & 489 & 755 & 474 & 327 & 328 & 489 & 742 & 474 & 297 & 421 & 489 & 743 & 410 \\
April & 540 & 383 & 540 & 784 & 456 & 376 & 377 & 540 & 778 & 456 & 323 & 393 & 540 & 777 & 456 \\
May & 562 & 497 & 562 & 791 & 383 & 386 & 386 & 562 & 750 & 383 & 323 & 393 & 562 & 757 & 383 \\
June & 378 & 226 & 378 & 656 & 380 & 153 & 155 & 378 & 621 & 380 & 296 & 297 & 377 & 628 & 380 \\
July & 649 & 604 & 649 & 844 & 562 & 438 & 438 & 649 & 840 & 562 & 312 & 379 & 649 & 838 & 562 \\
August & 482 & 379 & 482 & 779 & 387 & 424 & 424 & 482 & 770 & 387 & 281 & 345 & 482 & 770 & 387 \\
September & 540 & 447 & 540 & 780 & 430 & 406 & 407 & 540 & 774 & 430 & 320 & 370 & 540 & 772 & 430 \\
October & 563 & 481 & 563 & 830 & 458 & 461 & 462 & 563 & 823 & 458 & 352 & 435 & 563 & 822 & 458 \\
November & 632 & 538 & 632 & 800 & 515 & 467 & 468 & 632 & 795 & 515 & 390 & 458 & 632 & 790 & 515 \\
December & 498 & 268 & 498 & 704 & 493 & 317 & 321 & 498 & 666 & 501 & 316 & 417 & 498 & 667 & 503 \\
\midrule
\textbf{Yearly} & 6{,}393 & 4{,}986 & 6{,}393 & 9{,}136 & 5{,}549 & 4{,}263 & 4{,}276 & 6{,}392 & 8{,}928 & 5{,}585 & 3{,}920 & 4{,}843 & 6{,}391 & 8{,}919 & 5{,}522 \\
\bottomrule
\end{tabular}}
\end{table*}

\begin{table*}[!htpb]
\centering
\footnotesize
\caption{\new{Month-wise energy charges for the year 2019 under \emph{perfect} load and PV forecasts for each case and method. `Std' and `Track' abbreviate the Std Ref and Track Ref methods; `1st', `2nd' and `3rd' the correspondingly numbered proposed choices.}}
\label{table_breakdown_energy}
% Fixed font rather than \resizebox: scaling each table to \textwidth would
% set the narrow ones (OPDC) far larger than the wide ones (NCDC).
{\color{black}\scriptsize\setlength{\tabcolsep}{3.3pt}%
\begin{tabular}{l|rrrrr|rrrrr|rrrrr}
\toprule
 & \multicolumn{5}{c}{Case (i)} & \multicolumn{5}{c}{Case (ii)} & \multicolumn{5}{c}{Case (iii)} \\
\cmidrule(lr){2-6}\cmidrule(lr){7-11}\cmidrule(lr){12-16}
 & Std & 1st & 2nd & Track & 3rd & Std & 1st & 2nd & Track & 3rd & Std & 1st & 2nd & Track & 3rd \\
\midrule
January & 5{,}652 & 5{,}697 & 5{,}652 & 5{,}673 & 5{,}649 & 5{,}717 & 5{,}715 & 5{,}652 & 5{,}681 & 5{,}649 & 5{,}691 & 5{,}686 & 5{,}652 & 5{,}680 & 5{,}649 \\
February & 4{,}678 & 4{,}695 & 4{,}676 & 4{,}717 & 4{,}676 & 4{,}719 & 4{,}719 & 4{,}677 & 4{,}718 & 4{,}676 & 4{,}730 & 4{,}717 & 4{,}677 & 4{,}718 & 4{,}676 \\
March & 2{,}035 & 2{,}079 & 2{,}035 & 2{,}066 & 2{,}029 & 2{,}083 & 2{,}081 & 2{,}035 & 2{,}067 & 2{,}029 & 2{,}080 & 2{,}075 & 2{,}035 & 2{,}069 & 2{,}028 \\
April & 400 & 425 & 400 & 440 & 404 & 438 & 441 & 400 & 440 & 404 & 445 & 432 & 400 & 442 & 404 \\
May & 78 & 118 & 78 & 121 & 73 & 115 & 117 & 78 & 118 & 73 & 115 & 115 & 78 & 121 & 73 \\
June & 955 & 990 & 955 & 985 & 954 & 1{,}002 & 1{,}001 & 955 & 984 & 954 & 985 & 993 & 955 & 985 & 954 \\
July & -2{,}660 & -2{,}647 & -2{,}660 & -2{,}656 & -2{,}662 & -2{,}657 & -2{,}656 & -2{,}660 & -2{,}656 & -2{,}662 & -2{,}665 & -2{,}657 & -2{,}660 & -2{,}656 & -2{,}662 \\
August & -2{,}310 & -2{,}281 & -2{,}311 & -2{,}291 & -2{,}328 & -2{,}286 & -2{,}284 & -2{,}310 & -2{,}289 & -2{,}328 & -2{,}301 & -2{,}294 & -2{,}310 & -2{,}289 & -2{,}328 \\
September & -169 & -130 & -169 & -134 & -178 & -132 & -128 & -169 & -134 & -178 & -136 & -131 & -169 & -133 & -178 \\
October & -1{,}064 & -1{,}019 & -1{,}064 & -1{,}013 & -1{,}069 & -1{,}018 & -1{,}016 & -1{,}064 & -1{,}016 & -1{,}069 & -1{,}020 & -1{,}018 & -1{,}064 & -1{,}013 & -1{,}069 \\
November & 1{,}878 & 1{,}916 & 1{,}878 & 1{,}930 & 1{,}878 & 1{,}929 & 1{,}928 & 1{,}878 & 1{,}928 & 1{,}878 & 1{,}932 & 1{,}924 & 1{,}878 & 1{,}932 & 1{,}878 \\
December & 4{,}218 & 4{,}240 & 4{,}216 & 4{,}260 & 4{,}219 & 4{,}253 & 4{,}255 & 4{,}217 & 4{,}258 & 4{,}217 & 4{,}260 & 4{,}252 & 4{,}218 & 4{,}260 & 4{,}217 \\
\midrule
\textbf{Yearly} & 13{,}691 & 14{,}084 & 13{,}686 & 14{,}100 & 13{,}645 & 14{,}163 & 14{,}172 & 13{,}689 & 14{,}100 & 13{,}643 & 14{,}115 & 14{,}094 & 13{,}690 & 14{,}116 & 13{,}643 \\
\bottomrule
\end{tabular}}
\end{table*}

\begin{table*}[!htpb]
\centering
\footnotesize
\caption{\new{Month-wise total costs for the year 2019 under \emph{perfect} load and PV forecasts for each case and method. `Std' and `Track' abbreviate the Std Ref and Track Ref methods; `1st', `2nd' and `3rd' the correspondingly numbered proposed choices.}}
\label{table_breakdown_total}
% Fixed font rather than \resizebox: scaling each table to \textwidth would
% set the narrow ones (OPDC) far larger than the wide ones (NCDC).
{\color{black}\scriptsize\setlength{\tabcolsep}{3.3pt}%
\begin{tabular}{l|rrrrr|rrrrr|rrrrr}
\toprule
 & \multicolumn{5}{c}{Case (i)} & \multicolumn{5}{c}{Case (ii)} & \multicolumn{5}{c}{Case (iii)} \\
\cmidrule(lr){2-6}\cmidrule(lr){7-11}\cmidrule(lr){12-16}
 & Std & 1st & 2nd & Track & 3rd & Std & 1st & 2nd & Track & 3rd & Std & 1st & 2nd & Track & 3rd \\
\midrule
January & 10{,}595 & 11{,}733 & 10{,}594 & 11{,}047 & 11{,}200 & 12{,}925 & 12{,}910 & 10{,}594 & 11{,}394 & 10{,}704 & 12{,}093 & 10{,}964 & 10{,}594 & 11{,}472 & 10{,}705 \\
February & 10{,}067 & 11{,}018 & 10{,}066 & 10{,}478 & 10{,}106 & 11{,}303 & 11{,}222 & 10{,}066 & 10{,}723 & 10{,}109 & 12{,}271 & 10{,}946 & 10{,}066 & 10{,}968 & 10{,}115 \\
March & 6{,}938 & 7{,}468 & 6{,}938 & 7{,}384 & 7{,}208 & 8{,}384 & 8{,}335 & 6{,}938 & 7{,}374 & 7{,}208 & 9{,}509 & 8{,}027 & 6{,}937 & 7{,}377 & 8{,}362 \\
April & 5{,}004 & 6{,}088 & 5{,}003 & 5{,}533 & 5{,}967 & 6{,}230 & 6{,}235 & 5{,}003 & 5{,}527 & 5{,}967 & 7{,}244 & 5{,}756 & 5{,}004 & 5{,}527 & 5{,}967 \\
May & 3{,}608 & 4{,}130 & 3{,}608 & 4{,}276 & 5{,}517 & 4{,}906 & 4{,}908 & 3{,}608 & 4{,}753 & 5{,}517 & 5{,}433 & 4{,}886 & 3{,}608 & 4{,}649 & 5{,}517 \\
June & 5{,}279 & 6{,}639 & 5{,}279 & 5{,}766 & 5{,}333 & 7{,}264 & 7{,}239 & 5{,}279 & 5{,}959 & 5{,}331 & 6{,}626 & 6{,}050 & 5{,}279 & 6{,}024 & 5{,}331 \\
July & 1{,}729 & 2{,}297 & 1{,}729 & 2{,}061 & 1{,}726 & 2{,}098 & 2{,}098 & 1{,}729 & 2{,}056 & 1{,}726 & 3{,}383 & 2{,}402 & 1{,}729 & 2{,}055 & 1{,}726 \\
August & 1{,}616 & 2{,}526 & 1{,}616 & 2{,}001 & 2{,}142 & 2{,}056 & 2{,}057 & 1{,}616 & 1{,}994 & 2{,}142 & 3{,}368 & 2{,}501 & 1{,}616 & 1{,}993 & 2{,}142 \\
September & 3{,}950 & 4{,}594 & 3{,}950 & 4{,}400 & 4{,}516 & 4{,}788 & 4{,}793 & 3{,}950 & 4{,}395 & 4{,}516 & 5{,}699 & 5{,}089 & 3{,}950 & 4{,}394 & 4{,}516 \\
October & 2{,}092 & 2{,}850 & 2{,}092 & 2{,}504 & 3{,}151 & 2{,}908 & 2{,}912 & 2{,}092 & 2{,}495 & 3{,}151 & 3{,}829 & 2{,}931 & 2{,}092 & 2{,}496 & 3{,}151 \\
November & 5{,}474 & 6{,}068 & 5{,}474 & 5{,}908 & 5{,}397 & 6{,}338 & 6{,}338 & 5{,}474 & 5{,}998 & 5{,}397 & 6{,}931 & 6{,}239 & 5{,}474 & 5{,}961 & 5{,}397 \\
December & 8{,}693 & 9{,}696 & 8{,}691 & 8{,}953 & 9{,}307 & 9{,}726 & 9{,}673 & 8{,}691 & 9{,}244 & 8{,}701 & 10{,}068 & 8{,}953 & 8{,}692 & 9{,}372 & 8{,}696 \\
\midrule
\textbf{Yearly} & 65{,}046 & 75{,}107 & 65{,}040 & 70{,}311 & 71{,}571 & 78{,}925 & 78{,}719 & 65{,}041 & 71{,}912 & 70{,}472 & 86{,}455 & 74{,}742 & 65{,}041 & 72{,}289 & 71{,}626 \\
\bottomrule
\end{tabular}}
\end{table*}

\new{\section{Month-wise results under imperfect forecasts}\label{appendix_huge_tables_stochastic}

\begin{figure}[!htpb]
\centering
\includegraphics[width=\columnwidth]{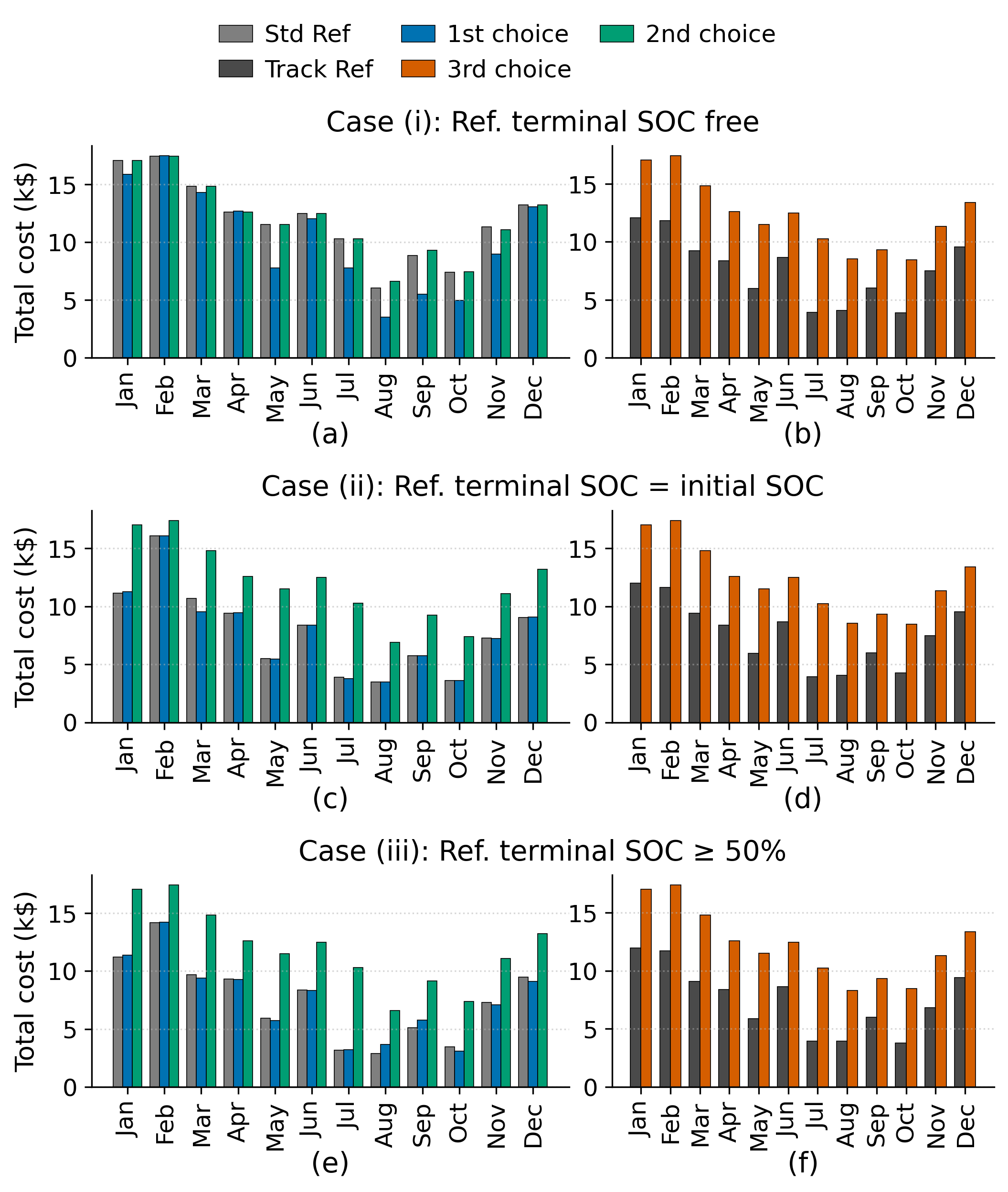}
\caption{\new{Month-wise total electricity costs for 2019 at the Port of San Diego MG under imperfect forecasts.
The left column ((a), (c), (e)) compares the Std Ref method with the $\supscr{1}{\rm st}$ and $\supscr{2}{\rm nd}$-choice methods, while the right column ((b), (d), (f)) compares the Track Ref method with the $\supscr{3}{\rm rd}$-choice method.}}
\label{fig_yearly_total_imperfect}
\end{figure}

Recursive feasibility under stochasticity of the terminal state for the $\supscr{1}{\rm st}$-choice terminal
region~\eqref{candidate_terminal_region_NCDC_OPDC} follows from Assumption~\ref{assumption_length_prediction_horizon},
on the length of the prediction horizon, together with the feasibility of the reference
trajectory (Assumption~\ref{assumption_sets}). The $\supscr{2}{\rm nd}$ and
$\supscr{3}{\rm rd}$-choice specifications take
$\mathcal{X}_f(t)=\mathcal{X}(t)$~\eqref{candidate_terminal_region_NCDC_OPDC_2}; and thus introduce no risk of infeasibility.

The SOC feasibility of the initial state to stay inside
$[\subscr{\rm SOC}{min},\subscr{\rm SOC}{max}]$ despite the forecast error is
ensured by the real-time postprocessing of~\cite[Assumption~6]{ghosh2025adaptive}: the BESS
dispatch $u_{1}$ and the grid import $u_{2}$ are mutually coupled control sources acting on
the SOC, the BESS being the primary source and the grid import the secondary one. Once the
net load is realized, the mismatch with respect to its forecast is absorbed first by the BESS until its SOC/power limits, and then by the grid
import, so the realized SOC is corrected back into
the admissible band at every time-step. Consequently, $x(t)\in\mathcal{X}(t)$, $\forall t \in
\mathcal{T}$, and the optimization is always initialized from an admissible state.}

\new{Fig.~\ref{fig_yearly_total_imperfect} reports the month-wise total electricity costs behind Table~\ref{table_yearly_savings_imperfect}, and Tables~\ref{table_breakdown_ncdc_stochastic}--\ref{table_breakdown_total_stochastic} break them down further, one table per cost component, under the imperfect load and PV forecasts of~\cite{ghosh2025adaptive}. The \emph{Yearly} row of Table~\ref{table_breakdown_total_stochastic} reproduces Table~\ref{table_yearly_savings_imperfect}. }

\begin{table*}[!htpb]
\centering
\footnotesize
\caption{\new{Month-wise NCDC for the year 2019 under \emph{imperfect} load and PV forecasts for each case and method. `Std' and `Track' abbreviate the Std Ref and Track Ref methods; `1st', `2nd' and `3rd' the correspondingly numbered proposed choices.}}
\label{table_breakdown_ncdc_stochastic}
% Fixed font rather than \resizebox: scaling each table to \textwidth would
% set the narrow ones (OPDC) far larger than the wide ones (NCDC).
{\color{black}\scriptsize\setlength{\tabcolsep}{3.3pt}%
\begin{tabular}{l|rrrrr|rrrrr|rrrrr}
\toprule
 & \multicolumn{5}{c}{Case (i)} & \multicolumn{5}{c}{Case (ii)} & \multicolumn{5}{c}{Case (iii)} \\
\cmidrule(lr){2-6}\cmidrule(lr){7-11}\cmidrule(lr){12-16}
 & Std & 1st & 2nd & Track & 3rd & Std & 1st & 2nd & Track & 3rd & Std & 1st & 2nd & Track & 3rd \\
\midrule
January & 6{,}272 & 5{,}952 & 6{,}272 & 4{,}857 & 6{,}272 & 4{,}705 & 4{,}705 & 6{,}272 & 5{,}210 & 6{,}272 & 4{,}730 & 4{,}821 & 6{,}272 & 4{,}705 & 6{,}272 \\
February & 7{,}055 & 7{,}055 & 7{,}055 & 6{,}094 & 7{,}055 & 6{,}272 & 6{,}272 & 7{,}055 & 6{,}037 & 7{,}055 & 6{,}272 & 6{,}272 & 7{,}055 & 6{,}272 & 7{,}055 \\
March & 7{,}055 & 7{,}055 & 7{,}055 & 6{,}426 & 7{,}055 & 7{,}055 & 7{,}055 & 7{,}055 & 6{,}573 & 7{,}055 & 7{,}055 & 6{,}878 & 7{,}055 & 6{,}272 & 7{,}055 \\
April & 7{,}055 & 7{,}055 & 7{,}055 & 7{,}055 & 7{,}055 & 7{,}055 & 7{,}055 & 7{,}055 & 7{,}055 & 7{,}055 & 7{,}055 & 7{,}055 & 7{,}055 & 7{,}055 & 7{,}055 \\
May & 6{,}272 & 5{,}833 & 6{,}272 & 5{,}194 & 6{,}272 & 4{,}998 & 4{,}952 & 6{,}272 & 5{,}196 & 6{,}272 & 5{,}447 & 5{,}227 & 6{,}272 & 5{,}099 & 6{,}272 \\
June & 7{,}055 & 7{,}055 & 7{,}055 & 7{,}055 & 7{,}055 & 7{,}055 & 7{,}055 & 7{,}055 & 7{,}055 & 7{,}055 & 7{,}055 & 6{,}982 & 7{,}055 & 7{,}055 & 7{,}055 \\
July & 7{,}838 & 7{,}838 & 7{,}838 & 5{,}869 & 7{,}838 & 5{,}059 & 4{,}851 & 7{,}838 & 5{,}869 & 7{,}838 & 5{,}488 & 5{,}488 & 7{,}838 & 5{,}893 & 7{,}838 \\
August & 6{,}272 & 5{,}431 & 6{,}272 & 5{,}798 & 6{,}272 & 5{,}431 & 5{,}431 & 6{,}272 & 5{,}779 & 6{,}272 & 4{,}853 & 5{,}605 & 6{,}272 & 5{,}637 & 6{,}272 \\
September & 6{,}272 & 5{,}254 & 6{,}272 & 5{,}476 & 6{,}272 & 5{,}024 & 5{,}024 & 6{,}272 & 5{,}476 & 6{,}272 & 4{,}846 & 5{,}476 & 6{,}272 & 5{,}476 & 6{,}272 \\
October & 4{,}705 & 4{,}705 & 4{,}705 & 4{,}152 & 5{,}488 & 3{,}447 & 3{,}447 & 4{,}705 & 4{,}533 & 5{,}488 & 4{,}097 & 3{,}669 & 4{,}705 & 4{,}065 & 5{,}488 \\
November & 5{,}488 & 4{,}705 & 5{,}238 & 4{,}705 & 5{,}488 & 4{,}705 & 4{,}705 & 5{,}279 & 4{,}705 & 5{,}488 & 4{,}705 & 4{,}705 & 5{,}240 & 4{,}133 & 5{,}488 \\
December & 5{,}097 & 5{,}097 & 5{,}097 & 4{,}469 & 5{,}293 & 4{,}294 & 4{,}333 & 5{,}097 & 4{,}474 & 5{,}293 & 4{,}645 & 4{,}378 & 5{,}097 & 4{,}370 & 5{,}293 \\
\midrule
\textbf{Yearly} & 76{,}436 & 73{,}035 & 76{,}185 & 67{,}151 & 77{,}415 & 65{,}101 & 64{,}885 & 76{,}226 & 67{,}962 & 77{,}415 & 66{,}249 & 66{,}557 & 76{,}187 & 66{,}032 & 77{,}415 \\
\bottomrule
\end{tabular}}
\end{table*}

\begin{table*}[!htpb]
\centering
\footnotesize
\caption{\new{Month-wise OPDC for the year 2019 under \emph{imperfect} load and PV forecasts for each case and method. `Std' and `Track' abbreviate the Std Ref and Track Ref methods; `1st', `2nd' and `3rd' the correspondingly numbered proposed choices.}}
\label{table_breakdown_opdc_stochastic}
% Fixed font rather than \resizebox: scaling each table to \textwidth would
% set the narrow ones (OPDC) far larger than the wide ones (NCDC).
{\color{black}\scriptsize\setlength{\tabcolsep}{3.3pt}%
\begin{tabular}{l|rrrrr|rrrrr|rrrrr}
\toprule
 & \multicolumn{5}{c}{Case (i)} & \multicolumn{5}{c}{Case (ii)} & \multicolumn{5}{c}{Case (iii)} \\
\cmidrule(lr){2-6}\cmidrule(lr){7-11}\cmidrule(lr){12-16}
 & Std & 1st & 2nd & Track & 3rd & Std & 1st & 2nd & Track & 3rd & Std & 1st & 2nd & Track & 3rd \\
\midrule
January & 4{,}916 & 3{,}914 & 4{,}916 & 837 & 4{,}916 & 305 & 429 & 4{,}916 & 431 & 4{,}916 & 345 & 399 & 4{,}916 & 882 & 4{,}916 \\
February & 5{,}531 & 5{,}531 & 5{,}531 & 373 & 5{,}531 & 4{,}916 & 4{,}916 & 5{,}531 & 265 & 5{,}531 & 2{,}910 & 2{,}951 & 5{,}531 & 93 & 5{,}531 \\
March & 5{,}531 & 4{,}916 & 5{,}531 & 82 & 5{,}531 & 1{,}201 & 0 & 5{,}531 & 121 & 5{,}531 & 128 & 0 & 5{,}523 & 89 & 5{,}531 \\
April & 4{,}916 & 4{,}916 & 4{,}916 & 203 & 4{,}916 & 1{,}484 & 1{,}508 & 4{,}916 & 203 & 4{,}916 & 1{,}377 & 1{,}310 & 4{,}916 & 203 & 4{,}916 \\
May & 4{,}916 & 1{,}492 & 4{,}916 & 0 & 4{,}916 & 0 & 0 & 4{,}916 & 0 & 4{,}916 & 0 & 0 & 4{,}916 & 0 & 4{,}916 \\
June & 4{,}302 & 3{,}765 & 4{,}302 & 90 & 4{,}302 & 0 & 0 & 4{,}302 & 86 & 4{,}302 & 0 & 0 & 4{,}302 & 77 & 4{,}302 \\
July & 4{,}916 & 2{,}265 & 4{,}916 & 0 & 4{,}916 & 1{,}110 & 1{,}183 & 4{,}916 & 0 & 4{,}916 & 0 & 0 & 4{,}916 & 0 & 4{,}916 \\
August & 1{,}907 & 87 & 2{,}494 & 0 & 4{,}424 & 0 & 0 & 2{,}768 & 0 & 4{,}424 & 0 & 0 & 2{,}448 & 0 & 4{,}223 \\
September & 2{,}481 & 0 & 2{,}951 & 0 & 3{,}033 & 476 & 460 & 2{,}903 & 0 & 3{,}033 & 0 & 0 & 2{,}799 & 0 & 3{,}033 \\
October & 3{,}392 & 911 & 3{,}417 & 0 & 3{,}688 & 771 & 769 & 3{,}410 & 0 & 3{,}688 & 0 & 0 & 3{,}386 & 0 & 3{,}688 \\
November & 3{,}688 & 2{,}003 & 3{,}688 & 124 & 3{,}688 & 160 & 145 & 3{,}688 & 122 & 3{,}688 & 214 & 0 & 3{,}688 & 50 & 3{,}688 \\
December & 3{,}688 & 3{,}420 & 3{,}688 & 119 & 3{,}688 & 0 & 0 & 3{,}688 & 114 & 3{,}688 & 118 & 0 & 3{,}688 & 121 & 3{,}688 \\
\midrule
\textbf{Yearly} & 50{,}185 & 33{,}220 & 51{,}268 & 1{,}829 & 53{,}551 & 10{,}425 & 9{,}410 & 51{,}486 & 1{,}342 & 53{,}551 & 5{,}092 & 4{,}659 & 51{,}031 & 1{,}514 & 53{,}351 \\
\bottomrule
\end{tabular}}
\end{table*}

\begin{table*}[!htpb]
\centering
\footnotesize
\caption{\new{Month-wise BESS usage losses for the year 2019 under \emph{imperfect} load and PV forecasts for each case and method. `Std' and `Track' abbreviate the Std Ref and Track Ref methods; `1st', `2nd' and `3rd' the correspondingly numbered proposed choices.}}
\label{table_breakdown_bess_stochastic}
% Fixed font rather than \resizebox: scaling each table to \textwidth would
% set the narrow ones (OPDC) far larger than the wide ones (NCDC).
{\color{black}\scriptsize\setlength{\tabcolsep}{3.3pt}%
\begin{tabular}{l|rrrrr|rrrrr|rrrrr}
\toprule
 & \multicolumn{5}{c}{Case (i)} & \multicolumn{5}{c}{Case (ii)} & \multicolumn{5}{c}{Case (iii)} \\
\cmidrule(lr){2-6}\cmidrule(lr){7-11}\cmidrule(lr){12-16}
 & Std & 1st & 2nd & Track & 3rd & Std & 1st & 2nd & Track & 3rd & Std & 1st & 2nd & Track & 3rd \\
\midrule
January & 227 & 367 & 227 & 726 & 206 & 499 & 495 & 221 & 715 & 206 & 476 & 489 & 227 & 725 & 203 \\
February & 146 & 164 & 144 & 624 & 126 & 189 & 192 & 142 & 623 & 126 & 224 & 225 & 143 & 626 & 126 \\
March & 201 & 282 & 202 & 645 & 204 & 361 & 435 & 202 & 644 & 207 & 408 & 435 & 202 & 652 & 205 \\
April & 192 & 266 & 194 & 653 & 192 & 419 & 422 & 194 & 656 & 192 & 395 & 417 & 194 & 654 & 192 \\
May & 242 & 349 & 245 & 655 & 234 & 400 & 403 & 241 & 644 & 234 & 372 & 391 & 241 & 658 & 234 \\
June & 202 & 248 & 203 & 544 & 196 & 352 & 353 & 202 & 555 & 196 & 335 & 346 & 202 & 556 & 196 \\
July & 232 & 368 & 233 & 732 & 193 & 399 & 402 & 232 & 719 & 193 & 349 & 402 & 233 & 723 & 193 \\
August & 210 & 322 & 207 & 599 & 170 & 330 & 337 & 206 & 589 & 170 & 337 & 335 & 207 & 608 & 170 \\
September & 249 & 399 & 248 & 657 & 183 & 363 & 369 & 249 & 650 & 183 & 385 & 394 & 249 & 655 & 183 \\
October & 332 & 341 & 333 & 753 & 320 & 329 & 331 & 332 & 740 & 320 & 357 & 383 & 333 & 743 & 320 \\
November & 251 & 349 & 255 & 719 & 250 & 439 & 443 & 253 & 713 & 251 & 413 & 443 & 253 & 717 & 250 \\
December & 194 & 277 & 196 & 682 & 177 & 470 & 467 & 195 & 670 & 177 & 431 & 465 & 195 & 669 & 177 \\
\midrule
\textbf{Yearly} & 2{,}678 & 3{,}732 & 2{,}687 & 7{,}990 & 2{,}452 & 4{,}551 & 4{,}650 & 2{,}670 & 7{,}919 & 2{,}456 & 4{,}480 & 4{,}726 & 2{,}680 & 7{,}985 & 2{,}448 \\
\bottomrule
\end{tabular}}
\end{table*}

\begin{table*}[!htpb]
\centering
\footnotesize
\caption{\new{Month-wise energy charges for the year 2019 under \emph{imperfect} load and PV forecasts for each case and method. `Std' and `Track' abbreviate the Std Ref and Track Ref methods; `1st', `2nd' and `3rd' the correspondingly numbered proposed choices.}}
\label{table_breakdown_energy_stochastic}
% Fixed font rather than \resizebox: scaling each table to \textwidth would
% set the narrow ones (OPDC) far larger than the wide ones (NCDC).
{\color{black}\scriptsize\setlength{\tabcolsep}{3.3pt}%
\begin{tabular}{l|rrrrr|rrrrr|rrrrr}
\toprule
 & \multicolumn{5}{c}{Case (i)} & \multicolumn{5}{c}{Case (ii)} & \multicolumn{5}{c}{Case (iii)} \\
\cmidrule(lr){2-6}\cmidrule(lr){7-11}\cmidrule(lr){12-16}
 & Std & 1st & 2nd & Track & 3rd & Std & 1st & 2nd & Track & 3rd & Std & 1st & 2nd & Track & 3rd \\
\midrule
January & 5{,}644 & 5{,}644 & 5{,}644 & 5{,}648 & 5{,}644 & 5{,}654 & 5{,}651 & 5{,}644 & 5{,}651 & 5{,}644 & 5{,}651 & 5{,}650 & 5{,}644 & 5{,}647 & 5{,}644 \\
February & 4{,}691 & 4{,}701 & 4{,}695 & 4{,}719 & 4{,}688 & 4{,}712 & 4{,}712 & 4{,}690 & 4{,}721 & 4{,}687 & 4{,}754 & 4{,}760 & 4{,}690 & 4{,}722 & 4{,}687 \\
March & 2{,}029 & 2{,}038 & 2{,}032 & 2{,}055 & 2{,}026 & 2{,}089 & 2{,}067 & 2{,}030 & 2{,}054 & 2{,}026 & 2{,}068 & 2{,}069 & 2{,}030 & 2{,}058 & 2{,}026 \\
April & 421 & 424 & 422 & 457 & 422 & 463 & 462 & 421 & 458 & 422 & 474 & 471 & 421 & 459 & 422 \\
May & 81 & 87 & 82 & 112 & 78 & 122 & 122 & 80 & 113 & 78 & 120 & 120 & 80 & 115 & 78 \\
June & 929 & 934 & 929 & 967 & 929 & 984 & 984 & 929 & 968 & 929 & 974 & 977 & 929 & 968 & 929 \\
July & -2{,}711 & -2{,}708 & -2{,}711 & -2{,}677 & -2{,}711 & -2{,}683 & -2{,}682 & -2{,}711 & -2{,}677 & -2{,}711 & -2{,}685 & -2{,}681 & -2{,}711 & -2{,}675 & -2{,}711 \\
August & -2{,}343 & -2{,}332 & -2{,}345 & -2{,}302 & -2{,}340 & -2{,}287 & -2{,}287 & -2{,}344 & -2{,}302 & -2{,}340 & -2{,}296 & -2{,}289 & -2{,}345 & -2{,}301 & -2{,}340 \\
September & -170 & -164 & -170 & -134 & -170 & -107 & -108 & -170 & -134 & -170 & -127 & -121 & -171 & -132 & -170 \\
October & -1{,}029 & -1{,}015 & -1{,}029 & -1{,}012 & -1{,}032 & -958 & -961 & -1{,}032 & -1{,}014 & -1{,}037 & -995 & -987 & -1{,}032 & -1{,}011 & -1{,}032 \\
November & 1{,}904 & 1{,}915 & 1{,}908 & 1{,}935 & 1{,}899 & 1{,}952 & 1{,}953 & 1{,}902 & 1{,}934 & 1{,}899 & 1{,}944 & 1{,}947 & 1{,}902 & 1{,}936 & 1{,}900 \\
December & 4{,}235 & 4{,}239 & 4{,}236 & 4{,}271 & 4{,}234 & 4{,}274 & 4{,}274 & 4{,}234 & 4{,}270 & 4{,}234 & 4{,}272 & 4{,}272 & 4{,}234 & 4{,}271 & 4{,}234 \\
\midrule
\textbf{Yearly} & 13{,}679 & 13{,}765 & 13{,}694 & 14{,}039 & 13{,}665 & 14{,}215 & 14{,}187 & 13{,}672 & 14{,}044 & 13{,}660 & 14{,}153 & 14{,}190 & 13{,}671 & 14{,}058 & 13{,}666 \\
\bottomrule
\end{tabular}}
\end{table*}

\begin{table*}[!htpb]
\centering
\footnotesize
\caption{\new{Month-wise total costs for the year 2019 under \emph{imperfect} load and PV forecasts for each case and method. `Std' and `Track' abbreviate the Std Ref and Track Ref methods; `1st', `2nd' and `3rd' the correspondingly numbered proposed choices.}}
\label{table_breakdown_total_stochastic}
% Fixed font rather than \resizebox: scaling each table to \textwidth would
% set the narrow ones (OPDC) far larger than the wide ones (NCDC).
{\color{black}\scriptsize\setlength{\tabcolsep}{2.2pt}% 6-digit totals: tighter than the sibling tables to stay inside \textwidth
\begin{tabular}{l|rrrrr|rrrrr|rrrrr}
\toprule
 & \multicolumn{5}{c}{Case (i)} & \multicolumn{5}{c}{Case (ii)} & \multicolumn{5}{c}{Case (iii)} \\
\cmidrule(lr){2-6}\cmidrule(lr){7-11}\cmidrule(lr){12-16}
 & Std & 1st & 2nd & Track & 3rd & Std & 1st & 2nd & Track & 3rd & Std & 1st & 2nd & Track & 3rd \\
\midrule
January & 17{,}059 & 15{,}878 & 17{,}059 & 12{,}067 & 17{,}038 & 11{,}163 & 11{,}280 & 17{,}053 & 12{,}007 & 17{,}038 & 11{,}202 & 11{,}359 & 17{,}059 & 11{,}959 & 17{,}035 \\
February & 17{,}422 & 17{,}450 & 17{,}425 & 11{,}810 & 17{,}399 & 16{,}089 & 16{,}092 & 17{,}417 & 11{,}646 & 17{,}399 & 14{,}159 & 14{,}208 & 17{,}418 & 11{,}713 & 17{,}399 \\
March & 14{,}816 & 14{,}292 & 14{,}820 & 9{,}208 & 14{,}815 & 10{,}707 & 9{,}557 & 14{,}817 & 9{,}393 & 14{,}818 & 9{,}658 & 9{,}383 & 14{,}810 & 9{,}071 & 14{,}816 \\
April & 12{,}585 & 12{,}662 & 12{,}587 & 8{,}369 & 12{,}585 & 9{,}421 & 9{,}447 & 12{,}586 & 8{,}373 & 12{,}585 & 9{,}301 & 9{,}253 & 12{,}586 & 8{,}372 & 12{,}585 \\
May & 11{,}511 & 7{,}761 & 11{,}515 & 5{,}961 & 11{,}500 & 5{,}521 & 5{,}478 & 11{,}509 & 5{,}953 & 11{,}500 & 5{,}939 & 5{,}738 & 11{,}509 & 5{,}871 & 11{,}500 \\
June & 12{,}489 & 12{,}002 & 12{,}489 & 8{,}657 & 12{,}483 & 8{,}391 & 8{,}392 & 12{,}489 & 8{,}664 & 12{,}482 & 8{,}364 & 8{,}305 & 12{,}489 & 8{,}656 & 12{,}482 \\
July & 10{,}275 & 7{,}763 & 10{,}277 & 3{,}925 & 10{,}237 & 3{,}885 & 3{,}755 & 10{,}276 & 3{,}911 & 10{,}237 & 3{,}152 & 3{,}209 & 10{,}277 & 3{,}941 & 10{,}237 \\
August & 6{,}044 & 3{,}509 & 6{,}628 & 4{,}094 & 8{,}527 & 3{,}474 & 3{,}480 & 6{,}902 & 4{,}067 & 8{,}527 & 2{,}894 & 3{,}651 & 6{,}582 & 3{,}944 & 8{,}326 \\
September & 8{,}832 & 5{,}489 & 9{,}301 & 5{,}999 & 9{,}318 & 5{,}756 & 5{,}745 & 9{,}254 & 5{,}992 & 9{,}318 & 5{,}103 & 5{,}749 & 9{,}149 & 5{,}998 & 9{,}318 \\
October & 7{,}400 & 4{,}943 & 7{,}426 & 3{,}894 & 8{,}464 & 3{,}590 & 3{,}586 & 7{,}415 & 4{,}259 & 8{,}460 & 3{,}459 & 3{,}065 & 7{,}391 & 3{,}797 & 8{,}464 \\
November & 11{,}331 & 8{,}972 & 11{,}089 & 7{,}483 & 11{,}326 & 7{,}257 & 7{,}246 & 11{,}123 & 7{,}474 & 11{,}326 & 7{,}276 & 7{,}095 & 11{,}083 & 6{,}836 & 11{,}326 \\
December & 13{,}214 & 13{,}033 & 13{,}217 & 9{,}541 & 13{,}392 & 9{,}038 & 9{,}073 & 13{,}214 & 9{,}527 & 13{,}392 & 9{,}466 & 9{,}116 & 13{,}214 & 9{,}431 & 13{,}392 \\
\midrule
\textbf{Yearly} & 142{,}978 & 123{,}752 & 143{,}834 & 91{,}009 & 147{,}084 & 94{,}291 & 93{,}132 & 144{,}054 & 91{,}266 & 147{,}083 & 89{,}974 & 90{,}132 & 143{,}568 & 89{,}589 & 146{,}880 \\
\bottomrule
\end{tabular}}
\end{table*}

\end{document}